\documentclass[12pt,a4]{report}
\usepackage{epsfig}
\usepackage{psfig}
\usepackage{axodraw}
\newcommand{\nc}{\newcommand}
\nc{\renc}{\renewcommand}
\newlength{\undereqskip}
\nc{\be}{\begin{equation}}
\nc{\bea}{\begin{eqnarray}}
\nc{\eea}{\vspace{\undereqskip}\end{eqnarray}}
\nc{\ee}{\vspace{\undereqskip}\end{equation}}
\nc{\bc}{\begin{center}}
\nc{\ec}{\end{center}}
\nc{\ba}{\begin{array}}
\nc{\ea}{\end{array}}
\nc{\inv}[1]{\frac{1}{#1}}
\nc{\dbar}[2]{\frac{d\,^{#1}#2}{(2\pi)^{#1}}}

\def\Im{{\rm Im}\hskip2pt}
\def\simleq{\; \raise0.3ex\hbox{$<$\kern-0.75em \raise-1.1ex\hbox{$\sim$}}\; }
\def\simgeq{\; \raise0.3ex\hbox{$>$\kern-0.75em \raise-1.1ex\hbox{$\sim$}}\; }
\def\dfrac#1#2{{\displaystyle\frac{#1}{#2}}}
\nc{\Tr}{{\rm Tr\,}}
\def\MeV{{\mathrm{MeV}}}
\def\GeV{{\mathrm{GeV}}}
\def\half{\frac 1 2}

\nc{\sign}{{\epsilon}} \nc{\g}{\gamma}
\nc{\mf}{magnetic field }
\nc{\mfs}{magnetic fields }
\nc{\aaa}[3]{{\ Astron.\ Astroph.\ }{{\bf #1},({#2}){#3}}}
\nc{\advp}[3]{{Adv.\ in\ Phys.\ }{{\bf #1}({#2}){#3}}}
\nc{\apl}[3]{{Appl. Phys. Lett. }{{\bf #1}{(#2)}{#3}}}
\nc{\apj}[3]{{Astrophys.\ J.\ }{{\bf #1} {(#2)} {#3}}}
\nc{\apjl}[3]{{Astrophys.\ J.\ Lett.\ }{{\bf #1} {(#2)} {#3}}}
\nc{\app}[3]{{\ Astrop.\ Phys..\ }{{\bf #1}, {(#2)} {#3}}}
\nc{\asp}[3]{{\ Astropart.\ Phys.\ }{{\bf #1} {(#2)} {#3}}}

\nc{\cmp}[3]{{  Comm.\ Math.\ Phys.\ }{{ \bf #1} {(#2)} {#3}}}
\nc{\cqg}[3]{{  Class.\ Quant.\ Grav.\ }{{\bf #1} {(#2)} {#3}}}
\nc{\epl}[3]{{  Europhys.\ Lett.\ }{{\bf #1} {(#2)} {#3}}}
\nc{\ijmp}[3]{{ Int.\ J.\ Mod.\ Phys.\ }{{\bf #1} {(#2)} {#3}}}
\nc{\ijtp}[3]{{ Int.\ J.\ Theor.\ Phys.\ }{{\bf #1} {(#2)} {#3}}}
\nc{\jhep}[3]{{ JHEP\ }{{\bf #1} {(#2)} {#3}}}
\nc{\jmp}[3]{{  J.\ Math.\ Phys.\ }{{ \bf #1} {(#2)} {#3}}}
\nc{\jpa}[3]{{  J.\ Phys.\ A\ }{{\bf #1} {(#2)} {#3}}}
\nc{\jpc}[3]{{  J.\ Phys.\ C\ }{{\bf #1} {(#2)} {#3}}}
\nc{\jpg}[3]{{ J.~Phys.~G:~Nucl.~Part.~Phys.~}{{\bf #1} {(#2)} {#3}}}
\nc{\jap}[3]{{ J.\ Appl.\ Phys.\ }{{\bf #1} {(#2)} {#3}}}
\nc{\jpsj}[3]{{ J.\ Phys.\ Soc.\ Japan\ }{{\bf #1} {(#2)} {#3}}}
\nc{\lmp}[3]{{ Lett.\ Math.\ Phys.\ }{{\bf #1} {(#2)} {#3}}}
\nc{\lncim}[3]{{ Lett.\ Nuov.\ Cim.\ }{{\bf #1} {(#2)} {#3}}}
\nc{\mpl}[3]{{ Mod.\ Phys.\ Lett.\ }{{\bf #1} {(#2)} {#3}}}
\nc{\nat}[3]{{  Nature \ }{{\bf #1} {(#2)} {#3}}}
\nc{\ncim}[3]{{  Nuov.\ Cim.\ }{{\bf #1} {(#2)} {#3}}}
\nc{\npb}[3]{{ Nucl.\ Phys.\ }{{\bf B#1} {(#2)} {#3}}}
\nc{\pr}[3]{{ Phys.\ Rev.\ }{{\bf #1} {(#2)} {#3}}}
\nc{\pra}[3]{{  Phys.\ Rev.\ }{{\bf A#1} {(#2)} {#3}}}
\nc{\prb}[3]{{  Phys.\ Rev.\ }{{\bf B#1} {(#2)} {#3}}}
\nc{\prc}[3]{{  Phys.\ Rev.\ }{{\bf C#1} {(#2)} {#3}}}
\nc{\prd}[3]{{  Phys.\ Rev.\ }{{\bf D#1} {(#2)} {#3}}}
\nc{\prl}[3]{{ Phys.\ Rev.\ Lett.\ }{{\bf #1} {(#2)} {#3}}}
\nc{\plb}[3]{{  Phys.\ Lett.\ }{{\bf B#1} {(#2)} {#3}}}
\nc{\prep}[3]{{ Phys.\ Rep.\ }{{\bf #1} {(#2)} {#3}}}
\nc{\prsl}[3]{{ Proc.\ R.\ Soc.\ London\ }{{\bf #1} {(#2)} {#3}}}
\nc{\ptp}[3]{{  Prog.\ Theor.\ Phys.\ }{{\bf #1} {(#2)} {#3}}}
\nc{\ptps}[3]{{ Prog\ Theor.\ Phys.\ suppl.\ }{{\bf #1} {(#2)} {#3}}}
\nc{\physa}[3]{{ Physica\ A\ }{{\bf #1} {(#2)} {#3}}}
\nc{\physb}[3]{{ Physica\ B\ }{{\bf #1} {(#2)} {#3}}}
\nc{\phys}[3]{{ Physica\ }{{\bf #1} {(#2)} {#3}}}
\nc{\rmp}[3]{{ Rev.\ Mod.\ Phys.\ }{{\bf #1} {(#2)} {#3}}}
\nc{\rpp}[3]{{ Rep.\ Prog.\ Phys.\ }{{\bf #1} {(#2)} {#3}}}
\nc{\sjnp}[3]{{ Sov.\ J.\ Nucl.\ Phys.\ }{{\bf #1} {(#2)} {#3}}}
\nc{\jetp}[3]{{ JETP\ }{{\bf #1} {(#2)} {#3}}}
\nc{\yf}[3]{{ Yad.\ Fiz.\ }{{\bf #1} {(#2)} {#3}}}
\nc{\zetp}[3]{{ Zh.\ Eksp.\ Teor.\ Fiz.\ }{{\bf #1} {(#2)} {#3}}}
\nc{\zp}[3]{{ Z.\ Phys.\ }{{\bf #1} {(#2)} {#3}}}
\nc{\zpc}[3]{{ Z.\ Phys.\ C\ }{{\bf #1} {(#2)} {#3}}}
\nc{\ibid}[3]{{\sl ibid.\ }{{\bf #1} {#2} {#3}}}

\begin{document}

\begin{titlepage}

\begin{center}
{\LARGE Magnetic Fields in the Early Universe}
\vskip2.cm
{\large Dario Grasso$^a$\footnote{dario.grasso@pd.infn.it}
and Hector R. Rubinstein$^b$\footnote{rub@physto.se}}
\vskip1.cm
$^1$ {\it Dipartimento di Fisica ``G.~Galilei'', Universit\`a di Padova, \\
Via Marzolo, 8,
I-35131 Padova, Italy,\\ and \\I.N.F.N. Sezione di Padova.}\\
$^2$ {\it Department of Theoretical Physics, Uppsala University,
\\Box 803, S-751 08 Uppsala, Sweden,\\ and \\
Fysikum, Stockholm University, Box 6730, 113 85  Stockholm, Sweden.}
\end{center}
\begin{abstract}
This review concerns the origin and the possible effects of \mfs
in the early Universe.
We start by providing to the reader with a short overview of the current
state of art of observations of cosmic magnetic fields.  We then illustrate
the arguments in favour of a primordial origin of magnetic fields in the
galaxies and in the clusters of galaxies.
We argue that the most promising way to test this hypothesis is to look
for possible imprints of magnetic fields on the temperature and polarization
anisotropies of the cosmic microwave background radiation (CMBR).
With this purpose in mind, we provide a review of the most relevant effects
of \mfs on the CMBR.
A long chapter of this review is dedicated to particle physics inspired models
which predict the generation of magnetic fields during the early Universe
evolution.
Although it is still unclear if any of these models can really explain the
origin of galactic and intergalactic magnetic fields, we show that interesting
effects may arise anyhow. Among these effects, we discuss the consequences of
strong magnetic fields on  the big-bang nucleosynthesis,
on the masses and couplings of the matter constituents, on the electroweak
phase transition, and on the baryon and lepton number violating sphaleron
processes.
Several intriguing common aspects, and possible interplay, of magnetogenesis
and baryogenesis are also dicussed.
\end{abstract}

\end{titlepage}
\newpage

\pagenumbering{arabic}

\tableofcontents

\newpage

\section*{Introduction}

\addcontentsline{toc}{chapter}{Introduction}

Magnetic fields are pervading. Planets, stars, galaxies and
clusters of galaxies have been observed that carry fields
that are large and extensive.
Though strong homogeneous fields are ruled out by the uniformity of
the cosmic background radiation, large domains with uniform fields
are possible.

A crucial ingredient for the survival of \mfs on astrophysical
scales is for them to live in a medium with a high electrical
conductivity. As we shall see in Chap.\ref{chap:chap1}, this
condition is comfortably fulfilled for the cosmic medium during
most of the evolution of the Universe. As a consequence, it is possible
for \mfs generated during the big-bang or later to have survived until
today as a relic.

To establish the existence and properties of primeval
\mfs would be of extreme importance for cosmology. Magnetic fields
may have affected a number of relevant processes which took place
in the early Universe as well as the Universe geometry itself.
Because of the Universe high conductivity, two important
quantities are almost conserved during Universe evolution: the
magnetic flux and the magnetic helicity (see
Sec.\ref{sec:evolution}). As we will see, valuable information
about fundamental physics which took place before the
recombination time may be encoded in these quantities.

In the past years a considerable amount of work has been done
about cosmic \mfs both on the astrophysical and the particle
physics point of view. The main motivations of such wide interest
are the following.
 \vskip0.5cm \noindent
The origin of the \mfs observed in the galaxies and in the
clusters of galaxies is unknown. This is an outstanding problem in
modern cosmology and, historically, it was the first motivation to
look for a primordial origin of magnetic fields. Some elaborated
magnetohydrodynamical (MHD) mechanisms have been proposed to
amplify very weak \mfs into the $\mu$G fields generally observed
in galaxies (see Sec.\ref{sec:observations}). These mechanisms,
known as {\it galactic dynamo}, are based on the conversion  of
the kinetic energy of the turbulent motion of the conductive
interstellar medium into magnetic energy. Today, the efficiency of
such a kind of MHD engines has been  put in question both by improved
theoretical work and new observations of \mfs in high redshift
galaxies (see Sec.\ref{sec:dynamo}). As a consequence, the
mechanism responsible for the origin of galactic \mfs has probably
to be looked back in the remote past, at least at a time
comparable to that of galaxy formation. Furthermore, even if the
galactic dynamo was effective, the origin of the seed fields which
initiated the processes has still to be identified.

Even more mysterious is the origin of \mfs in galaxy clusters.
These fields have been observed to have strength and coherence
size comparable to, and in some cases larger than,
galactic fields. In the standard cold dark matter (CDM)
scenario of structure formation clusters form by aggregation of
galaxies. It is now understood that \mfs in the
inter-cluster medium (ICM) cannot form from ejection of the
galactic fields (see Sec.\ref{sec:dynamo}). Therefore, a common astrophysical
origin of both types of fields seems to be excluded. Although,
independent astrophysical mechanisms have been proposed for the
generation of \mfs in galaxies and clusters, a more economical,
and conceptually satisfying solution would be to look for a
common cosmological origin.
 \vskip0.5cm \noindent
Magnetic fields could have played a significant role in
structure formation. It may be not a coincidence that
primordial \mfs as those required to explain  galactic fields,
without having to appeal to a MHD amplification, would also produce
pre-recombination density perturbations on protogalactic scales.
These effects go in the right direction to
solve one of the major problems of the CDM scenario of structure
formation (see Sec.\ref{sec:structures}). Furthermore, if
primordial \mfs affected structure formation they also probably
left detectable imprints in the temperature and polarization
anisotropies, or the thermal spectrum, of the cosmic microwave
background radiation (CMBR) (see Chap.\ref{chap:cmb}).
 \vskip0.5cm \noindent
 Field theory provides several clues about the physical
mechanisms which may have  produced  \mfs  in the early Universe.
Typically, magnetogenesis requires an out-of-thermal equilibrium
condition and a macroscopic parity violation. These conditions
could be naturally provided by those phase transitions which
presumably took place during the big-bang. Well known examples are
the QCD (or quark confinement) phase transition, the electroweak
(EW) phase transition,  the GUT phase transition. During these
transitions magnetic fields can be either generated by the
turbulent motion induced in the ambient plasma by the rapid
variation of some thermodynamic quantities (if the transition is
first order) or by the dynamics of the Higgs and gauge fields. In
the latter case the mechanism leading to magnetogenesis shares some
interesting common aspects with the mechanism which have been proposed
for the formation of topological defects. On the other hand, if
cosmic strings  were produced in
the early Universe they could also generate cosmic magnetic fields
in several ways. Inflation, which provides a consistent solution
to many cosmological puzzles, has also several
features which make it interesting in the present context (see
Sec.\ref{sec:inflation}). Although to implement an inflationary
scenario of magnetogenesis requires some nontrivial extensions of
the particle physics standard model, recent independent
developments in field theory may provide the required ingredients.
Magnetic fields might also be produced by a preexisting lepton
asymmetry by means of the Abelian anomaly (see
Sec.\ref{sec:helicity}). Since the predictions about the strength
and the spatial distribution of the \mfs are different for
different models, the possible detection of primeval \mfs may shed
light on fundamental physical processes which could, otherwise, be
unaccessible.
 \vskip0.5cm \noindent
Even if primordial \mfs did not produce any relevant effect after
recombination, they may still have played a
significant role in several fundamental processes which occurred
in the first 100,000 years. For example, we shall show that \mfs
may have affected the big-bang nucleosynthesis, the dynamics of
some phase transitions, and baryogenesis. Since big-bang
nucleosynthesis (BBN) has been often used to derive constraints
on cosmological and particle physics parameters, the reader may
be not surprised to learn here that BBN also provides interesting
limits on the strength of primordial magnetic fields (see
Chap.\ref{chap:bbn}). Even more interesting is the interplay which
may exist between baryogenesis and magnetogenesis. Magnetic fields
might have influenced baryogenesis either by affecting the
dynamics of the electroweak phase transition or by  changing the
rate of baryon number violating sphaleron processes (see
Chap.\ref{chap:stability}). Another intriguing possibility is that
the hypercharge component of primeval \mfs possessed a net
helicity (Chern-Simon number) which may have been converted into
baryons and leptons by the Abelian anomaly (see
Chap.\ref{chap:generation}). In other words, primordial \mfs may
provide a novel scenario for the production of the observed
matter-antimatter asymmetry of the Universe.
 \vskip0.5cm \noindent
An interesting aspect of \mfs is their effect on the constituents
of matter. This in turn is of importance on many aspects of
the processes that took place in the early times.
Masses of hadrons get changed so that protons are heavier than neutrons.
The very nature of chirality could get changed see
(Chap.\ref{chap:stability}).
However the characteristic field for this to happen is $H=m_{\pi}^2$
which is about $10^{18}$ G. These fields cannot exist at times when
hadrons are already existing and therefore are probably not relevant.
Near cosmic superconductive strings the story may be different.
 \vskip0.5cm \noindent
 Clearly, this is a quite rich and interdisciplinary subject and
we will not be able to cover with the same accuracy all its
different aspects. Our review is manly focused on the production
mechanism and the effects of \mfs before, or during, the
photon decoupling from matter.

In Chap.\ref{chap:chap1} we shortly review the current status of
the observations. In order to establish some relation between
recent time and primeval magnetic fields we
also  provide a short description of some of the mechanisms which
are supposed to control the evolution of magnetic fields in the
galaxies and in the intergalactic medium. We only give a very
short and incomplete description of the effect of \mfs on
structure formation. Some basic aspects of this subject are,
anyhow, presented
in Chap.\ref{chap:cmb} where we discuss the effect of
\mfs on the anisotropies of the cosmic microwave background
radiation. From a phenomenological point of view
Chap.\ref{chap:cmb} is certainly the most interesting of our
review. The rapid determination of the CMBR acoustic peaks
at the level of a few percent will constrain these fields significantly.
We briefly touch upon the recent determination of the second acoustic
peak.
In Chap.\ref{chap:bbn} we describe several effects of
strong \mfs on the BBN and present some constraints which can be
derived by comparing the
theoretical predictions of the light elements relic abundances
with observations. Since it can be of some relevance for BBN,
propagation of neutrinos in magnetized media is also shortly
discussed at the end of that chapter. In
Chap.\ref{chap:generation} we review several models which predict
the generation of \mfs in the early Universe. In the same chapter
some possible mutual effects of magnetogenesis and baryogenesis
are also discussed. Some aspects of the effects which are
described in Chapts.\ref{chap:bbn} and \ref{chap:generation},
which concern the stability of strong \mfs and the effect that
they may produce on matter and gauge fields, are discussed in
more details in Chap.\ref{chap:stability}.
At the end we report our conclusions.


\chapter{The recent history of cosmic magnetic
fields}\label{chap:chap1}

\section{Observations}\label{sec:observations}

The main observational tracers of galactic and extra-galactic magnetic
fields are (comprehensive reviews of the subject can be found in
Refs.\cite{Kronberg, Zweibel}): the Zeeman splitting  of spectral lines;
the intensity and the polarization of synchrotron emission from free
relativistic electrons; the Faraday rotation measurements (RMs) of polarized
electromagnetic radiation passing through a ionized medium.

Typically the Zeeman splitting, though direct, is too small to be useful
outside our galaxy. Unfortunately, although the synchrotron emission and RMs
allows to trace \mfs in very distant objects, both kind of measurements
requires an independent determination of the local electron density $n_e$.
This is sometimes possible, e.g. by studying the  X-ray emission from the
electron gas when this is very hot, typically when this is confined in a
galaxy cluster.
Otherwise $n_e$ may be not always easy to determine, especially for
very rarefied media like the intergalactic medium (IGM).
In the case of synchrotron emission, which intensity is proportional
to $n_eB^2$, an estimation of $B$ is sometimes derived by assuming
equipartition between the magnetic and the plasma energy densities.

If the \mf to be measured is far away one relies on Faraday rotation.
The agreement generally found between the
strength of the field determined by RMs and
that inferred from the analysis of the synchrotron emission in relatively
close objects gives reasonable confidence on the reliability of the first
method also for far away systems. It should be noted, however, that
observations of synchrotron emission and RMs are sensitive to different
spatial components of the magnetic field \cite{Zweibel}.
The RM of the radiation emitted by a source with redshift $z_s$ is
given by
\begin{equation}
\label{FarRot}
RM(z_s) \equiv \frac{\Delta(\kappa)}{\Delta(\lambda^2)}= 8.1 \times 10^5
\int_0^{z_s} n_e B_\parallel(z)(1+z)^{-2}dl(z)~~~~\frac{\rm rad}{{\rm m}^2}
\end{equation}
where $B_\parallel $ is the field strength along the line of sight and
\begin{equation}
dl(z) = ~10^{-6}H_0^{-1}(1+z)(1+\Omega z)^{-\frac{1}{2}} dz~~{\rm Mpc}~.
\end{equation}
$H_0$ is the Hubble constant. The previous expression holds for a vanishing
cosmological constant and modification for finite $\Lambda$ is straightforward.
This method requires knowledge of the electron column and
possibility of field reversals. For nearby measurements in our own
galaxy pulsar frequency and their decays can pin down these
effects. Otherwise these stars are too far to help.
For this reason to determine the \mf of the IGM by Faraday RMs is quite hard
and only model dependent upper limits are available.

We now briefly summarize the observational situation.
\vskip0.5cm \noindent
{\bf Magnetic fields in galaxies.} The interstellar \mf in the Milky
Way has been determined using several methods which allowed to
obtain valuable information about the amplitude and spatial structure of the
field. The average field strength is  $3-4\mu$G. Such a strength
corresponds to an approximate energy equipartition between the magnetic
field, the cosmic rays confined in the Galaxy,  and the
small-scale turbulent motion \cite{Kronberg}
\begin{equation}
\rho_m = \frac{B^2}{8\pi} \approx \rho_t \approx \rho_{CR}~.
\end{equation}
Remarkably, the magnetic energy density almost coincides with
energy density of the cosmic microwave background radiation (CMBR).
The field keeps its orientation on scales of the order of few kiloparsecs
(kpc), comparable with the galactic size, and two reversals have been observed
between the galactic arms, suggesting that the Galaxy field morphology may be
symmetrical. Magnetic fields of similar intensity have been observed in a
number of other spiral galaxies. Although equipartition fields were observed
in some galaxies, e.g. M33, in some others, like the Magellanic Clouds and
M82, the field seems to be stronger than the equipartition threshold.
Concerning the spatial structure of the galactic fields, the
observational situation is, again, quite confused with some galaxies
presenting an axially symmetrical geometry, some other a symmetrical one, and
others with no recognizable field structure \cite{Zweibel}.
\vskip0.5cm \noindent
{\bf Magnetic fields in galaxy clusters.}
Observations on a large number of Abel clusters \cite{Kim}, some of which
have a measured X-ray emission,  give valuable information on
fields in clusters of galaxies. The magnetic field strength in the inter
cluster medium (ICM) is well described by the phenomenological equation
\begin{equation}
B_{ICM} \sim
2~\mu{\rm G}~\left(\frac{L}{10 kpc}\right)^{-\frac{1}{2}}(h_{50})^{-1}
\end{equation}
where $L$ is the reversal field length and $h_{50}$ is the
reduced Hubble constant. Typical values of $L$ are $10-100$ kpc
which correspond to field amplitudes of $1-10~\mu$G.
The concrete case of the
Coma cluster \cite{Feretti} can be fitted with a core magnetic
field $B~\sim~ 8.3 h_{100}^\frac{1}{2}~$G tangled at scales of about
1 kpc.
A particular example of clusters with a strong field is the Hydra A cluster
for which the RMs imply a 6$~\mu$G field
coherent over 100 kpc superimposed with a tangled field of
strength $\sim~30~\mu$G \cite{Taylor}.
A rich set of high resolution images of radio sources embedded in
galaxy clusters shows evidence of strong \mfs in the cluster central regions
\cite{Eilek99}. The typical central field strength $\sim~10~-~30~\mu$G with
peak  values as large as $\sim 70~\mu$G. It is noticeable that for such large
fields the magnetic pressure exceeds the gas pressure derived from X-ray data
suggesting that magnetic fields may play a significant role in the cluster
dynamics.It is interesting, as it has been shown by Loeb and Mao
\cite{LoeMao} that
a discrepancy exists between the estimate of the mass of the Abell cluster
2218 derived from
gravitational lensing and that inferred from X-ray observations which can be
well explained by the pressure support produced by a \mfs with strength
$\sim 50~\mu$G.
It is still not clear if the apparent decrease of the \mf
strength in the external region of clusters is due to the intrinsic field
structure or it is a spurious effect due to the decrease of the gas density.
Observations show also evidence for a filamentary spatial
structure of the field. According to Eilek \cite{Eilek99} the filaments
are presumably structured as a {\it flux rope}, that is a twisted field
structure in which the field lies along the axis in the center of the tube,
and becomes helical going away from the axis.

It seems quite plausible that all galaxy clusters are magnetized.
As we will discuss in the next section, these observations are a serious
challenge to most of the models proposed to explain
the origin of galactic and cluster magnetic fields.
\vskip0.5cm \noindent
{\bf Magnetic fields in high redshift objects.}
High resolution RMs of very far quasars have allowed to probe \mfs in
the distant past. The most significative measurements are due to
Kronberg and collaborators (see Ref.\cite{Kronberg} and Refs. therein).
RMs of the radio emission of the quasar 3C191,
at $z = 1.945$, presumably  due a magnetized shell of gas at the same
redshift, are consistent with a field strength in the range $0.4-4~\mu$G.
The field was found to maintain its prevailing direction over at least
$\sim 15$ kpc, which is comparable with a typical galaxy size.
The \mf of a relatively young spiral galaxy at  $z = 0.395$ was determined
by RMs of the radio emission of the quasar PKS 1229-021
lying behind the galaxy at $z = 1.038$. The \mf amplitude was firmly
estimated to be in the range $1-4~\mu$G. Even more interesting was the
observation of field reversals with distance roughly equal to the
spiral arm separation, in a way quite similar to that observed in
the Milky Way.
\vskip0.5cm \noindent
{\bf Intergalactic magnetic fields.}
The radio emission of distant quasars is also used to constrain the
intensity of \mfs in the IGM which we may suppose to pervade the
entire Universe.  As we discussed, to translate RMs into an estimation of the
field strength is quite  difficult for rarefied media in which ionized gas
density and field coherence length are poorly known.
Nevertheless, some interesting limits can be derived on the basis
of well known estimates of the Universe ionization fraction and adopting
some reasonable values of the magnetic coherence length.
For example, assuming a cosmologically aligned magnetic field, as well as
$\Omega = 1$, $\Lambda = 0$, and $h = 0.75$, the RMs of distant quasar
imply   $B_{IGM} \simleq 10^{-11}$ G \cite{Kronberg}.
A field which is aligned on cosmological
scales is, however, unlikely. As we have seen in the above, in galaxy clusters
the largest reversal scale is at most 1 Mpc. Adopting this scale as the
typical cosmic \mf coherence length and applying the RM($z_s$) up to $z_s \sim
2.5$, Kronberg found the less stringent limit $B_{IGM} \simleq 10^{-9}$ G
for the \mf strength at present time.

A method to determine the power spectrum of cosmic magnetic fields
from RMs of a large number of extragalactic sources has been proposed by
Kolatt \cite{Kolatt}.
The result of this kind of analysis would be of great help to determine
the origin and the time evolution of these fields.

Another interesting idea proposed by Plaga \cite{Plaga} is
unfortunately not correct. The idea here is to look to photons
from an instantaneous cosmological source, like a gamma burst or
a supernova, and check for the existence of a delayed component of
the signal.
This new component would be due to an original photon creating an
electron positron pair and in turn the charged particle sending a
photon in the original direction by inverse Compton scattering.
For sources at cosmological distances the delay would be sensitive to a small
B field, say $10^{-11}$ G that would affect the motion of the
charged intermediate particle. Unfortunately, the uncontrollable
opening of the pair will produce a similar delay that cannot be
disentangled from the time delay produced by the magnetic field.

\section{The alternative: dynamo or primordial ?}\label{sec:dynamo}

For a long time the preferred mechanism to explain the
aforementioned observations was the dynamo
mechanism \cite{Zeldovich}. Today, however, new
observational and theoretical results seems to point to a different
scenario.
Before trying to summarize the present status of art, a short, though
incomplete, synthesis of what is a dynamo mechanism may be useful to some
of our readers.
More complete treatments of this subject can be found e.g. in Refs.
\cite{Kronberg,Parker,Dolginov,AssSol,Kulsrudrep}.

A dynamo is a mechanism responsible for the conversion of kinetic
energy of an electrically conducting fluid into magnetic energy.
It takes place when in the time evolution equation of the \mf (see e.g.
Ref.\cite{Jackson})
\begin{equation}\label{diffusion}
    {\partial{\bf B}\over\partial t} =
  \mbox{\boldmath$\nabla$}\times({\bf v}\times{\bf B})+
  \frac {1}{4\pi \sigma} \nabla^2{\bf B}~,
\end{equation}
where $\sigma$ is the electric conductivity, the first term on the
RHS of Eq.(\ref{diffusion}) (frozen-in term)  dominates the second
one which account for magnetic diffusion. As we will see in
Sec.\ref{sec:evolution} this statement can be reformulated in terms of
the magnetic Reynolds number which has to be much larger than unity.
As it is clear from Eq.(\ref{diffusion}), a novanishing seed field is
needed to initiate  the dynamo process.
Other three key ingredients are generally required. They are hydrodynamic
turbulence, differential rotation and fast reconnection of magnetic lines.
In the frozen-in limit magnetic lines are distorted and stretched by turbulent
motion. It can be shown \cite{AssSol} that in the same limit the ratio
$B/\rho$ of the magnetic field strength with the fluid density behaves like
the distance between two fluid elements. As a consequence, a stretching of
the field lines result in an increase of $B$. However, this effect alone
would not be sufficient to explain the exponential amplification of the field
generally predicted by the dynamo advocates.  In fact, turbulence and
global rotation of the fluid (e.g. by  Coriolis force)  may
produce twisting of closed flux tubes and put both part of the twisted
loop together, restoring the initial single-loop configuration but with a
double flux (see Fig.2 in Ref.\cite{Dolginov}). The process can be iterated
leading to a $2^n$-amplification of the \mf after the $n$-th cycle.
The merging of magnetic loops, which produce a change in the topology
(quantified by the so called magnetic helicity, see Sec.\ref{sec:evolution})
of  the magnetic field lines, requires a finite, though small, resistivity of
the medium.  This process occurs in regions of small extension where
the field is more tangled and the diffusion time is smaller (see
Sec.\ref{sec:evolution}). As a consequence, the entire magnetic configuration
evolves from a small-scale tangled structure towards a mean ordered one.

The most common approach to magnetic dynamo is the so called mean field
dynamo. It is based on the assumption that fluctuations in the magnetic and
velocity fields are much smaller of the mean slowly varying components of the
corresponding quantities. Clearly, mean field dynamo is suitable to explore
the amplification of large scale magnetic structures starting from small
scale seed fields in the presence of a turbulent fluid. The temporal evolution
of the mean component of the \mf is obtained by a suitable averaging of
Eq.(\ref{diffusion}) (below, mean quantities are labelled by a 0 and random
quantities by a 1)
\begin{equation}
\label{dynamo}
    {\partial{\bf B}_0\over\partial t} =
  {\bf \nabla}\times\left( \alpha {\bf B}_0
  +  {\bf v}_0 \times{\bf B}_0 \right) -
   {\bf \nabla} \times
  \left[ (\eta + \beta) {\bf \nabla} \times {\bf B}_0
  \right]~,
\end{equation}
where
\begin{equation}
  \alpha = - \frac 1 3 \tau_c \langle {\bf v}_1 \cdot {\bf \nabla}
  \times {\bf v}_1 \rangle \qquad
  \beta = \frac 1 3 \tau_c \langle {\bf v}^2_1 \rangle ~,
\end{equation}
$\eta = 1/ 4 \pi \sigma$ is the magnetic diffusivity, and $\tau_c$
is the correlation time for the ensemble of random velocities.
The coefficient $\alpha$ is proportional to the helicity
$h = \langle {\bf v}_1 \cdot {\bf \nabla} \times {\bf v}_1 \rangle$
of the flow; $h$ measures the degree to which streamlines are twisted.
A macroscopic parity violation is required to have $\alpha \propto h \neq 0$.
One of the possible sources of this violation can be the Coriolis force
produced by the rotation of the galaxy \cite{Parker}.
The term $ {\bf \nabla} \times
  \left( \beta {\bf \nabla} \times {\bf B}_0\right)$ describes
the additional field dissipation due to turbulent motion.
Turbulence plays another crucial role in the generation of a toroidal component
of the large scale magnetic fields which is essential for the stability of
the entire field configuration \cite{AssSol}. Indeed the helicity, through the
$\alpha$-term, is responsible for the generation of an electric field parallel
to ${\bf B}_0$ \footnote{Our readers with some experience in field theory
may recognize that by producing parallel electric and magnetic fields the
$\alpha$ term is responsible of a sort of macroscopic CP violation.}. This
field provides a mode for conversion of toroidal into poloidal \mf components.
This is the so called  $\alpha$-effect. To complete the ``dynamo cycle''
$B_T \rightleftharpoons B_P$, another mechanism is required to convert the
poloidal component into a toroidal one. This mechanism is provided  by the
differential rotation of the galactic disk which will wrap up the field line
producing a toroidal field starting form a poloidal component; this is the
$\omega$-effect. The combination of the $\alpha$ and $\omega$ effects gives
rise to the, so called, $\alpha-\omega$ galactic dynamo.
As a result of the coexistence of the poloidal and toroidal magnetic
components, one of the main prediction of the  of  $\alpha-\omega$ dynamo is
the generation of an axially symmetric mean fields.

In the case the $\beta$ term can be neglected,
the solution of the mean field dynamo equation (\ref{dynamo}) can be written
in the form \cite{Zeldovich}
\begin{equation}
{\bf B}_0 = \left(\pm \sin kz, \cos kz, 0 \right) e^{\gamma t}~,
\end{equation}
where $z$ is the coordinate along the galaxy rotation axis, and
$\gamma = - \eta k^2 \pm \alpha k$, $k \sim 1/L$ being the wavenumber.
The field grows exponentially with time for non-zero helicity and if the
scale $L$ is sufficiently large.
A general prediction of a dynamo mechanism is that amplification ends
when equipartition is reached between the kinetic energy density of the
small-scale turbulent fluid motion and the magnetic energy density.
This correspond to a magnetic field strength in the range of $2-8~~\mu{\rm G}$.
Depending on the details of the model and of the local properties of the
medium, the time required to reach saturation, starting from a seed magnetic
fields with intensity as low as $10^{-20}$ G, may be of $10^8-10^9$ years.
It should be noted that such an estimation holds under the assumption that the
Universe is dominated by CDM with no cosmological constant. If, however,
as recent observations of distant type-IA supernovae \cite{Perlmutter}
and CMB anisotropy measurments \cite{boomerang} suggest, the Universe
posses a sizeable cosmological constant, the available time for the dynamo
amplification increases and a smaller initial seed field may be required.
This point has been recently rised by Davis, Lilley and T\"ornkvist
\cite{DavLil} who showed as the required seed field might be as low as
$10^{-30}$ G.
\vskip0.5cm
In the last decade the effectiveness of the mean field dynamo has been
questioned by several experts of the field (for a recent review see
Ref.\cite{Kulsrudrep}).
One of the main arguments rised by these authors against
this kind of dynamo is that it neglects the strong amplification of
small-scale
\mfs which reach equipartition, stopping the process, before a coherent field
may develop on galactic scales.

The main, though not the unique, alternative to the galactic dynamo is to
assume that the galactic field results directly from a
primordial field which gets adiabatically compressed when the protogalactic
cloud collapse. Indeed,  due to the large conductivity of
the intergalactic medium (see Sec.\ref{sec:evolution}),
magnetic flux is conserved in the intergalactic medium which implies that
the \mf has to increase
like the square of the size of the system $l$. It follows that
\begin{equation}
\label{fieldcompression}
B_{\rm prim,~0} = B_{\rm gal} ~\left(\frac{\rho_{\rm cosmic}}{\rho_{\rm gal}}
\right)^{2/3}~.
\end{equation}
Since the present time ratio between the interstellar medium density in the
galaxies and the density of the IGM is
$\rho_{\rm IGM}/\rho_{\rm gal} \simeq 10^{-6}$, and $B_{\rm gal} \sim 10^{-6}~$
G, we see that the required strength of the cosmic \mf at the galaxy formation
time ($z \sim 5$), adiabatically rescaled to present time, is
\begin{equation}
\label{primordialfield}
 B_{\rm prim,~0} \simeq 10^{-10}~~{\rm G}.
\end{equation}
This value is compatible with the observational limit on the field in the IGM
derived by RMs, with the big-bang nucleosynthesis constraints (see
Chap.\ref{chap:bbn}), and may produce observable effects
on the anisotropies of the cosmic microwave background radiation
(see Chap.\ref{chap:cmb}).
Concerning the spatial structure of the galactic field produced by this
mechanism, differential rotation should wrap the field into a symmetric
spiral with field reversal along the galactic disk diameter and no reversal
across the galactic plane \cite{Zweibel}.

To decide between the dynamo and the primordial options astrophysicists
have to their disposal three kind of information. They are:
\begin{itemize}
\item the observations of intensity and spatial distribution of the
galactic magnetic fields;
\item the observations of intensity and spatial distribution of the
intergalactic magnetic fields;
\item the observations of \mfs in objects at high redshift.
\end{itemize}
Observations of the magnetic field intensity in a some galaxies,
including the Milky Way,
show evidence of approximate equipartition between  turbulent motion and
magnetic energies, which is in agreement with the prediction of linear dynamo.
There are however some exceptions, like the M82 galaxy and the Magellanic
Clouds, where the field strength exceed the equipartition field.
An important test concerns the parity properties of the field with respect
to the rotations by $\pi$ about the galactic center. As we have discussed
above, the primordial theory predicts odd parity and the presence of
reversals with radius (a symmetric spiral field), whereas most dynamo models
predict even parity (axially symmetric spiral) with no reversal.
Although most galaxies exhibit no recognizable large-scale pattern, reversals
are observed between the arms in the Milky Way, M81 and the high redshift
galaxy discussed in the previous section, though not in M31
and IC342. Given the low statistical significance of the sample any
conclusions are, at the moment, quite premature \cite{Zweibel}.
\vskip0.5cm
As we reported in the previous section only upper limits are available for
the intensity of \mfs in the intergalactic medium. Much richer is
the information that astrophysicists collected in the recent years about
the \mfs in the inter-cluster medium (ICM). As we have seen, magnetic fields of
the order of $1-10~\mu{\rm G}$ seems to be a common features of
galaxy clusters.
The strength of these fields is comparable to that of galactic fields.
This occurs in spite of the lower matter density of the ICM
with respect to the density of interstellar medium in the galaxies.
It seems quite difficult to explain the origin of the inter-cluster \mfs
by simple ejection of the galactic fields. Some kind of dynamo process
produced by the turbulent wakes behind galaxies moving in the ICM has been
proposed by some authors  but criticized by some others  (for a review
see Ref.\cite{Kronberg}). This problem has become even more critical
in light of recent high-precision Faraday RMs which showed evidence of
\mfs with strength exceeding $10~\mu$G in the cluster central regions.
According to Kronberg \cite{Kronberg}, the observed
independence of the field strength from the local matter density  seems to
suggest that galactic systems have
evolved in a magnetic environment where $B \simgeq 1~~\mu{\rm G}$.
This hypothesis seems to be corroborated by the measurements of the
Faraday rotations produced by high redshift protogalactic clouds. As we wrote
in the previous section such measurements show evidence for magnetic fields
of the order of $1~~\mu{\rm G}$ in clouds with redshift larger than 1.
Since, at that time galaxies should have rotated few times, these
observations pose a challenge to the galactic dynamo advocates.
We should keep in mind, however,  that galaxy formation in the presence of
\mfs with strength $\simgeq 10^{-8}~$G  may be problematic
due to the magnetic pressure which inhibits the collapse \cite{Rees}.

It is worthwhile to observe that primordial (or pre-galactic) \mfs
are not necessarily produced in the early Universe, i.e.
before recombination time. Several alternative astrophysical mechanisms have
been proposed like the generation of the fields by a Biermann battery effect
\cite{Biermann} (see also Ref.\cite{Kronberg}). It has been suggested that the
Biermann battery may produce seed fields which are successively  amplified on
galactic scale by a dynamo powered by the turbulence in the protogalactic cloud
 \cite{Kulsrudrep,LesChi95}. This mechanism, however,  can hardly account for
the \mfs observed in the galaxy clusters. Therefore, such a scenario
would lead us to face an unnatural
situation where two different mechanisms are invoked  for the generation
of \mfs in galaxies and clusters, which have quite similar characteristics
and  presumably merge continuously  at the border of the
galactic halos.

Another possibility is that \mfs may have been generated by batteries
powered by starbursts or jet-lobe radio sources (AGNs).
In a scenario recently proposed by Colgate and Li \cite{Colgate}
strong cluster \mfs are produced by a dynamo operating in the accretion
disk of massive black holes powering AGNs.
We note, however, that the dynamics of the process leading to the formation of
massive black holes is still unclear and that preexisting magnetic fields may
be required to carry away the huge angular moment of the in-falling matter
(see e.g. Ref.\cite{Rees}).
For the same reason, preexisting \mfs may also be required to trigger
starbursts (see the end of next section).
This suggests that seed fields produced before recombination time
may anyway be required.

In conclusion, although the data available today do not allow to answer yet
to the question raised in this section, it seems that recent observations
and improved theoretical work are  putting in question the old wisdom in
favour of a dynamo origin of galactic magnetic fields.
Especially the recent observations of strong \mfs in galaxy clusters
suggest that the origin of these fields may indeed be primordial.

Furthermore, \mfs with strength as large as that required for the primordial
origin of the galactic fields through gravitational compression of the
magnetized fluid, should give rise to interesting, and perhaps necessary,
effects for structure formation. This will be the subject of the next section.

\section{Magnetic fields and structure formation}\label{sec:structures}

The idea that cosmic \mfs may have played some role in the
formation of galaxies is not new. Some early work has been done on
this subject, e.g. by Peblees \cite{Peebles71}, Rees and
Rheinhardt \cite{ReeReh} and especially by Wasserman
\cite{Wasserman}. A considerable amount of recent papers
testify the growing interest around this issue. A
detailed review of this particular aspect of cosmology is,
however, beyond the purposes of this report.  We
only summarize here few main points with the hope to convince the
reader of relevance of this subject.
 \vskip 0.5cm
Large scale \mfs modify standard equations of linear density
perturbations in a gas of charged particles by adding the effect
of the Lorentz force. In the presence of the field the set of
Euler, continuity and Poisson equations become respectively \cite{Wasserman}
\begin{eqnarray}
   \rho\left( \frac {\partial{\bf v}}{\partial t} + {\dot{a}\over a}{\bf v} +
   \frac{{\bf v}\cdot {\bf \nabla v}}{a} \right) &=&
   - \frac{\bf \nabla p}{a} - \rho \frac {{\bf \nabla} \phi}{a}
   + \frac{\left({\bf \nabla}\times {\bf B}\right) \times {\bf B} }
   {4\pi a}~, \label{euler1}\\
  \frac {\partial\delta}{\partial t} + 3\frac{\dot{a}}{a}\rho +
  \frac {{\bf \nabla} \cdot (\rho {\bf v})}{a} &=&
  0~,\label{continuity1}\\
  {\mathbf \nabla}^{2} \phi &=& 4\pi G a^2\left( \rho - \rho_0(t)\right)~.
      \label{poisson1}
\end{eqnarray}
Here $a$ is the scale factor and other symbols are obvious.
This set of equations is completed by the Faraday equation
\begin{equation}
  \frac {\partial (a^2 {\bf B})}{\partial t} =
        \frac {{\mathbf \nabla} \times \left( {\bf v} \times
        a^2 {\bf B} \right)}{a}~,
        \label{faraday1}
\end{equation}
and
\begin{equation}
{\bf \nabla}\cdot {\bf B} = 0~.
\end{equation}
The term due to the Lorentz force is clearly visible in the right
hand side of the Euler equation.
It is clear that, due to this term,  an inhomogeneous \mf becomes
itself a source of density, velocity and gravitational perturbations in
the electrically conducting fluid. It has been estimated \cite{Wasserman} that
the \mf needed to produce a  density contrast $\delta \sim 1$, as required to
induce structure formation on a scale $l$, is
\begin{equation}
B_0(l) \sim 10^{-9}~ \left( \frac {l}{1~{\rm Mpc}} \right) h^2 \Omega~~~{\rm G}
~.
\end{equation}
In his recent book Peebles Ref.\cite{Peeblesbook}
pointed-out a significabt coincidence:
the primordial \mf required to explain galactic fields without invoking
dynamo amplification (see Eq.\ref{primordialfield}) would also play a relevant
dynamical role in the galaxy formation process.

The reader may wonder if such a dynamical role of \mfs is really required.
To assume that \mfs were the dominant dynamical factor at the time of
galaxy formation and that they were the
main source of initial density perturbations is perhaps  too extreme and
probably incompatible with recent measurements of the CMBR anisotropies.
A more reasonable possibility is that \mfs are an important missing ingredient
in the current theories on large scale structure formation (for a recent
review  on this subject see Ref.\cite{BatLes}). It has been argued by Coles
\cite{Coles}
that an inhomogeneous \mf could modulate galaxy formation in the cold dark
matter picture (CDM) by giving the baryons a streaming velocity relative to
the dark matter. In this way, in some places the baryons may be prevented from
falling into the potential wells and the formation of luminous galaxies on
small scales may be inhibited. Such an effect could help to reconcile the
well know discrepancy of the CDM model with clustering observations
without invoking more exotic scenarios.

Such a scenario received some support from a paper by Kim, Olinto and Rosner
\cite{KimOli} which extended Wasserman's \cite{Wasserman} pioneering work.
Kim et al. determined the power spectrum of density
perturbation due to a primordial inhomogeneous magnetic field.
They showed that a present time {\it rms} \mf of $10^{-10}~$G may have
produced perturbations on galactic scale which should have entered the
non-linear grow stage at $z \sim 6$, which is compatible with observations.
Although, Kim et al. showed that \mfs alone cannot be responsible of the
observed galaxy power spectrum on large scales, according to the authors
it seems quite plausible that in a CDM scenario \mfs played a not minor role
by introducing a bias for the formation of galaxy sized objects.

A systematic study of the effects of \mfs on structure formation was
recently undertaken by Battaner, Florido and Jimenez-Vicente \cite{Battaner1},
Florido and Battaner \cite{Battaner2}, and Battaner, Florido and Garcia-Ruiz
\cite{Battaner3}. Their results show that primordial \mfs with strength
$B_0 \simleq 10^{-9}$  in the pre-recombination era are able to produce
significant anisotropic density inhomogeneities in the baryon-photon plasma
and in the metric. In particular, Battaner at al. showed that \mfs tend
to organize themselves and the ambient plasma into filamentary structures.
This prediction seems to be confirmed by recent observations of \mfs
in galaxy clusters \cite{Eilek99}. Battaner et al. suggest that such a
behavior may be common to the entire Universe and  be responsible
for the very regular spider-like structure observed in the local supercluster
\cite{Einasto} as for the filaments frequently observed in the large scale
structure of the Universe \cite{BatLes}. Araujo and Opher
\cite{AraOph} have considered the formation of voids by the magnetic
pressure.

An interesting hypothesis has been recently rised by Totani \cite{Totani}.
He suggested that spheroidal galaxy formation occurs has a consequence of
starbursts triggered by strong magnetic fields. Totani argument is based on two
main observational facts. The first is that \mfs strengths observed in spiral
galaxies sharply concentrate at few microgauss (see Sec.\ref{sec:observations}),
quite independently on the galaxy luminosity and morphology.
The second point on which Totani based his argument, is that star
formation activity has been observed to be correlated to the strength of
local \mf \cite{Valle94}. A clear example is given by the spiral galaxy
M82, which has an abnormally large \mf of $\sim 10~\mu$G and is known as an
archetypal starburst galaxy. Such a correlation is theoretical motivated
by the so-called {\it magnetic braking} \cite{Rees}: in order for a
protostellar gas
cloud to collapse into a star a significant amount of angular moment must be
transported outwards. Magnetic fields provide a way to fulfill this requirement
by allowing the presence of Alfv\'en waves (see Sec.\ref{sec:peaks}) which
carry away the excess of angular moment.
Whereas it is generally agreed that galaxy bulges and elliptical galaxies
have formed by intense starburst activity at high redshift, the trigger
mechanism leading to this phenomenon is poorly known.
According to Totani, starbursts, hence massive galaxy formation, take
place only where the \mf is stronger of a given threshold, which would explain
the apparent uniformity in the \mf amplitude in most of the observed galaxies.
The value of the threshold field depends on the generation mechanism of
the galactic magnetic field. Totani assumed that a seed field may have been
produced by a battery mechanism followed by a dynamo amplification
period. Such an assumption, however, looks not necessary and a primordial field
may very well have produced the same final effect.

\section{The evolution of primordial magnetic fields}\label{sec:evolution}

A crucial issue for the investigation of a possible primordial
origin of present time galactic and intergalactic \mfs is that concerning
the time evolution of the magnetic fields in the cosmic medium. Three
conditions are needed for the persistence of large static fields:
\begin{itemize}
\item[a)] intrinsic stability of the field;
\item[b)] the absence of free charges which could screen the
field;
\item[c)] to have a small diffusion time of the field with respect to the age
of the Universe.
\end{itemize}
 \noindent
 Condition a) does not hold for strong electric fields. It is a
firm prediction of QED that an electric field decays by converting
its energy in electron-positron pairs if $e \vert E\vert \geq
m_e^2$ \cite{Schwinger,ItzZub}. This, however, is a purely
electric phenomenon. Although, at the end of the sixties, there
was a claim that strong magnetic fields may decay through a
similar phenomenon \cite{OConnell68} the argument was proved to be
incorrect. Only very strong fields may produce nontrivial
instabilities in the QCD (if $B > 10^{17}~$G) and the electroweak vacuum
(if $B > 10^{23}~$G) which may give rise to a partial screening of the field.
These effects (see Chap.\ref{chap:stability})
may have some relevance for processes which occurred at very
early times and, perhaps, for the physics of very peculiar collapsed objects
like  magnetars \cite{magnetars}.
They are, however, irrelevant for the evolution of cosmic \mfs after BBN time.
The same conclusion holds for finite temperature and densities effects
which may induce screening of static \mfs (see e.g. Ref.\cite{DanGra}).

Condition b) is probably trivially fulfilled for \mfs  due to the
apparent absence of magnetic monopoles in nature.
It is interesting to observe that even a small
abundance of magnetic monopoles at present time would have
dramatic consequences for the survival of galactic and
intergalactic magnetic fields which would lose energy by
accelerating the monopoles. This argument was first used by Parker
\cite{Parkerlimit} to put a severe constraint on the present time
monopole flux, which is $F_M \simleq 10^{-15}~{\rm cm}^{-2} s^{-1}
{\rm sr}^{-1}$. It was recently proposed by Kephart and Weiler \cite{Weiler}
that magnetic monopoles accelerated by galactic \mfs could give
rise to the highest energy cosmic rays ($E \simleq 10^{19}$ eV)
and explain the violation of the famous Greisen-Zatsepin-Kuzmin
cut-off \cite{GZK}.

Also the condition c) does not represent a too serious problem for
the survival of primordial magnetic fields.  The time evolution
law of a \mf in a conducting medium has been already written in
Eq.(\ref{diffusion}).

Neglecting fluid velocity this equation reduces to the diffusion
equation which implies that an initial magnetic configuration will
decay away in a time
\begin{equation}\label{decaytime}
  \tau_{\rm diff}(L) = {4\pi \sigma L^2}~,
\end{equation}
where $L$ is the characteristic length scale of the spatial
variation of $\bf B$. In a cosmological framework, this means that
a magnetic configuration with coherence length $L_0$ will survive
until the present time $t_0$ ($t = 0$ corresponds to the big-bang
time) only if $\tau_{\rm diff}(L_0) > t_0$.  In our convention, $L_0$
corresponds to the present time length scale determined by the
Hubble law
\begin{equation}\label{Lscaling}
  L_0 = L(t_i) ~\displaystyle \frac {a(t_0)}{a(t_i)}~,
\end{equation}
where $a(t)$ is the Universe scale factor and $L(t_i)$ is the
length scale at the time at which the magnetic configuration was
formed. Note that $L_0$ may not coincide with the actual size of
the magnetic configuration since other effects (see below) may
come-in to change the comoving coherence length. As we see from
Eq.(\ref{decaytime}) the relevant quantity controlling the decay
time of a magnetic configuration is the electric conductivity of
the medium. This quantity changes in time depending on the varying
population of the available charge carriers and on their kinetics
energies. However, since most of the Universe evolution takes
place in a matter dominated regime, during which all charge
carriers are non-relativistic, our estimate of the magnetic
diffusion length is simpler. In general, electric
conductivity can be determined by comparing Ohm law ${\bf J} =
\sigma {\bf E}$ with the electric current density definition ${\bf
J} =  n e {\bf v}$, where for simplicity we considered a single
charge carrier type with charge $e$, number density $n$ and
velocity $\bf v$. The mean drift velocity in the presence of the
electric field $\bf E$ is ${\bf v} \sim e{\bf E}\tau/m$ where $m$
is the charge carrier mass and $\tau$ is the average time between
particle collisions. Therefore the general expression for the
electron conductivity is \footnote{In the case the average
collision time of the charge carrier is larger than the Universe
age $\tau_U$, the latter has to be used in place of $\tau$ in
Eq.\ref{sigma} \cite{Widrow}.}
\begin{equation}\label{sigma}
  \sigma = \frac{n e^2 \tau}{m}~.
\end{equation}
After recombination of electron and ions into stable atoms
Universe conductivity is dominated by residual free electrons.
Their relative abundance is roughly determined by the value that
this quantity took at the time when the rate of the reaction $p +
e \leftrightarrow H + \gamma$ became smaller than the Universe
expansion rate. In agreement with the results reported in
Ref.\cite{KolTur}, we use
\begin{equation}\label{efree}
  n_e(z) \simeq 3\times 10^{-10}~{\rm cm}^{-3}~\Omega_0 h~(1 +
  z)^3~,
\end{equation}
where $\Omega_0$ is the present time density parameter and $h$ is
Hubble parameter. Electron resistivity is dominated by Thomson
scattering off cosmic background photons. Therefore $\tau \simeq
1/n_\gamma \sigma_T$, where $\sigma_T = \displaystyle \frac{e^4}
{6 \pi m_e^2}$ is the Thomson cross section, and $n_\gamma = 4.2
\times 10^2 (1 + z)^3$. Substituting these expressions in
Eq.(\ref{sigma}) we get
\begin{equation}
  \sigma = \frac{n e^2 }{m_e n_\gamma \sigma_T} \simeq
  10^{11} \Omega_0 h~~s^{-1}.
\end{equation}
It is noticeable that after recombination time Universe
conductivity is a constant. Finally, the cosmic diffusion length,
i.e. the minimal size of a magnetic configuration which can
survive diffusion during the Universe time-life $t_0$, is found
by substituting $t_0 = 2 \times (\Omega_0 h^2)^{-1/2}~~{\rm
s}^{-1}$ into Eq.(\ref{decaytime}) which, adopting $\Omega_0 = 1$
and $h = 0.6$,  gives
\begin{equation}\label{Ldiff}
  L_{\rm diff} \simeq 2 \times 10^{13}~{\rm cm} \simeq 1~{\rm
  A.U.}
\end{equation}
It follows from this result that magnetic diffusion is negligible
on galactic and cosmological scales.

The high conductivity of the cosmic medium has other relevant
consequences for the evolution of magnetic fields. Indeed, as we
already mentioned in the Introduction, the magnetic flux through
any loop moving with fluid is a conserved quantity in the limit
$\sigma \rightarrow \infty$. More precisely, it follows from the
diffusion equation (\ref{diffusion}) and few vector algebra
operations (see Ref.\cite{Jackson})that
\begin{equation}\label{fluxcons}
  \frac{d\Phi_S(B)}{d t} = - \frac{1}{4\pi \sigma} \int_S {\mathbf
  \nabla} \times \left({\mathbf \nabla} \times {\bf B} \right)
  \cdot d{\bf S}~,
\end{equation}
where $S$ is any surface delimited by the loop. On scale where
diffusion can be neglected the field is said to be  {\it
frozen-in}, in the sense that lines of force move together with
the fluid. Assuming that the Universe expands isotropically
\footnote{In the Sec.\ref{sec:anisotropy} we shall discuss under
which hypothesis such an assumption is consistent with the
presence of  a cosmic magnetic field.}, and no other effects
come-in, magnetic flux conservation implies
\begin{equation}\label{scaling}
  B(t) = B(t_i)\;\left(\frac{a(t_i)}{a(t)}\right)^2~.
\end{equation}
This will be one of the most relevant equations in our review. It
should be noted by the reader that $B(t)$ represents the local
strength of the \mf obtained by disregarding any effect that may
be produced by  spatial variations in its intensity and direction.
Eq.(\ref{scaling}) is only slightly modified in the case the a
uniform magnetic field produce a significative anisotropic
component in the Universe expansion (see Sec.\ref{sec:anisotropy}).
\vskip0.5cm
 Another quantity which is almost conserved due to the
high conductivity of the cosmic medium is the, so-called, {\it
magnetic helicity}, defined by
\begin{equation}\label{mhelicity}
  {\cal H} \equiv  \int_V d^3x\, {\bf B}
  \cdot {\bf A} ~,
\end{equation}
where ${\bf A}$ is the vector potential.
Helicity is closely analogous to vorticity in fluid dynamics. In a
field theory language, ${\cal H}$ can be identified with the
Chern-Simon number which is known to be related to the topological
properties of the field. Indeed, it is known from
Magnetohydrodynamics (MHD) that $\cal H$ is proportional to the
sum of the number of links and twists of the \mf lines
\cite{Biskamp}. As it follows from Eq.(\ref{diffusion}), the time
evolution of the magnetic helicity is determined by
\begin{equation}\label{helevo}
  \frac{d{\cal H}}{d t} = - \frac{1}{4\pi \sigma}\int_V d^3x {\bf B}
  \cdot \left( {\mathbf \nabla} \times {\bf B} \right)~.
\end{equation}
As we shall show in Chap.\ref{chap:generation}, several models
proposed to explain the origin of primordial \mfs predict
these fields to have some relevant amount of helicity.

In the previous section we have already mentioned the important
role played by magnetic helicity in some MHD dynamo mechanisms
driving an instability of small scale \mfs into large scale
fields. A similar effect may take place at a cosmological level
leading to significative corrections to the simple scaling laws
expressed by the Eqs.(\ref{Lscaling},\ref{scaling}). Generally, these
kind of MHD effects occur in the presence of some, turbulent
motion of the conductive medium  (note that
Eqs.(\ref{Lscaling},\ref{scaling}) has been derived under the assumption
of vanishing velocity of the fluid $v = 0$). Hydrodynamic
turbulence is generally parameterized in terms of the Reynolds
number, defined by
\begin{equation}\label{reynold}
  Re = \frac{v L}{\nu}~,
\end{equation}
where $\nu$ is the kinematic viscosity. Fluid motion is said to be
turbulent if $Re \gg 1$. In the presence of a \mf another
parameter turns-out to be quite useful. This is the {\it magnetic
Reynolds number} defined by
\begin{equation}\label{mreynold}
   Re_M = \frac{v L}{\eta}~,
\end{equation}
where $\eta = \displaystyle \frac {1}{4\pi \sigma}$. When $Re_M
\gg 1$ transport of the magnetic lines with the fluid dominates
over diffusion. In this case hydrodynamic turbulence in a
conducting medium give rise to {\it magnetic turbulence}. It is
often assumed in MHD that a fully developed magnetic turbulence
give rise to equipartion between the kinetic and the magnetic
energy of the fluid. Whether equipartition hypothesis is valid or not
is a controversial issue.

Both the hydrodynamic and magnetic Reynolds numbers can be very
large in the early Universe. This is a consequence of the
high electric conductivity and low viscosity of the medium
and, especially, of the large scales which are involved. The
electric conductivity of the early Universe has been computed by
several authors. A first simple estimation of $\sigma$ in the
radiation dominated era was performed by Turner and Widrow
\cite{Widrow}. In terms of the resistivity $\eta$ their result is
$\eta \sim \displaystyle \frac \alpha T $. A more exact series of
calculations can be found in Ref.\cite{Hosoya} which include
logarithmic corrections  due to Debye and dynamical screening. As
a result a more correct expression for $\eta$ is
\begin{equation}
  \eta  \sim \frac \alpha T \ln(1/\alpha)~.
\end{equation}
Other detailed computations of the Universe conductivity close to
the QCD and the electroweak phase transitions are available in
Refs.\cite{conductivity}. The kinematic viscosity follows the
behavior \cite{Son}
\begin{equation}
  \nu  \sim \frac {1}{\alpha T \ln(1/\alpha)}~.
\end{equation}
In the early Universe $\nu \gg \eta$, i.e. $Re_M \gg Re$.
Concerning  the absolute value of these parameters, using the
previous expressions it is easy to verify that for a reasonable
choice of the velocity field that may be produced by a phase
transition, $v \simleq 10^{-3}$, both $Re$ and $Re_M$ are much
larger than unity by several orders of magnitude, even for very
small scales  (for more details see Chap.\ref{chap:generation}).
It seems that the early Universe was a quite ``turbulent child" !
Turbulence is expected to cease after $e^+ e^-$ annihilation since
this process reduces the plasma electron population and therefore
increases the photon diffusion length hence also the kinematic
viscosity. This should happen at a temperature around 1 MeV.

Turbulence is expected to produce substantial modification in the
scaling laws of cosmological magnetic fields. This issue has been
considered by several authors. Brandenburg, Enqvist and Olesen
\cite{BraEO}  first consider MHD in an expanding Universe in the
presence of hydro-magnetic turbulence. MHD equations were written
in a covariant form and solved numerically under some simplifying
assumptions. The magnetic field was assumed to be distributed
randomly either in two or three spatial dimensions. In the latter
case a cascade (shell) model was used to reduce the number of
degree of freedom. In both cases a transfer of magnetic energy
from small to large magnetic configurations
was observed in the simulations.
In hydrodynamics this phenomenon is know as an {\it inverse cascade}.
Cascade processes are know to be related to certain conservation
properties that the basic equations obey \cite{Pouquet}. In the
two-dimensional inverse cascade, the relevant conserved quantity
is the volume integral of the vector potential squared, $\int d^2x
{\bf A}^2$,  whereas in the three-dimensional cases it is the
magnetic helicity. It was recently showed by Son \cite{Son} that
no inverse cascade can develop in 3d if the mean value of $\cal H$
vanishes. If this is the case, i.e. in the presence of non-helical
MHD turbulence, there is still an anomalous grow of the magnetic
correlation length with respect to the scaling given in
Eq.(\ref{Lscaling}) but this is just an effect of a {\it selective
decay} mechanism: modes with larger wavenumbers decay faster than
those whose wavenumbers are smaller. Assuming Universe expansion
is negligible with respect to the decay time, which is given by
the eddy turnover time $\tau_L \sim L/v_L$, and the decay of the
large wavenumber modes does not affect those with smaller
wavenumbers, Son found that the correlation length scale with time
as
\begin{equation}\label{turbLscal}
  L(t) \sim \left(\frac t t_i \right)^{2/5}~L_i
\end{equation}
where $\tau_i = L_i/v_i$ is the eddy turnover time at $t = 0$.
Assuming equipartition  of the kinetic and magnetic energies, that
is $v_L \sim B_L$, it follows that the energy decay with time like
$t^{-6/5}$. When the Universe expansion becomes not negligible,
i.e. when $t> t_0$, one has to take into account that the
correlation length grows as $\tau^{2/5}$, where $\tau$ is the
conformal time. Since $\tau \sim T^{-1}$, it follows the
$T^{-2/5}$ law. In the real situation, the final correlation
length at the present epoch is,
\begin{equation}
  L_0 = L_i \cdot \biggl({t_0 v_i\over L_i}\biggr)^{2/5}
  \cdot \biggl({{T_i} \over T_{\rm nt}}\biggr)^{2/5}\cdot
  {T_i \over T_0} \, .
  \label{lnow}
\end{equation}
In the above, the first factor comes from the growth of the
correlation length in the time interval $0 < t < t_0$ when eddy
decay is faster than Universe expansion; the second factor comes
from the growth of $L$ in the $t > t_0$ period; the last factor
comes from trivial redshift due to the expansion of the Universe.
$T_{\rm nt}$ is the temperature of the Universe when the fluid
becomes non-turbulent. As we discussed $T_{\rm nt} \sim 1$ MeV.
If, for example, we assume that turbulence was produced at the
electroweak phase transition, so that $T_i = T_{EW} \sim 100$ GeV,
that $v_i \sim 0.1$ and $L_i \sim 10^{-2} r_H(T_{EW}) \sim
10^{-2}$ cm, one find $L_0 \sim 100$ AU. This result has to be
compared with the scale one would have if the only mechanism of
dissipation of magnetic energy is resistive diffusion which, as we
got in Eq.(\ref{Ldiff}) is $\sim 1$ AU.

A larger coherence length can be obtained by accounting for
the magnetic helicity which is probably produced during a
primordial phase transition. The conservation of $\cal H$ has an
important consequence for the evolution of the magnetic field.
When $\cal H$ is non-vanishing, the short-scale modes are not
simply washed out during the decay: their magnetic helicity must
be transferred to the long-scale ones.  Along with the magnetic
helicity, some magnetic energy is also saved from turbulent decay.
In other words, an inverse cascade is taking place. Assuming
maximal helicity, i.e. that ${\bf B}\cdot ({\mathbf \nabla} \times
{\bf B}) \sim L B^2$, the conservation of this quantity during the
decay of turbulence implies the scaling law
\[ B_L\sim B_i\left({L\over L_i}\right)^{-1/2}~.\]
This corresponds to ``line averaging", which gives a much larger
amplitude of the \mf than the usual ``volume averaging". Equipartition
between magnetic and kinetic energy implies
\[
  v_L\sim v_i\left({L\over L_i}\right)^{-1/2}~.
\]
This relation together with the expression for the eddy decay
time, $\tau_L = L/v_L$, leads to the following scaling law for the
correlation length of helical magnetic structures
\begin{equation}\label{helLscal}
      L\sim L_i\left({t\over t_i}\right)^{2/3}\ .
\end{equation}
Comparing this result with Eq.(\ref{turbLscal}), we see that in
the helical case the correlation length grow faster than it does
in the turbulent non-helical case. The complete expression for the
scaling of $L$ is finally obtained by including trivial redshift
into Eq.(\ref{helLscal}). Since in the radiation dominated era
$T^{-1} \sim a \sim t^{1/2}$, we have \cite{Son}
\begin{equation}\label{helLscalT}
  L_0 = \left(\frac {T_0} {T_{\rm nt}} \right)
  \left(\frac {T_i} {T_0} \right)^{5/3} L_i~,
\end{equation}
and
\begin{equation}\label{helBscalT}
  B_0 = \left(\frac {T_{\rm nt}}{T_0} \right)^{-2}
  \left(\frac {T_i} {T_0}\right)^{-7/3} ~ B(T_i)~.
\end{equation}
According to Son, \cite{Son} helical hydromagnetic turbulence
survives longer than non-helical turbulence allowing $T_{\rm nt}$
to be as low as 100 eV. If again we assume that helical magnetic
turbulence is generated at the electroweak phase transition
(which will be justified in Chap.\ref{chap:generation}) we
find
\begin{equation}
  L_0 \sim L_i \left(\frac {T_{EW}} {T_{\rm nt}} \right)^{5/3}
  \left(\frac {T_{\rm nt}} {T_0}\right) \sim 100~~{\rm pc}~,
\end{equation}
which is much larger than the result obtained in the non-helical
case.
It is worthwhile to observe that,as the scale derived in the previous
expression is also considerably larger than the cosmological magnetic
diffusion length scale given in Eq.(\ref{Ldiff}), that \mf fields produced
by the EW phase transition may indeed survive until present.


\chapter{Effects on the Cosmic Microwave Background}\label{chap:cmb}

\section{The effect of a homogeneous magnetic field}\label{sec:anisotropy}

It is well know from General Relativity that electromagnetic
fields can affect the geometry of the Universe. The energy
momentum tensor
\begin{equation}\label{Tem}
  T_{\mathrm{em}}^{\alpha\beta}  = \frac 1 {4\pi} \left(-
  F^{\alpha\mu}F^{\beta}_\mu + \frac 1 4 g^{\alpha\beta}
  F_{\mu\nu}F^{\mu\nu}\right)~,
\end{equation}
where $F^{\mu\nu}$ is the electromagnetic field tensor, acts as a
source term in the Einstein equations. In the case of a
homogeneous \mf  directed along the $z$-axis
\begin{equation}\label{Tcomponents}
  T^{00} = T^{11} = T^{22} = - T^{33} = \rho_B = \frac {B^2}{8\pi}~~~~T^{0i}
  = 0~.
\end{equation}
Clearly, the energy-momentum tensor becomes anisotropic due to the presence
of the magnetic field. There is a
positive pressure term along the $x$ and $y$-axes but a ``negative
pressure" along the field direction. It is known that an isotropic
positive pressure increases the deceleration of the universe
expansion while a negative pressure tends to produce an
acceleration. As a consequence, an anisotropic pressure must give
rise to an anisotropy expansion law \cite{ZelNov}.

Cosmological models with a homogeneous \mf  have been considered by
several authors (see e.g. \cite{Thorne67}). To discuss
such models it is beyond the purposes of this review. Rather, we
are more interested here in the possible signature that the
peculiar properties of the space-time in the presence of a cosmic \mf
may leave on the Cosmic Microwave Background Radiation (CMBR).

Following Zeldovich and Novikov \cite{ZelNov} we shall consider
the most general axially symmetric model with the metric
\begin{equation}\label{aximetric}
  ds^2 = dt^2 - a^2(t)(dx^2 + dy^2) - b^2(t)dz^2~.
\end{equation}
It is convenient to define $ \alpha =\dfrac{\dot a}a;~~\beta
 =\dfrac{\dot b}b;$ and
\begin{equation}
r \equiv \frac {\rho_B}{\rho_{\mathrm{rad}}}~~~~~~\sigma \equiv
 {\alpha - \beta}~.
\end{equation}
Then, assuming $r, \sigma < 1$, the Einstein equations are well
approximated by
\begin{eqnarray}\label{anisEin1}
  \frac{d}{dt}\left(\frac{\sigma}{H}\right) &=& -  \left(
  \frac{\sigma}{H}\right)\frac{\gamma - 2}{\gamma t} + \frac
  {4r}{\gamma t}\\
  \label{anisEin2} \frac{dr}{dt} &=& - \frac {2r}{9 \gamma t}
  \left(4 \frac{\sigma}{H} + 9\gamma - 12\right)~,
\end{eqnarray}
where $H = (2\alpha + \beta)$ and $\gamma$ are defined by the
equation of state $p = (\gamma - 1)\rho$. It is easy to infer from
the first of the previous equations that the \mf  acts so as to
conserve the anisotropy that would otherwise decay with time
in the case $r = 0$. By
substituting the asymptotic value of the anisotropy, i.e. $\sigma
\rightarrow 6r$, into the evolution equation for $r$ in the RD era
one finds
\begin{equation}\label{logrho}
  r(t) = \frac {q}{1 + 4q \ln(t/t_0)}~,
\end{equation}
where $q$ is a constant.
Therefore, in the case the cosmic \mf  is homogeneous, the ratio of
the magnetic and blackbody radiation densities is not a constant,
but falls logarithmically  during the radiation era.

In order to determine the temperature anisotropy of the CMBR we
assume that at the recombination time $t_{\mathrm{rec}}$ the
temperature is everywhere $T_{\mathrm{rec}}$. Then, at the present
time, $t_0$, the temperature of relic photons coming  from the $x$
(or $y$) and $z$ directions will be respectively
\begin{equation}
  T_{x,y} = T_{\mathrm{rec}} \frac a a_0 = T_{\mathrm{rec}}
  \exp\left({-\int_{t_{\mathrm{rec}}}^{t_0}} \alpha dt \right)\qquad
 T_z = T_{\mathrm{rec}} \frac b b_0 = T_{\mathrm{rec}}
  \exp\left({-\int_{t_{\mathrm{rec}}}^{t_0}} \beta dt \right)~.
\end{equation}
Consequently, the expected temperature anisotropy is
\begin{eqnarray}
  \frac {\Delta T}{T} &=& \frac{T_x - T_z}{T_{\mathrm{rec}}} = 1 -
  \exp\left({\int_{t_{\mathrm{rec}}}^{t_0}} (\alpha - \beta) dt \right)
  \nonumber \\
  &\approx& \int_{t_{\mathrm{rec}}}^{t_0} (\beta - \alpha) dt
  = - \half \int_{t_{\mathrm{rec}}}^{t_0} \sigma d\ln t ~.\label{deltaTZel}
\end{eqnarray}
By using this expression, Zeldovich and Novikov estimated that a
cosmological \mf  having today the strength of $10^{-9}\div
10^{-10}$ Gauss would produce a temperature anisotropy ${\delta
T}/ {T} \simleq 10^{-6}$.

The previous analysis has been recently updated by Barrow,
Ferreira and Silk \cite{BarFS}. In that work the authors derived an upper
limit on the strength of a homogeneous  magnetic field at the recombination
time on the basis of the 4-year Cosmic Background Explorer (COBE)
microwave background isotropy measurements \cite{COBE}. As it is
well know COBE detected quadrupole anisotropies at a level
${\delta T}/ {T} \sim 10^{-5}$ at an angular scale of few degrees.
By performing a suitable statistical average of the data and
assuming that the field keeps frozen-in since the recombination
till today, Barrow at al. obtained the limit
\begin{equation}\label{barrowlim}
B(t_0) < 3.5 \times 10^{-9} f^{1/2} (\Omega_0
h_{50}^2)^{1/2}~\mathrm{G}~.
\end{equation}
In the above $f$ is a $O(1)$ shape factor accounting for possible
non-Gaussian characteristics of the COBE data set.

From this results we see that COBE data are not incompatible with
a primordial origin of the galactic \mf  even without invoking a
dynamo amplification. \\

\section{The effect on the acoustic peaks}\label{sec:peaks}

We will now focus our attention on possible effects of primordial
\mfs on small angular scales. That is, temperature, as well polarization,
anisotropies of the CMBR. By small angular scale  ($< 1^o$) we mean
angles which correspond to a distance smaller than the Hubble
horizon radius at the last scattering surface. Therefore, what we
are concerning about here are anisotropies that are produced by
causal physical mechanisms which are not related to the large
scale structure of the space-time.

Primordial density fluctuations, which are necessary to explain the
observed structures in the Universe, give rise to acoustic oscillations
of the primordial plasma when they enter the horizon some time before the
last scattering. The oscillations distort the primordial spectrum of
anisotropies by the following primary effects \cite{Hu97}:
{\it a}) they produce temperature fluctuations
in the plasma, {\it b}) they induce a velocity Doppler shift of photons,
{\it c}) they give rise to a gravitational Doppler shift of photons when they
climb-out or fall-in the gravitational potential well produced by the density
fluctuations (Sachs-Wolfe effect).

In the linear regime, acoustic plasma oscillations are well described by
standard fluid-dynamics (continuity + Euler equations) and Newtonian
gravity (Poisson's equation).
In the presence of a \mf  the nature of plasma oscillations can be
radically modified as Magneto-Hydro-Dynamics (MHD) has to be taken
into account.

To be pedagogical, we will first consider a single component
plasma and neglect any dissipative effect, due for example to a
finite viscosity and heat conductivity. We will also assume that
the \mf  is  homogeneous on scales larger than the plasma
oscillations wavelength. This choice allows us to treat the
background magnetic field ${\bf B}_0$ as a uniform field in our
equations (in the following symbols with the 0 subscript stand for
background quantities whereas the subscript 1 is used for
perturbations). Within these assumptions the linearized equations
of MHD in comoving coordinates are \cite{AdamsDGR}
{\footnote{Similar equations were derived by Wasserman \cite{Wasserman}
to the purpose to study the possible effect of primordial \mfs on galaxy
formation.}}:

\begin{equation}
  \dot \delta + {{\bf \nabla \cdot v_{1}}\over a} = 0~,
        \label{continuity}
\end{equation}
where a is the scale factor.

\begin{equation}
   {\bf \dot{v}}_{1} + {\dot{a}\over a}{\bf v}_{1} +
   {c_{S}^{2}\over a}
   {\bf \nabla} \delta + {{\bf \nabla} \phi_{1}\over a} +
   {{\bf \hat{B}}_{0} \times \left(\dot{\bf v}_{1} \times
   {\bf \hat{B}}_{0}\right)
   \over 4\pi a^{4}} +
   {{\bf \hat{B}}_{0} \times \left({\bf \nabla} \times
   {\bf \hat{B}}_{1}\right)
   \over 4\pi\rho_{0} a^{5}} =0~,
        \label{euler}
\end{equation}

\begin{equation}
        \partial_{t}{\bf \hat{B}}_{1} =
        {{ \bf \nabla} \times \left(\bf{v}_{1} \times
        {\bf \hat{B}}_{0}\right)
        \over a}~,
        \label{faraday}
\end{equation}

\begin{equation}
{\bf \nabla}^{2} \phi_{1} = 4\pi G \rho_{0} \left( \delta +
{{\bf \hat{B}}_{0}\cdot {\bf \hat{B}}_{1} \over 4\pi \rho_{0} a^{4}}
\right)
    \label{phi}
\end{equation}
and
\begin{equation}
        {\bf \nabla \cdot \hat{B}}_{1} = 0~,
        \label{divergenceless}
\end{equation}
where ${\bf \hat{B}} \equiv {\bf B}a^{2}$ and
$\delta=\dfrac{\rho_1 }{\rho_0}$, $\phi_{1}$ and $v_{1}$ are
small perturbations
on the background density, gravitational potential and velocity
respectively. $c_{S}$ is the sound velocity.
Neglecting its direct gravitational influence, the magnetic field
couples to fluid dynamics only through the last two terms in
Eq.(\ref{euler}).
The first of these terms is due to the displacement current
contribution to $\mathbf{\nabla} \times {B}$, whereas the latter
account for the magnetic force of the current density.
The displacement current term can be neglected provided that
\be
\label{Alfen-vel}
 v_{A} \equiv \frac{B_{0}}{\sqrt{4\pi(\rho+p)}} \ll c_{S}~,
\ee
where $v_{A}~$ is the, so called, Alfv\'en velocity.

Let us now discuss the basic properties of the solutions of
these equations, ignoring for the moment the expansion of the Universe.
In the absence of the magnetic field there are only ordinary sound waves
involving density fluctuations and longitudinal velocity fluctuations
(i.e. along the wave vector).
By breaking the rotational invariance, the presence of a \mf  allows
new kind of solutions that we list below (useful references on
this subject are \cite{Akhiezer,KapTsy}).

\begin{enumerate}

\item {\it Fast magnetosonic waves}\\
\noindent
In the limit of small magnetic fields these waves become the ordinary
sound waves. Their velocity,
$c_+$, is given by
\begin{equation}
c_+^2 \sim c_S^2 + v_A^2 \sin ^2 \theta~, \label{c+}
\end{equation}
where $\theta$ is the angle between $\mathbf k$ and ${\mathbf B}_0$.
Fast magnetosonic waves involve fluctuations in the velocity,
density, magnetic field and gravitational field. The velocity
and density fluctuations are out-of-phase by $\pi/2$.
Eq. (\ref{c+}) is valid for
$v_A << c_S$. For such fields the wave is approximatively
longitudinal.

\item {\it Slow magnetosonic waves}\\
\noindent
Like the fast waves, the slow waves  involve both density and
velocity fluctuations. The velocity is however fluctuating both
longitudinally and transversely even for small fields.
The velocity of the slow waves is approximatively
\begin{equation}
c_-^2 \sim v_A^2 \cos ^2 \theta~.
\end{equation}

\item {\it Alfv\'en waves}\\
\noindent
For this kind of waves ${\mathbf B}_1$ and ${\mathbf v}_{1}$ lie in a plane
perpendicular to the plane through $\mathbf k$ and ${\mathbf B}_0$.
In contrast
to the magnetosonic waves, the Alfv\'en waves are purely
rotational, thus they involve no density fluctuations.
Alfv\'en waves are linearly polarized.
Their velocity of propagation is
\begin{equation}
c_A^2 = v_A^2 \cos ^2 \theta~.
\end{equation}
\end{enumerate}
Detailed treatments of the evolution of MHD modes in the
matter dominated and radiation dominated eras of the Universe can
be found in Refs.\cite{Gailis,TsaMaa99}.

The possible effects of MHD waves on the temperature anisotropies of
the CMBR has been first investigated by Adams et al. \cite{AdamsDGR}
In the simplest case of magnetosonic waves, they found that the linearized
equations of fluctuations in the Fourier space are
\begin{equation}
\dot{\delta} _b + V_b - 3 \dot{\phi} =0 ,
\end{equation}
\begin{equation}
\dot{V}_b + \frac{\dot{a}}{a} V_b - c_b^2 k^2 \delta _b + k^2 \psi
+\frac{a n_e \sigma _T (V_b - V_{\gamma})}{R}
-\frac{1}{4 \pi \hat{\rho} _b a} {\bf k} \cdot \left({\bf \hat{B}} _0
\times ({\bf k} \times {\bf \hat{B}} _1)\right) = 0~, \label{tjo}
\end{equation}
for the baryon component of the plasma and
\begin{equation}
\dot{\delta}_{\gamma } +\frac{4}{3} V_{\gamma} - 4 \dot{\phi}=0
\end{equation}
\begin{equation}
\dot{V} _{\gamma} - k^2 (\frac{1}{4} \delta _\gamma -
\sigma _\gamma )
- k^2 \psi - a n_e \sigma _T (V_b - V_ \gamma ) = 0~,
\end{equation}
for the photon component.
In the above $V= i {\bf k} \cdot {\bf v}$, $R= \dfrac{p_b + \rho_b}
{p_\gamma + \rho_\gamma} = \dfrac{3 \rho_b}{4\rho_\gamma}$ and  $c_b$
is the baryon sound velocity in the absence
of interactions with the photon gas.
As it is evident from the previous equations, the coupling between the
baryon and the photons fluids is supplied by Thomson scattering
with cross section $\sigma_T$.

In the tight coupling limit ($V_b \sim V_{\gamma}$)
the photons provide the baryon fluid with a pressure term and a
non-zero sound velocity.
The magnetic field, through the last term in Eq.(\ref{tjo}),
gives rise to an additional contribution to the effective
baryon sound velocity. In the case of longitudinal waves this
amounts to the change
\begin{equation}
c_b^2 \rightarrow c_b^2 + v_A^2 \sin ^2 \theta~.
\end{equation}
In other words, the effect of the field can be somewhat mimicked by
a variation of the baryon density. A complication arises due to
the fact that the velocity of the fast waves depends on the angle
between the wave-vector and the magnetic field. As we mentioned
previously, we are assuming that the magnetic field direction changes
on scales larger than the scale of the fluctuation.
Different patches of the sky might therefore show different
fluctuation spectra depending on this angle.
\begin{figure}[t]
\vskip -1cm
\centerline{\protect \hbox{
\psfig{file=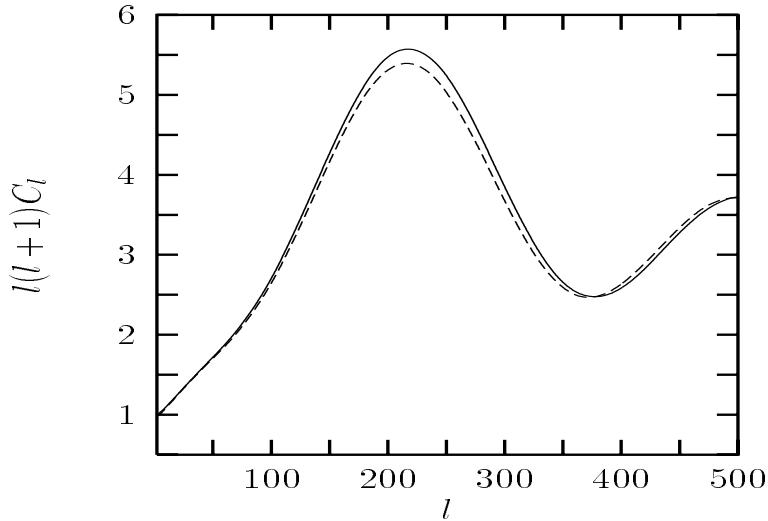,
height=13.cm,
width=13.cm,angle=0}}}
\vskip -3.0cm
\caption{The effect of a cosmic magnetic field on the multipole
moments. The solid line shows the prediction of a standard CDM cosmology
($\Omega=1$,$h=0.5$, $\Omega_{\rm B}=0.05$) with an $n=1$ primordial
spectrum of adiabatic fluctuations. The dashed line shows the effect
of adding a magnetic field equivalent to $2 \times 10^{-7}$ Gauss today.
From Ref.\cite{AdamsDGR}}
\label{fig:peaks}
\end{figure}

The authors of Ref.\cite{AdamsDGR} performed an all-sky average
summing also over the angle between the field and the
line-of-sight. The effect on the CMBR temperature power spectrum
was determined by a straightforward modification  of the CMBFAST
\cite{SelZal} numerical code. From the Fig.\ref{fig:peaks} the reader can
see the effect of a field $B_0 = 2\times 10^{-7}$ G on the first
acoustic peak. The amplitude of the peak is reduced with respect
to the free field case. This is a consequence of the magnetic
pressure which opposes the in-fall of the photon-baryon fluid in
the potential well of the fluctuation. Although this is not
clearly visible from the figure, the variation of the sound
velocity, hence of the sound horizon,  should also produce a
displacement of the acoustic peaks. The combination of these two
effects may help to disentangle the signature of the \mf from
other cosmological effects (for a comprehensive review see
\cite{KamKos}) once more precise observations of the CMBR power
spectrum will be available. Adams at al. derived an estimate of the
sensitivity to
$B$ which MAP \cite{MAP} and PLANCK \cite{PLANCK} satellites observations
should allow to reach by translating the predicted sensitivity of these
observations to $\Omega_b$. They found that a \mf with strength today
$B_0 > 5 \times 10^{-8}$ G should be detectable.

It is interesting to observe that a magnetic
field cannot lower the ratio of the first to second acoustic peak as
showed by recent observations \cite{Edsjo}.
\vskip 0.6cm
\noindent
{\bf Alfv\'en waves} may also leave a signature on the CMBR anisotropies.
There are at least three main reasons which make this kind of wave of
considerable interest. The first is that Alfv\'en waves should
leave a quite peculiar imprint on the CMBR power spectrum.
In fact, as we discussed in the above, these waves do not
involve fluctuations in the density of the photon-baryon fluid.
Rather, they consist only of oscillations of the fluid velocity and of
the magnetic field. Indeed, by assuming that the wavelength is smaller than
the Hubble radius and that relativistic effects are negligible,
the equations describing Alfv\'en waves are \cite{AdamsDGR}
\begin{eqnarray}
        &\delta_{b}& = \delta_{\gamma} = 0~,\\
        &{\bf \dot{v}}_{b}& + \frac{\dot a}{a}{\bf v}_{b}
        + \frac {a n_e \sigma _T
        ({\bf v}_b - {\bf v}_{\gamma})}{R} -
        i \frac {({\bf k \cdot \hat{B}_{0}})} {4\pi \hat{\rho}_{b} a}
        {\bf \hat{B}}_{1} =0~,\\
        &{\bf \dot{v}}_{\gamma}& -a n_e \sigma _T ({\bf v}_b -
        {\bf v}_{\gamma})
        = 0\\
        &\phi& = 0~.
\end{eqnarray}
Since the gravitational Doppler shift (Sachs-Wolfe effect) is
absent in this case, the cancellation against the velocity Doppler
shift which occurs for the acoustic modes \cite{Hu97} does not
take place for the Alfv\'en waves.
This could provide a more clear signature of the presence of
magnetic fields at the last scattering surface \cite{AdamsDGR}.

The second reason why Alfv\'en waves are so interesting in this contest is
that they are vector (or rotational) perturbations.
As a consequence they are well suited to probe peculiar initial condition
such as those that might be generated from primordial phase-transitions.
It is remarkable that whereas vector perturbations are suppressed
by universe expansion and cannot arise from small deviations from the
isotropic Friedmann Universe for $t \rightarrow 0$ \cite{ZelNov},
this is not true in the presence of a cosmic \mf
{\footnote{ Collisionless matter, like e.g. gravitons after the Planck era,
may however support nonzero vorticity even with initial conditions compatible
with an isotropic universe \cite{Rebhan92}. }}.

The third reason of our interest for Alfv\`en waves is that for this kind
of waves the effect of dissipation is less serious than what it is
for sound and fast magnetosonic waves. This issue will be touched upon in
the next section.

A detailed study of the possible effects of Alfv\'en waves on the CMBR
anisotropies has been independently performed by Subramanian and Barrow
\cite{SubBar98b} and Durrer at al.\cite{DurrerKY} who reached similar
results. We summarize here the main points of the derivation as given in
Ref.\cite{DurrerKY}.

In general, vector perturbations of the metric have the form
\be
        \left(h_{\mu\nu}\right) =\left(\begin{array}{cc}
        0 & B_i \\
        B_j & H_{i,j} + H_{j,i}\end{array} \right)~,
\ee
where  ${\bf B}$ and ${\bf H}$ are divergence-free, 3d vector fields
supposed to vanish at infinity.
Two gauge-invariant quantities \cite{Bardeen80} are conveniently introduced
by the authors of Ref.\cite{DurrerKY}:
\be
 \mathbf{\sigma} = \dot{\bf H}-{\bf B} ~~~\mbox{ and }~~~~
  {\bf \Omega} ={\bf v-B} ~.
\ee
which represents the vector contribution
to the perturbation of the extrinsic curvature and the vorticity.
In the absence of the magnetic field, and assuming a perfect fluid equation
of state,
the vorticity equation of motion is
\be
\dot{\bf\Omega} + (1-3c_s^2){\dot{a}\over a} {\bf \Omega} = 0~.
\ee
In the radiation dominated era the solution of this equation is
$\Omega = \mathrm{const}.$ which clearly does not describe waves and,
as we mentioned, is incompatible with an isotropic universe when
$t \rightarrow 0$. In the presence of the magnetic field, Durrer
et al. found
\be
\ddot{{\bf \Omega}} ={({\bf B}_0\cdot {\bf k})^2 \over 4\pi(\rho_r+p_r)}
        {\bf \Omega}~~ \mbox{ and}
\ee
\be
\dot{{\bf \Omega}} ={i{\bf B}_0\cdot {\bf k} \over 4\pi(\rho_r+p_r)}
        {\bf B}_1~.
\ee
These equations describe Alfv\'en waves propagating at the velocity
$v_A({\bf e}\cdot\hat{\bf k})$, where $v_A$ is the Alfv\'en velocity
and ${\bf e}$ is the unit vector in the direction of the magnetic
field {\footnote{Differently form the authors of Ref.\cite{AdamsDGR},
Durrer at al. assumed a homogeneous background \mf. This however is not a
necessary condition for the validity of the present considerations.}}.
In this case some amount of initial vorticity is allowed which is
connected to the amplitude of the \mf perturbation ${\bf B}_1$
\be
|\mathbf{\Omega_0}|=(v_A/B_0)|{\bf B}_1| ~.
\ee

The general form of the CMBR temperature anisotropy produced by vector
perturbations is
\be
\left({\Delta T\over T}\right)^{(\mathrm{vec})} =
        -\left.{\bf V\cdot n}\right|_{t_{dec}}^{t_0} +
  \int_{t_{dec}}^{t_0}\dot{{\bf \sigma}}\cdot {\bf n}d\lambda   ~,
\ee
where ${\bf V} =\mathbf{\Omega-\sigma}$ is a gauge-invariant generalization
of the velocity field. We see from the previous equation that besides the
Doppler effect Alfv\`en waves gives rise to an integrated Sachs-Wolfe term.
However, since the geometric perturbation $\mathbf{\sigma}$ is decaying
with time, the integrated term is dominated by its lower boundary and just
cancels ${\bf \sigma}$ in ${\bf V}$.
Neglecting a possible dipole contribution from vector perturbations
today, Durrer at al. obtained
\be
{\delta T \over T} ({\bf n}, {\bf k}) \simeq {\bf n}\cdot{\bf \Omega}
({\bf k},t_{dec})=  {\bf n}\cdot{\bf \Omega}_0
\sin\left( v_A kt_{dec} ({\bf e}\cdot \widehat{\bf k}) \right)~.
\ee
As predicted in Ref.\cite{AdamsDGR}, Alfv\'en waves produce Doppler peaks
with a periodicity which is determined by the Alfv\'en velocity.
Since, for reasonable values of the \mf strength, $v_A \ll 1$ this
peaks will be quite difficult to detect.

Durrer et al. argued that  Alfv\'en waves may leave a phenomenologically more
interesting signature on the statistical properties of the CMBR anisotropies.
In the absence of the \mf all the relevant information is encoded in the
$C_\ell$'s coefficients
defined by
\be
 \left.\left\langle{\delta T\over T}({\bf n}){\delta T\over T}({\bf n}')
\right\rangle\right|_{{~}_{\!\!({\bf n} \cdot {\bf n}'=\mu)}} =
  {1\over 4\pi}\sum_\ell(2\ell+1)C_\ell P_\ell(\mu)~,
\ee
where $\mu \equiv {\bf n} \cdot {\bf n}'$. By introducing the usual
spherical harmonics decomposition
\be
  {\delta T\over T}({\bf n})=\sum_{\ell,m}a_{\ell m}Y_{\ell m}({\bf n}),
\ee
the $C_\ell$'s  are just
\be
C_\ell = \langle a^{}_{\ell m}a^*_{\ell m}\rangle~.
\ee
Because of its  spin-1 nature, the vorticity vector field induces
transitions $\ell\rightarrow\ell\pm 1$ hence a correlation between
the multipole amplitudes $a_{\ell+1, m}$ and $a_{\ell-1, m}$.
This new kind of correlation is encoded in the coefficients
\be
D_{\ell}(m) = \langle a^{}_{\ell-1, m}a^*_{\ell+1, m}\rangle =
\langle a^{}_{\ell+1, m}a^*_{\ell-1, m}\rangle~.
\ee

Durrer at al. \cite{DurrerKY} determined the form of the $C_\ell$ and
$D_\ell$ coefficients for the case of a homogeneous background \mf
in the range $-7 < n < - 1$, where $n$ determine the vorticity power
spectrum according to
\begin{eqnarray}
        \left\langle\Omega_{0i}({\bf k})\Omega_{0j}({\bf k})\right\rangle
        &=&(\delta_{ij}-\hat{k}_i\hat{k}_j)A(|{\bf k}|) \\
A(k)=&A_0& {k^n\over k_0^{(n+3)}}, \qquad k < k_0~.
\end{eqnarray}
On the basis of these considerations they found that 4-year COBE data
allow to obtain a limit on the \mf amplitude in the range $-7 < n < - 3$
on the order of $(2-7)\times10^{-9}$Gauss.

\section{Dissipative effects on the MHD modes}\label{sec:dissipation}

In the previous section we neglected any dissipative effect which may
possibly affect the evolution of the MHD modes. However, similar to the damping
 of baryon-photon sound waves by photon shear viscosity and heat conductivity,
damping of MHD perturbations may also occur. This issue was studied in
detail by Jedamzik, Katalini\'c and Olinto \cite{JedKO98} who first
determined the damping rates of fast and slow magnetosonic waves as well as
of Alfv\'en waves. Furthermore, it was shown in Refs.\cite{JedKO98,SubBar98a}
that dissipation of MHD modes produce an effective damping of inhomogeneous
magnetic fields. The dissipation process goes as follows. A spatially tangled
magnetic field produces Lorentz forces which accelerate the plasma and set up
oscillations. Since the radiation-baryon pressure is much larger than the
magnetic pressure, as long as the photon mean-free-path is smaller than
the scale of the magnetic tangle, the motions can be
considered as being largely incompressible. In this situation mainly
 Alfv\'en waves, which do not involve density fluctuations, are excited.
In the absence of dissipation, this process will continue until, for all
scales $\lambda$ with magnetic field relaxation time
$\tau \sim \lambda/v_A$ shorter than the  Hubble time $t_H$, an approximate
equipartition between magnetic and kinetic energies is produced.
If the fluid is non-ideal, however, shear viscosity
will induce dissipation of kinetic energy, hence also of magnetic energy, into
heat.  In this case dissipation will end only when the magnetic field reaches
a force-free state.

In the absence of magnetic fields it is known that in the diffusive regime
(i.e. when the perturbation wavelength is much larger than the mean free path
of photon or neutrinos) acoustic density fluctuations are effectively damped
because of the finite viscosity and heat conductivity
(Silk damping \cite{Silkdamp}). At recombination time, dissipation occurs for modes smaller
than the approximate photon diffusion length,
$d_\gamma \sim (l_\gamma t_H)^{1/2}$, where $l_\gamma$ is photon mean free path.
The dissipation of fast magnetosonic waves proceeds in a quite similar way.
Indeed, it was showed in Ref.\cite{JedKO98} that the dissipation length scale
of these kind of waves coincide with the Silk damping scale.
More interesting is the result found in Refs.\cite{JedKO98,SubBar98a} which
shows that damping of Alfv\'en and slow magnetosonic waves is
significantly different from damping of sound and fast magnetosonic waves.
The reason for such a different behavior is that, for a small background
magnetic field $v_A \ll 1$ so that the oscillation frequency of an Alfv\'en
mode ($v_A k/a$) is much smaller than the oscillation frequency of a fast
magnetosonic mode with the same wavelength ($v_{\rm sound} k/a$).
While all magnetosonic modes of
interest satisfy the condition for damping in the oscillatory regime
($v_{\rm sound}\ll l_\gamma k/a$), an Alfv\'en mode can become
{\it overdamped} when the photon (or neutrino) mean-free-path becomes large
enough for dissipative effects to overcome the oscillations
($v_A \cos\theta\simeq l_\gamma(T) k/a$, where $\theta$ is the angle between
the background \mf and the wave vector). Because of the strong viscosity, that
prevent fluid acceleration by the magnetic forces, damping is quite
inefficient for non-oscillating overdamped Alfv\'en modes with
\begin{equation}
\lambda \le \lambda_{od} \simeq \frac{2\pi l_\gamma(T)}{v_A \cos\theta}~.
\end{equation}
As a result, the damping scale of overdamped Alfv\'en modes at the end of
the diffusion regime is smaller than the damping scale of sound and
fast magnetosonic modes (Silk damping scale) by a factor which depends
on the strength of the background \mf and the $\theta$ angle,
$L_A \sim v_A \cos \theta d_\gamma$.

From the previous considerations it follows that
the results discussed in the previous section hold only under the
assumption that the magnetic field  coherence length is not much smaller than
the comoving Silk damping  scale ($L_S \sim 10$ Mpc), in the case of fast
magnetosonic waves, and not smaller than $L_A$ for Alfv\'en waves.
\vskip0.4cm
Some other interesting work has been recently done by Jedamzik, Katalini\'c
and Olinto \cite{JedKO99} concerning the effects of dissipation of small-scale
\mfs on the CMBR. The main idea developed in the paper by Jedamzik et al.
is that the dissipation of tangled \mfs before the recombination epoch
should give rise to a nonthermal injection of energy into the heat-bath
which may distort the thermal spectrum of CMBR.
It was showed by the authors of Ref.\cite{JedKO99}
that once photon equilibration has occurred, mainly via photon-electron
scattering and double-Compton scattering, the resultant distribution should
be of Bose-Einstein type with a non-vanishing chemical potential.
The evolution of the chemical potential distortions at large frequencies may
be well approximated by \cite{HuSilk}
\begin{equation}
\label{muevol}
\frac{d\mu}{dt} = - \frac{\mu}{t_{DC}(z)} + 1.4 \frac{Q_B}{\rho_\gamma}~,
\end{equation}
where, in our case, $Q_B = d\rho_B/dt$ is the dissipation rate of the magnetic
field and $t_{DC} = 2.06\times 10^{33}~{\rm s}~(\Omega_b h^2)^{-1} z^{-9/2}$
is a characteristic time scale for double-Compton scattering.
Jedamzik et al. assumed a statistically isotropic \mf configuration with the
following power spectrum
\begin{equation}
\vert {\tilde{b}}_k \vert^2 = B_0^2\left(\frac {k}{k_N} \right)^n
\frac{(n + 3)}{4\pi}
\qquad {\rm for} ~~~~~~ k < k_N
\end{equation}
and zero otherwise, normalized such that $\langle {\tilde b}^2 \rangle = B_0^2$.
The energy dissipation rate was determined by substituting this spectrum in
the following Fourier integral
\begin{equation}
\label{QB}
Q_B = {1\over 8\pi k_N^3}\int d^3k\, {d|{{\tilde b}_k}|^2\over dt}
= {1\over 8\pi k_N^3}\int d^3k\,|{{\tilde b}_k}|^2\,\,
(2 \Im \omega)\,
\exp\biggl(-2\int \Im\omega dt\biggr)~,
\end{equation}
together with the  mode frequencies for Alfv\'en and slow magnetosonic waves
determined in Ref.(\cite{JedKO98})
\begin{equation}
\label{dispersion}
\omega_{\rm osc}^{\rm SM,A}=v_A{\rm cos}\theta\biggl({k\over a}\biggr)
+{3\over 2}i{\eta'\over (1+R)}\biggl({k\over a}\biggr)^2\, ,
\end{equation}
where  $3(\rho_{\gamma} + p_{\gamma})\eta' =\eta$, and $\eta$ is the shear
viscosity. For  $k_N\gg k_D^0z_{\mu}^{3/2}$, where
$k_D^0= (15n_e^0\sigma_{Th}/2.39\times10^19~{\rm s})^{1/2}$,
an analytic solution of Eq.(\ref{muevol}) was then found to be
\begin{equation}
\label{mu}
\mu = K{B_0^2\over 8\pi \rho_{\gamma}^0}\biggl({k_D^0\over k_N}
z_{\mu}^{3/2}\biggr)^{(n+3)}~.
\end{equation}
In the above $K$ is a numerical factor of order 1, the precise value depending
on the spectral index $n$ and $z_\mu$ is the characteristic redshift for
``freze-out'' from double-Compton scattering. This redshift equals
$z_\mu = 2.5 \times 10^6$ for typical values $\Omega_b h^2 = 0.0125$, and
$Y_p=0.24$.
The scale $k_D^0z_{\mu}^{3/2}$ has a simple interpretation. It is the
scale which at redshift $z_{\mu}$ is damped by one e-fold. For the
above values of $\Omega_b h^2$ and $Y_p$ the corresponding comoving
wavelength is $\lambda_D = (2\pi)/(k_D^0z_{\mu}^{3/2}) = 395 {\rm
pc}$.

The present upper limit on chemical potential distortion of the CMBR come
from the COBE/FIRAS data: $|\mu |< 9\times 10^{-5}$ at
$95\%$ confidence level \cite{FIRAS}.
Comparing this limit with the prediction of Eq.(\ref{mu})
it follows that primordial magnetic fields of strength ${}^>_{\sim}\,
3\times 10^{-8}$ G, and comoving
coherence length $\approx 400$ pc are probably excluded.
On slightly larger scales, dissipation of spatially tangled \mfs may
give to a different kind of CMBR distortion which may be described by a
superposition of blackbodies of different temperature, i.e. a Compton $y$
distortion \cite{ZelIS}. The absence of this kind of distortions in the
observed CMBR thermal spectrum disallow \mfs of $\simgeq
3\times 10^{-8}$Gauss on scales $\sim 0.6$ Mpc.

\section{Effects on the CMBR polarization}\label{sec:polarization}

Thomson scattering is a natural polarizing mechanism for the CMBR.
It is enough if the photon distribution function seen by the electrons has
a quadrupole anisotropy  to obtain polarization.
At early times, the tight coupling between the photons and the
electron-baryon fluid prevents the development of any photon anisotropy in
the baryon's rest frame, hence the polarization vanishes.
As decoupling proceeds, the photons begin
to free-stream and temperature quadrupole anisotropies can source
a space dependent polarization.
For this reason temperature and polarization anisotropies are expected to
be correlated (for a comprehensive review on the subject see \cite{Kosowsky96}).

The expected polarization anisotropy is not large, perhaps about $10^{-6}$.
Currently the best polarization limit  comes from the
Saskatoon experiment \cite{Saskatoon}, with a 95\% confidence level upper
limit of $25~\mu{\mathrm K}$  at angular scales of about a degree,
corresponding to $9 \times 10^{-6}$ of the mean temperature.
Future balloons and satellites observations, like e.g. the PLANCK
\cite{PLANCK} mission to be launched in 2007, are expected to have a good
chance to measure the CMBR polarization power spectrum.

Kosowsky and Loeb \cite{KosLoe96} first observed that
the possible presence of \mfs at the decoupling time may induce a
sizeable Faraday rotation in the CMBR.  Since the rotation angle depend on
the wavelength, it is possible to estimate this effect by comparing the
polarization vector on a given direction at two different frequencies.
The basic formula is \cite{Jackson}
\begin{equation}
\label{Faraday}
\phi = \frac{e^3n_ex_e {\mathbf B}\cdot{\hat{\bf q}} \lambda^2L}{8\pi^2m^2c^2}
\end{equation}
where $\phi$ is the amount by which the plane of polarization of linearly
polarized radiation has been rotated, after traversing a distance $L$ in a
homogeneous magnetic field $B$ in a direction $\hat{\bf q}$.~
$x_e$ is the ionized fraction of the total electron density $n_e$ and $m$ is
the electron mass. Finally $\lambda$ is the radiation wavelength.

Although the \mf strength is expected to be larger at early times, the
induced Faraday  rotation depends also on the free electron density
(see Eq.(\ref{Faraday})) which drops to negligible values as recombination
ends. Therefore, rotation is generated during the brief
period of time when the free electron density has dropped enough to end
the tight coupling but not so much that Faraday rotation ceases.
A detailed computation requires the solution of the radiative transport
equations in comoving coordinates \cite{KosLoe96}
\begin{eqnarray}
\dot\Delta_T + ik\mu(\Delta_T+\Psi) &=& -\dot\Phi-\dot\kappa[\Delta_T-
\Delta_T(0)- \mu V_b +\frac{1}{2}P_2(\mu)S_P] \label{DeltaT}\\
\dot\Delta_Q ik\mu\Delta_Q &=& -\dot\kappa[\Delta_Q-\frac{1}{2}
(1-P_2(\mu)S_P +2\omega_B\Delta_U] \label{DeltaQ}\\
\dot\Delta_U +ik\mu\Delta_U &=& -\dot\kappa\Delta_U - 2\omega_B\Delta_Q~.
\label{DeltaU}
\end{eqnarray}
In the above $\Delta_T$, $\Delta_Q$ and $\Delta_U$ respectively represent
the fluctuations of temperature and of the
of the Stokes parameters Q and U \cite{Jackson}.
The linear polarization is $\Delta_P=\sqrt{\Delta_Q^2 + \Delta_U^2}$.
The numerical subscripts on the radiation brightnesses $\Delta_X$
indicate moments defined by an expansion of the
directional dependence in Legendre polynomials $P_\ell(\mu)$:
\begin{equation}
\Delta_{X\,\ell}(k) \equiv {1\over 2}\int_{-1}^1 d\mu\,
P_\ell(\mu) \Delta_X(k,\mu).
\label{moments}
\end{equation}
Therefore, the subscripts $0,~1,~2$ label respectively monopole, dipole and
quadrupole moments.
~ $\displaystyle \dot\kappa=x_en_e\sigma_T\frac{\dot a}{a}$
is the differential optical depth and the quantities $V_b$ and $R$
have been defined in the previous section.
Derivatives are respect to conformal time.
Finally,  $\omega_B$ is the Faraday rotation rate
\begin{equation}
\omega_B= \frac{d\phi}{d\tau}= \frac{e^3n_ex_e{\bf B}\cdot{\bf {\hat q}}}
{8\pi^2m^2\nu^2}\frac{a}{a_0}~.
\end{equation}
From Eqs.({\ref{DeltaQ},\ref{DeltaU}) wee see that Faraday rotation mixes $Q$
and $U$ Stokes parameters.
The polarization brightness $\Delta_Q$ is induced by the
function $S_P= -\Delta_T(2)-\Delta_Q(2) + \Delta_Q(0)$
and $\Delta_U$ is generated as $\Delta_Q$ and $\Delta_U$ are
rotated into each other. In the absence of magnetic
fields, $\Delta_U$ retains its tight-coupling value of zero.
The set of Eqs.(\ref{DeltaT}-\ref{DeltaU}) is not easily solved. A convenient
approximation is the tight coupling approximation which is an expansion
in powers of $k\tau_C$ with $\tau_C=\dot\kappa^{-1}$.
This parameter measures the average conformal time between collisions.
At decoupling the photon mean free
path grows rapidly and the approximation breaksdown, except for long
wavelength as measured with respect to the thickness of the last scattering
surface.
For these frequencies the approximation is still accurate.

Kosowsky and Loeb assumed a uniform magnetic field on the scale of the width
of the last scattering surface, a comoving scale of about 5 Mpc. This
assumption is natural if the coherent magnetic field observed in galaxies
comes from a primordial origin, since galaxies were assembled from a comoving
scale of a few Mpc. The mean results was obtained by averaging over the
entire sky. Therefore the equations still depend only on $k$ and
$\mu = \cos({\bf\hat k\cdot\hat q})$ and not on the line-of-sight vector
$\bf\hat q$ and the perturbation wave-vector $\bf k$ separately.
The evolution of the polarization brightnesses, for given values of
$k$ and $\mu$ is represented in Fig.\ref{fig:KosLob}
as a function of the redshift.
\begin{figure}[t]
\centerline{\protect\hbox{
\psfig{file=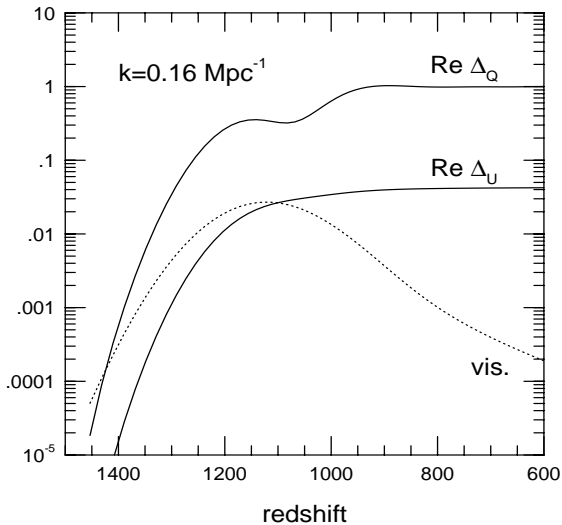,
height=10.cm,
width=14.cm,angle=90}}}
\vskip -1.5cm
\caption{The evolution of the polarization brightnesses, for
$k=0.16\,{\rm Mpc}^{-1}$ and $\mu = 0.5$ (in
arbitrary units). Also plotted as a dotted line is
the differential visibility function $\dot\tau e^{-\tau}$
in units of ${\rm Mpc^{-1}}$. From Ref.\cite{KosLoe96}.}
\label{fig:KosLob}
\end{figure}

By following the approach described in the above Kosowsky and Loeb estimated
the polarization angle produced by a \mf on CMB photons with
frequency $\nu_1 < \nu_2$ to be
\begin{equation}
\langle \varphi_{12}^2\rangle^{1/2} =  1.1^{\circ} \left(1-{\nu_1^2\over
\nu_2^2}\right)
                  \left({B_0\over 10^{-9}\,{\rm G}}\right)
                  \left({30 \,{\rm GHz}\over \nu_1}\right)^2~.
\label{rotation}
\end{equation}
A $10~\%$ correction may apply to this expression to account for the effects
$\Omega_b h^2$ and $\Omega_0 h^2$ (in the range $\Omega_b h^2>0.007$
and $\Omega_0h^2<0.3$).

For a primordial field of $B_0 \sim 10^{-9}~{\rm G}$ which
could result in the observed galactic field without dynamo amplification,
one can therefore expect a
rotation measure of order $1.6~{\rm deg~cm^{-2}}=280~{\rm rad~m^{-2}}$.
This rotation is considerable by astrophysical standards and could
in principle be measured.
\vskip0.4cm
We noticed at the beginning of this section that temperature and polarization
anisotropies of the CMBR are generally expected to be correlated. The
statistical properties of such correlation may be affected by the presence
of a \mf at the decoupling time in a peculiar way. In fact, it was
showed by Scannapieco and Ferreira \cite{ScaFer} that
such a field may induce an observable parity odd cross correlation between
polarization and temperature anisotropies.
Any polarization pattern on the sky can be separated into ``electric'' ($E$)
and ``magnetic'' ($B$) components. The nomenclature reflects the global parity
property.  Like multipole radiation, the harmonics of an $E$-mode have
$(-1)^\ell$ parity on the sphere, whereas those of a $B$-mode have
$(-1)^{\ell + 1}$ parity.
Indeed, given a measurement of the Stokes parameters $Q$ and $U$, this data can
be decomposed into a sum over spin $\pm 2$ spherical harmonics
\begin{equation}
(Q \pm iU)({\bf n}) = \sum_{\ell m} a^{\pm 2}_{\ell m}~
{}_{\pm 2}Y_{\ell m}({\bf n})~.
\end{equation}
Under parity inversion, ${}_sY_{\ell m}\rightarrow(-1)^\ell~{}_sY_{\ell m}$
so that ${}_{2}Y_{\ell m} \pm {}_{-2}Y_{\ell m}$ are parity eigenstates.
It is then convenient to define the coefficients
\begin{equation}
a^{E}_{\ell m} \equiv - \frac 1 2 (a^{2}_{\ell m} + a^{- 2}_{\ell m} )
\quad
a^{B}_{\ell m} \equiv  \frac i 2 (a^{2}_{\ell m} - a^{- 2}_{\ell m} )
\end{equation}
so that the $E$-mode  remains unchanged under parity inversion
for even $\ell$, whereas the $B$-mode changes sign.

In an isotropic Universe, cross correlation between the $B$ and $E$
polarizations is forbidden as this would imply parity violation.
Magnetic fields, however, are maximally parity violating and
therefore they may reveal their presence by producing such a cross correlation,
Faraday rotation being the physical process which is responsible for this
effect.
The authors of Ref.\cite{ScaFer} determined the expected cross correlation
between temperature and $E$ and $B$ polarization modes. On the basis of such
result they concluded that magnetic fields strengths as low as $10^{-9}$
(present time value obtained assuming adiabatic scaling) could be detectable
by the PLANCK satellite mission.
It is worthwhile to note that Scannapieco and Ferreira only considered
homogeneous magnetic fields. We note, however, that most of their
considerations  should apply also to the case of \mfs with a finite coherence
length. In this case measurements taken in different patches of the sky
should present different temperature-polarization cross correlation depending
on the \mf and the line-of-sight direction angle.
\vskip0.4cm
The consequences of Faraday rotation may go beyond the
effect they produce on the CMBR polarization.
Indeed, Harari, Hayward and Zaldarriaga \cite{Harari97}
observed that Faraday rotation may also perturb the temperature power
spectrum of CMBR. The effect mainly comes as a back-reaction of the radiation
depolarization which induces a larger photon diffusion length reducing
the viscous damping of temperature anisotropies.

In the absence of the magnetic field ($\omega_B=0$), to the first order
in the tight-coupling approximation one finds
\begin{equation}
\Delta_U=0,\quad     \Delta_Q= \frac{3}{4}S_P sin^2(\theta)
\end{equation}
and
\begin{equation}
S_P= - \frac{5}{2}\Delta_T(2)=\frac{4}{3}ik\tau_C \Delta_T(1)=
-\frac{4}{3} \tau_C {\dot \Delta}_0
\end{equation}
$\Delta_0= \Delta_T(0)+ \Phi$. Obviously, all multipoles with
$l > 3$ vanish to this order. Replacing all quantities in terms of
$\Delta_0$ one obtains \cite{Hu}:
\begin{equation}
\ddot \Delta + \left(\frac{\dot R}{1+R}+\frac{16}{45}\frac{k^2\tau_C}{1+R}
\right)\dot \Delta_0 +\frac{k^2}{3(1+R)}\Delta_0 =
\frac{k^2}{3(1+R)}\left(\Phi-(1+R)\Psi\right)
\end{equation}
that can be interpreted as the equation of a forced oscillator in the presence
of damping.

In the presence of the \mf $\omega_B \neq 0$.
The depolarization depends upon two angles: a) the angle between the
magnetic field and wave propagation and b) the angle of the field with the
wave vector ${\bf k}$. Since we assume that the vector ${\bf k}$ is
determined by stochastic Gaussian fluctuations, its spectrum will have no
preferred  direction.
Therefore this dependence will average out when integrated.
It is also assumed that for evolution purposes, the magnetic field has no
component perpendicular to ${\bf k}$.
This imposed axial symmetry is compatible with the derivation of the above
written Boltzmann equations.
Under these assumptions Harari et al. found \cite{Harari97}
\begin{equation}
\label{DUDQ}
\Delta_U=-F\cos\theta\Delta_Q\quad\quad
\Delta_Q=\frac 34 \frac{S_P \sin^2\theta}{(1+F^2\cos^2\theta)}
\end{equation}
where the coefficient $F$ was defined by
\begin{equation}
F\cos\theta\equiv  2\omega_B\tau_C
\end{equation}
which gives
\begin{equation}
F= \frac{e^3}{4\pi^2m^2\sigma_T}\frac{B}{\nu^2}\sim 0.7
\left(\frac{B_*}{10^{-3}{\rm G}}\right)
\left(\frac{10~ {\rm Ghz}}{\nu_0}\right)^2~.
\end{equation}
Physically, $F$ represents the average Faraday rotation between two
photon-electron scattering. Note that assuming perfect conductivity
\begin{equation}
B(t)=B(t_*)\left(\frac{a(t_*)}{a(t)}\right)^2
\end{equation}
and therefore $F$ is a time independent quantity.
Faraday rotation between collisions becomes considerably large at frequencies
around and below $\nu_d$. This quantity is implicitly defined by
\be
F \equiv  \left(\frac {\nu_d}{\nu_0} \right)^2
\ee
which gives
\begin{equation}
\nu_d\sim 8.4~{\rm Ghz}~ 9 \left(\frac{B_*}{10^{-3}~{\rm G}}\right)^\frac{1}{2}
\sim 27~{\rm Ghz}~\left(\frac{B_*} {10^{-2}~{\rm G}}\right)^\frac{1}{2}~.
\end{equation}
From Eqs.(\ref{DUDQ}) and, the definition of $S_P$ given in the first part of
this section, one can extract
\begin{equation}
\Delta_{Q_0}=\frac 12 d_0(F)S_P,\quad \quad
\Delta_{Q_2}=-\frac 1{10} d_2(F) S_P ,
\end{equation}
\begin{equation}
\Delta_{T_2}=-S_P(1-\frac 35 d)
\label{DT}
\end{equation}
and, from the equation for $\Delta_T$ in the
tight coupling,
\begin{equation}
S_P=\frac 4{3(3-2d)} ik\tau_C\Delta_{T_1}
=-\frac 4{3(3-2d)}\tau_C\dot\Delta_0~ .
\label{SP}
\end{equation}
In the above  the coefficients are defined so that $d_i\approx 1+ O(F^2)$
for small $F$, i.e. small Faraday rotation, while $d_i \rightarrow O(1/F)$
as $F\rightarrow\infty$ (for the exact definition see Ref.\cite{Harari97}).
Equations (\ref{DUDQ},\ref{DT}) and (\ref{SP}) condense the main effects of
a magnetic field upon polarization.
When there is no magnetic field $(F=0,d=1)$
$\Delta_U=0$ and $\Delta_{Q}=-\frac {15}{8}\Delta_{T_2}\sin^2\theta$. A
magnetic field generates $\Delta_U$, through Faraday rotation,
 and reduces $\Delta_{Q}$. In the limit of very large $F$
(large Faraday rotation between collisions) the polarization vanishes.
The quadrupole anisotropy $\Delta_{T_2}$ is also reduced by the
depolarizing effect of the
magnetic field, by a factor 5/6 in the large $F$ limit, because
of the feedback of $\Delta_Q$ upon the anisotropy or, in
other words, because of the polarization dependence of
Thomson scattering. The dipole $\Delta_{T_1}$ and monopole
$\Delta_{T_0}$ are affected by the magnetic field only
through its incidence upon the damping mechanism due to photon
diffusion for small wavelengths.
Indeed, the equation for $\Delta_0=\Delta_{T_0}+\Phi$,
neglecting $O(R^2)$ contributions, now reads
\begin{equation}
\label{hararieq}
{\ddot \Delta}_0+
\left(\frac {\dot R}{1+R}
+\frac {16}{90}\frac{(5-3d)}{(3-2d)}\frac {k^2\tau_C}{(1+R)}\right)
{\dot \Delta}_0+
\frac {k^2}{3(1+R)}\Delta_0=
\frac {k^2}{3(1+R)}\left(\Phi-(1+R)\Psi\right)
\end{equation}
which is the equation of a damped harmonic oscillator.
\begin{figure}[t]
\centerline{\protect\hbox{
\psfig{file=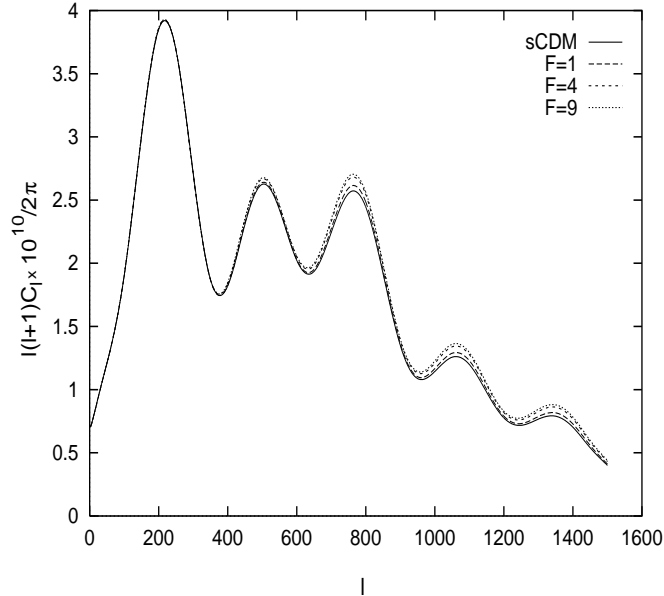,
height=8.cm,
width=9.cm,angle=0}}}
\caption{Numerical integration for the
multipoles of the anisotropy correlation function
in a standard CDM model without a primordial
magnetic field $(F=0)$, and with $F=1,\ 4,\ 9$,
which correspond
to $\nu_0=\nu_{\rm d},\ \nu_{\rm d}/2,\ \nu_{\rm d}/3$
respectively, with $\nu_{\rm d}\approx$ 27 GHz
$(B_*/0.01{\rm Gauss})^{1/2}$. From Ref.\cite{Harari97}.}
\label{fig:Harari}
\end{figure}

The damping of the temperature anisotropies on small angular scales
can be determined by solving the radiative transfer  equation to
second order in the tight-coupling approximation. By assuming solutions of the
form
\begin{equation}
\Delta_{X}(\tau)=\Delta_X e^{i\omega\tau}
\end{equation}
for $X=T$, $Q$, and $U$, and similarly for the baryon velocity $V_b$,
Harari et al. \cite{Harari97} found the following solution for
Eq.(\ref{hararieq})
\begin{equation}
\omega = \frac {k}{\sqrt{3(1+R)}} + i \gamma~,
\end{equation}
where the photon-diffusion damping length-scale is
\begin{equation}
\gamma (d)\equiv \frac{k^2}{k_D^{2}}=
\frac {k^2\tau_C}{6(1+R)} \Big(\frac {8}{15}
\frac {(5-3d)}{(3-2d)} + \frac {R^2}{1+R}\Big)~.
\end{equation}
The damping affects the multipole coefficients of the anisotropy power spectrum
which are defined by
\begin{equation}
C_l=(4\pi)^2 \int k^2 dk P(k)|\Delta_{T_l}(k,\tau_0)|^2~.
\end{equation}
The average damping factor due to photon diffusion upon the $C_l$'s
is given by an integral of $e^{-2\gamma}$ times the visibility function
across the last scattering surface \cite{Hu,HuSug95}. It depends upon
cosmological parameters, notably $R$, and upon the recombination history.

In the Fig.\ref{fig:Harari} it is represented the correction to the
temperature power spectrum expected for several values of the
parameter $F$.
We see from that figure that on small angular scales the effect of the \mf
is to increase the temperature anisotropies. The magnitude of this
effect was estimated to be up to 7.5\% in a CDM Universe
on small angular scales ($l \approx 1000$) at a level that should be
reachable from future CMBR satellite experiments like MAP \cite{MAP}
and PLANCK \cite{PLANCK}.
The frequency at which the effect should be detectable will,  however,
depend on the strength and coherence length of the \mf at the
recombination time. Both experiment should be sensitive to \mfs
around $B_{z = 1000} = 0.1{\rm G}$ or, equivalently, $B_0 = 10^{-7}{\rm G}$
a level that is comparable to be BBN limit (see Chap.\ref{chap:bbn}).


\chapter{Constraints from the Big Bang Nucleosynthesis}\label{chap:bbn}

The study of the influence of magnetic fields on the Big Bang
Nucleosynthesis (BBN) began with the pioneering works of Matese
and O'Connell \cite{MatOco69b,MatOco69,MatOco70} and Greenstein
\cite{Greenstein}. It is remarkable that most of the more relevant
effects were already pointed-out in those early papers.

In their first paper on the subject Matese and O'Connell
\cite{MatOco69b} showed that in the presence of  very strong
magnetic fields, $B > B_c \equiv eB/m_e^2 = 4.4 \times 10^{13}$ G
(above this field strength quantized magnetic levels, ``cyclotron
lines", appear), the $\beta$ decay rate of neutrons is
significantly increased. This is mainly a consequence of the
periodicity of the electron wave function in the plane normal to
the field which turns into an enlarging of the electron available
phase-space. Since the \mfs required to obtain any sizeable effect
cannot be reached in the laboratory in the  foreseeable future, Matese
and O'Connell addressed their attention to the early Universe. The
effects of primordial \mfs on the production of $^4$He during BBN
was first considered in Ref.\cite{MatOco69}. On the basis of the
results obtained in their previous work \cite{MatOco69b}, Matese
and O'Connell argued that strong \mfs should suppress the $^4$He
relic abundance with respect to the standard case. Briefly, their
argument was the following. Since, after the neutron to proton
ratio has been frozen, it takes some time for neutrons to be
bounded into composite nuclei, a faster neutron decay due to the
\mf implies smaller relic abundances of $^4$He and of the heavier
elements.

In Ref. \cite{MatOco69} two other possible effects of a \mf on BBN
were shortly considered. The first of these effects consists in
the variation that a strong \mf induces on the energy density of
the electron-positron gas. This effect is a consequence of the
grow of the electron and positron phase-space in the presence of
over-critical ($B > B_c$) magnetic fields. Below we shall show how such an
effect may have relevant consequences on the BBN through its
action on the expansion rate of the Universe and the entropy
transfer from the $e^+e^-$ gas to the photons. The second effect
touched by Matese and O'Connell concerns the influence of a
uniform \mf on the Universe geometry and its consequences on the
BBN \footnote{This issue was previously considered by Thorne
\cite{Thorne67}}.  Matese and O'Connell analysis of these two
effects was only qualitative and, as far as we know, no further
work was published by these authors about these issues.

In spite of the large number of effects considered in
Ref.\cite{MatOco69} Matese and O'Connell did not include
in their analysis a
simpler and quantitatively  more relevant effect of \mfs on the BBN,
namely the direct contribution of the \mf energy density to the expansion
rate of the Universe. The relevance of such effect was realized by
Greenstein \cite{Greenstein} shortly after the publication of the
 Matese and O'Connell paper. Greenstein showed that by increasing
the Universe expansion rate the presence of the \mf also increases
the temperature at which the neutron-proton equilibrium ratio is
frozen. Since this ratio is roughly given by \cite{KolTur}
\be
\label{n/p} (n/p)_{eq} = \left(\frac{m_{n}}{m_{p}}\right)^{{3/2}}
\exp(- Q/T)~,
 \ee where $Q \equiv m_n - m_p$, a small change in the
freezing temperature gives rise to a large variation in the
neutron relative abundance hence in the  relic abundance of
the light elements. In his paper Greenstein also noted that if the \mf is
sufficiently tangled over distances small compared to the events
horizon, it will have no effect on the Universe geometry. An
explicit calculations of the $^4$He relic abundance as a function
of the \mf strength were reported in a previous paper by the same
author \cite{Greenstein68}. Greenstein concluded that the effect
of the \mf energy density overcomes that of the \mf on the neutron
decay discussed by Matese and O'Connell. Furthermore, from the
requirement that the relic $^4$He mass fraction does not exceed
the 28\%, he inferred the upper limit $B \simleq 10^{12}$ Gauss at
the time when $T = 5\times 10^9~^o$K.

In a following paper by Matese and O'Connell \cite{MatOco70}, the
authors performed a more careful analysis of the effects of a \mf
on the weak reactions which keep neutron and protons in thermal
equilibrium considering, this time, also the direct effect of the
\mf on the Universe expansion rate. Their final conclusions were
in agreement with Greenstein's result.

The recent activity about the origin of \mfs during phase
transitions in the early Universe (see Chap.\ref{chap:generation}) renewed
the interest on the BBN bounds on primordial \mfs and induced
several authors to reconsider the work of Matese and O'Connell and
Greenstein. It is remarkable that after about twenty years and a
large number of new astrophysical observations Greenstein's and
Matese and O'Connell upper limit remains today roughly unchanged.
Moreover, this is the case in spite of important developments
of the BBN numerical computations codes.

We shall now abandon our historical approach to this section and
proceed to give a more detailed description of the subject.

\section{The effect of a \mf on the neutron-proton conversion
rate}\label{sec:weakrates}

The reactions which are responsible for the chemical equilibrium
of neutrons and protons in the early Universe are the weak
processes
 \bea \label{ne-pnu}
 n \,\, + \,\, e^+ \,\, &\leftrightarrow& \,\, p\,\, +
\,\,\overline\nu_e \\
\label{nnu-pe}
 n \,\, + \,\, \nu_e \,\, &\leftrightarrow& \,\, p\,\, + \,\,e^- \\
\label{n-penu}
n \,\, &\leftrightarrow& \,\, p\,\, + \,\,e^- \,\,
+ \,\, \overline\nu_e ~.
 \eea
  In the absence of the magnetic field and in the presence of a heat-bath,
the rate of each of the previous processes takes the generic form
\be
\Gamma(12 \rightarrow 34) =
\left(\prod_i \frac{\int d^3 {\bf p}_i}{(2 \pi)^3 2 E_i} \right)
  (2 \pi)^4 \delta^4(\hbox{$\sum_i$} p_i) |{\cal{M}}|^2\,
    f_1 f_2 (1 - f_3) (1 - f_4),
\ee
 where $p_i$ is the four momentum, $E_i$ is the energy and
$f_i$ is the distribution function of the $i$-th particle species
involved in the equilibrium processes. All processes
(\ref{ne-pnu},\ref{nnu-pe},\ref{n-penu}) share the same amplitude
${\cal{M}}$ determined by the standard electroweak theory.

The total neutrons to protons conversion rate is
\be
\label{ndecay}
 \displaystyle \Gamma_{n\to p}(B=0) = {1\over \tau}
\int_1^\infty d\epsilon\,{\epsilon \sqrt{\epsilon^2 - 1} \over
1+e^{{m_e\epsilon\over T}} } \left[{(q+\epsilon)^2 e^{(\epsilon +
q )m_e\over T_\nu}\over 1+e^{(\epsilon + q) m_e\over T_\nu}} +
{(\epsilon-q)^2 e^{{\epsilon m_e\over T} } \over
1+e^{(\epsilon-q)m_e\over T_\nu}}\right] \label{rate0}
 \ee
 where $q$ and $\epsilon$  are respectively the
neutron-proton mass difference and the electron, or positron,
energy, both expressed in units of the electron mass $m_e$.
We neglect here the
electron and neutrino electron chemical potentials as these are
supposed to be small quantities during BBN \cite{KolTur}.
The rate $1/\tau$ is defined by
 \be
  {1\over \tau} \equiv {G^2 (1 + 3 \alpha^2)
 m_e^5 \over 2 \pi^3}
 \ee
  where $G$ is the Fermi constant and
 $\alpha \equiv g_A/g_V \simeq - 1.262$.
For $T \rightarrow 0$ the integral in Eq.(\ref{ndecay}) reduces to
 \be
 I = \int_1^q d\epsilon \epsilon (\epsilon - q)^2\sqrt{\epsilon^2 -
 1} \simeq 1.63
 \ee
 and $\tau_n = \tau/I$ is the neutron time-life.

The total rate for the inverse processes ($p \to n$) can be
obtained reversing the sign of $q$ in Eq.(\ref{rate0}). It is
assumed here that the neutrino chemical potential is vanishing
(at the end of Sec.\ref{sec:bbnconst} it will also be discussed the case where
such an assumption is relaxed). Since, at the BBN time temperature is much
lower than the nucleon masses, neutrons and protons are assumed to be at
rest.

As pointed out by Matese and O'Connell \cite{MatOco69b,MatOco70},
the main effect of a  \mf stronger than the critical value $B_c$
on the weak processes (\ref{ne-pnu}-\ref{n-penu}) comes-in through
the effect of the field on the electron, and positron, wave
function which becomes periodic in the plane orthogonal to the
field \cite{ItzZub}. As a consequence, the components of the
electron momentum  in that plane are discretized and the electron
energy takes the form
\begin{equation}\label{edisp}
 E_n(B) = \left[p_{z}^2 + |e|B(2n+1+s) + {m_e}^2
 \right]^{1\over 2}
\end{equation}
where we assumed ${\bf B}$ to be directed along the ${\bf {z}}$
axis. In the above, $n$ denotes the Landau level, and $s=\pm 1$
if, respectively, the electron spin is along or opposed to the
field direction.  Besides the effect on the electron dispersion
relation, the discretization of the electron momentum due to the
\mf  has also a crucial effect on the phase-space volume occupied
by these particles. Indeed, in the presence of a field with strength
larger than $B_{c}$ the substitution
\be
\label{phase-space} \int {d^3p\over (2\pi)^3 } f_{FD}(E_0)\ \
\longrightarrow \ \ |e|B \sum_{n=0}^{\infty} (2-\delta_{n0}) \int
{dp_e \over (2\pi)^{2}}
 f_{FD}(E_n(B),T)~,
 \ee
 has to be performed \cite{LandauSM}. Since we only
consider here \mfs which are much weaker than the proton critical
value ($eB \ll m_{p}^{2}$), we can safely disregard any effect
related to the periodicity of the proton wave function.

The squared matrix element for each of the reactions
(\ref{ne-pnu}-\ref{n-penu}) is the same when the spin of the
initial nucleon is averaged and the spins of the remaining
particles are summed. Neglecting neutron polarization, which is
very small for $B < 10^{17}$ G, we have \cite{MatOco69b}
\begin{equation}\label{matele}
  \sum_{\rm spins} \vert {\cal M}(n) \vert = \frac \gamma \tau
  \left[1 - \delta_{n0}\left(1 - \frac {p_z}{E_n} \right)\right]~.
\end{equation}
It is interesting to observe the singular behaviour when a new
Landau level opens up ($E_n = p_z$). Such an effect is smoothed-out
when temperature is increased \cite{Esteban}.

Expressions (\ref{edisp}) and (\ref{phase-space}) can be used to
determined the rate of the processes (\ref{ne-pnu}-\ref{n-penu})
in a heat-bath and in the presence of an over-critical \mf. We
start considering the neutron $\beta$-decay. One finds
 \bea
\label{rate-n-penu} \displaystyle \Gamma_{n\to p e {\bar
\nu}}(\gamma) =
 {\gamma\over \tau}\sum_{n=0}^{n_{max}} (2 - \delta_{n0})
 \times \int_{\sqrt{1+2(n+1)\gamma}}^q \;d\epsilon
{\epsilon\over \sqrt{\epsilon^2-1-2(n+1)\gamma}} \nonumber
\\
 \times {e^{{m_e\epsilon \over T}} \over {1+e^{{m_e\epsilon \over T}} }}
{(q - \epsilon)^2 e^{m_e(q - \epsilon)\over T_\nu} \over
1+e^{{m_e(q - \epsilon)} \over {T_\nu} }}
 \eea
  where $\gamma \equiv B/B_c$ and $n_{max}$
 is the maximum Landau level accessible to the final state electron
determined by the requirement $p_{z}(n)^2 = q^2 - m_e^2 - 2n eB >
0 $. It is noticeable that for $\gamma > \frac{1}{2} (q^{2} -
1)^{2} = 2.7$ only the $n = 0$ term survives in the sum. As a
consequence the $\beta$-decay rate increases linearly with
$\gamma$ above such a value. The computation leading to
(\ref{rate-n-penu}) can be readily generalized to determine the
rate of the reactions (\ref{ne-pnu}) and (\ref{nnu-pe}) for $\gamma
\neq 0$
 \bea
 \Gamma_{n e\to p {\bar \nu}}(\gamma) =
 {\gamma\over \tau}\sum_{n=0}^{\infty} (2 - \delta_{n0})
 \times \int_{\sqrt{1+2(n+1)\gamma}}^\infty \;d\epsilon
{\epsilon\over \sqrt{\epsilon^2-1-2(n+1)\gamma}} \nonumber
\\
\displaystyle
 \times {1\over 1+e^{{m_e\epsilon\over T}}}
{(q + \epsilon)^2 e^{m_e(q + \epsilon)\over T_\nu} \over
1+e^{{m_e(q + \epsilon)} \over {T_\nu}}}~, \label{rate-ne-pnu}
\eea
 and
\begin{eqnarray}
   \Gamma_{n \nu\to p e}(\gamma) =
 {\gamma\over \tau}\left[\sum_{n=0}^{\infty} (2 - \delta_{n0})
 \times \int_{\sqrt{1+2(n+1)\gamma}}^\infty \;d\epsilon
{\epsilon \over \sqrt{(\epsilon -\kappa)^2-1-2(n+1)\gamma}}\right.
\nonumber \\ \left. \displaystyle \times {e^{{m_e\epsilon \over
T}}\over 1+e^{{m_e\epsilon\over T}}} {(\epsilon - q)^2
e^{m_e(q + \epsilon)\over T} \over 1+e^{m_e(\epsilon - q) \over
T_\nu}} \right] \nonumber \\
 - \left[\sum_{n=0}^{n_{max}} (2 - \delta_{n0})
 \times \int_{\sqrt{1+2(n+1)\gamma}}^q\;d\epsilon
{\epsilon \over \sqrt{\epsilon^2-1-2(n+1)\gamma}}\right.
\nonumber \\ \left. \times {e^{m_e\epsilon \over T}\over
1+e^{m_e\epsilon\over T} {(\epsilon - q)^2 e^{m_e(q -
\epsilon)\over T} \over 1+e^{m_e(\epsilon - q) \over
T_\nu}}}\right]~.\label{rate-nnu-pe}
\end{eqnarray}
By using the well know expression of the Euler-MacLaurin sum (see
e.g. Ref.\cite{LandauSM}) it is possible to show
that in the limit $B\rightarrow 0$
Eqs.(\ref{rate-n-penu}-\ref{rate-nnu-pe}) reduces to the standard
expressions derived in the absence of the \mf \footnote{For a
different approach see Ref.\cite{CheST93}}.

The global neutron to proton conversion rate is obtained by
summing the last three equations
\begin{eqnarray}
\Gamma_{n\,\to\,p}(\gamma) = {\gamma\over \tau}\sum_{n=0}^{\infty}
(2 - \delta_{n0})
 \times \int_{\sqrt{1+2(n+1)\gamma}}^\infty \;d\epsilon
{\epsilon\over \sqrt{(\epsilon -\kappa)^2-1-2(n+1)\gamma}}
\nonumber \\ \displaystyle
 \times \frac{1}{1+e^{{m_e\epsilon\over T}}} \left[
{(\epsilon+q)^2 e^{m_e(\epsilon+q)\over T_\nu} \over
1+e^{{m_e(\epsilon + q) \over T_\nu}}}
 + {(\epsilon-q)^2 e^{{m_e\epsilon \over T}}\over
1+e^{m_e(\epsilon-q)\over T_\nu}} \right]~. \label{rate-n-p}
\end{eqnarray}
It is noticeable that the contribution of Eq.(\ref{rate-n-penu}) to
the total rate (\ref{rate-n-p}) is canceled by the second term of
(\ref{rate-nnu-pe}). As a consequence it follows that
Eq.(\ref{rate-n-p}) does not depend on $n_{max}$ and the ${n\,\to\,p}$
conversion grows linearly with the field strength above
$B_{c}$.  From Fig.\ref{fig:rates}  the reader can  observe that, in the
range considered for the field strength, the neutron depletion rate
drops quickly to the free-field when the temperature grows above
few MeV's. Such a behaviour is due to the suppression of the
relative population of the lowest Landau level when $eB \gg
T^{2}$.
\begin{figure}[t]
\vskip -1cm
\centerline{\protect \hbox{
\psfig{file=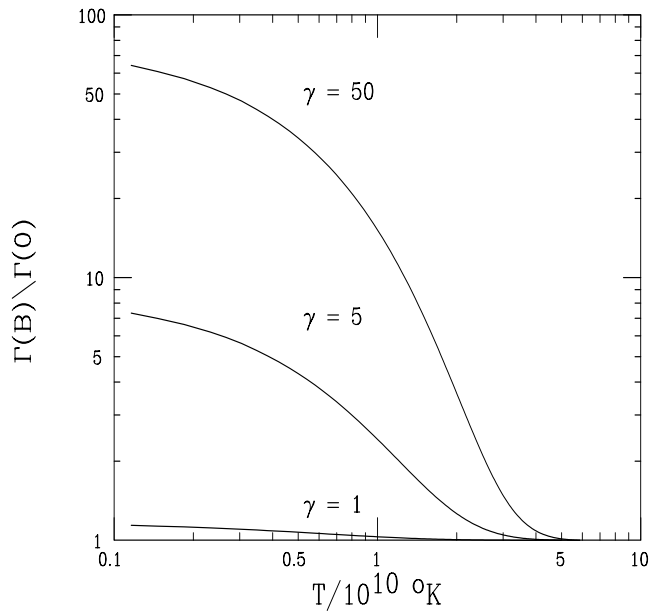,
height=10.cm,
width=10.cm,angle=0}}}
\vskip -1.0cm
\caption{The neutron-depletion rate $\Gamma_{n \to p}$,
normalized to the free-field rate, is plotted as a function of the
temperature for several values of $\gamma$.
From Ref.\cite{GraRub95}}
\label{fig:rates}
\end{figure}

In the absence of other effects, the consequence of the
amplification of $\Gamma_{n\,\to\,p}$ due to the \mf
would be to decrease the relic abundance of
$^4$He. In fact, a larger $\Gamma_{n~\to p}$ implies a lower value
of the temperature ($T_F$) at which the neutron to proton
equilibrium ratio is frozen because of the expansion of the
Universe. It is evident from (\ref{n/p}) that the final value of
$(n/p)$ drops exponentially as $T_F$ is increased. Furthermore,
once $n/p$ has been frozen, occasional neutron $\beta$-decays can
still reduce the relic neutron abundance \cite{KolTur}. As it
follows from Eq.(\ref{rate-n-penu}), the presence of a strong \mf
accelerates the process which may give rise to a further
suppression of the $n/p$ ratio. In practice, however, neutron
decay takes place at a times when the \mf  strength has already
decreased significantly due to the Universe expansion so that the
effect is negligible.

The result of Matese and O'Connell has been confirmed by Cheng et
{\it al.} \cite{CheST93} and by Grasso and Rubinstein
\cite{GraRub95}. Among other effects, the authors of
Ref.\cite{GraRub95} considered also QED and QCD corrections in the
presence of strong \mfs. In principle these corrections may not be
negligible in the presence of over-critical magnetic fields and
their computation requires a proper treatment. In order to give to
the reader a feeling of the relevance of this issue, we remind him
the wrong result which was derived by O'Connell
\cite{OConnell68} by neglecting QED radiative correction to the
electron Dirac equation in the presence of a strong magnetic
field. By assuming the electron anomalous magnetic moment to be
independent on the external field O'Connell found
\begin{equation}
  E_n = \left[p_{z}^2 + |e|B(2n+1+s) + {m_e}^2\right]^{\frac 1 2}
  + s \frac{\alpha}{2\pi} m_e \gamma~.
\end{equation}
For $B > (4\pi/\alpha) B_c$ this expression give rise to negative
values of the ground state energy which, according to O'Connell,
is the manifestation of the instability of the vacuum to spontaneous
production of electron-positron pairs. This conclusion, however, is in
contradiction with standard electrodynamics from which we know
that a constant \mf cannot transfer energy. This problem was
solved by several authors (see e.g. Ref.\cite{Schwinger}) by
showing that by properly accounting for QED radiative corrections
to the Dirac equation no negative value of the electron energy
appear. The effect can be parametrized by a field dependent
correction to the electron mass, $m_e \rightarrow m_e + M$, where
\begin{equation}
M = \left\{
\begin{array}{cc}
\displaystyle
- \frac{\alpha}{2\pi} \frac{eB}{2m_e} \left[ 1 - \frac {8}{3} \frac{eB}{m_e^2}
 \left( \log \frac {m_e^2}{2eB} - \frac{13}{24}\right)\right]
 & \;\;\;\;{B\ll
 B_c} \\
\displaystyle \frac{\alpha}{4\pi} m_e \left( \log \frac {2eB}{m_e^2}\right)^2~.
 & \;\;\;\;{B\gg B_c}
\end{array}
\right.
\end{equation}
Such a correction was included in Refs.\cite{GraRub95,GraRub96}.
It is interesting to observe that although pair production cannot
occur at the expense of the \mf, this phenomenon can take place in
a situation of thermodynamic equilibrium where pair production can
be viewed as a chemical reaction $e^+ + e^- \leftrightarrow
\gamma$,  the \mf playing the role of a catalysts agent
\cite{Dittrich81}. We will return on this issue in
Sec.\ref{sec:bbnthermo}.

Even more interesting are the corrections due to QCD. In fact,
Bander and Rubinstein showed that in the presence of very strong
\mfs the neutron-proton effective mass difference $q$ becomes
\cite{BanRub93} (for a more detailed discussion of this issue see
Chap.\ref{chap:stability})
\be
Q(B) =  0.12\mu_N B - m_n+ m_p + f(B)~~.
 \ee
 The function $f(B)$ gives the rate of mass change due to colour forces
being affected by the field. $\mu_N$ is the nucleon magnetic
moment. For nucleons the main change is produced by the chiral
condensate growth, which because of the different quark content of
protons and neutrons makes the proton mass to grow
faster\cite{BanRub93}. Although, as a matter of principle, the
correction to $Q$ should be accounted in the computation of the
rates that we reported above, in practice however, the effect on
the final result is always negligible. More subtle it is
the effect of the correction to $Q$ on the neutron-to-proton
equilibrium ratio. In fact, as it is evident from Eq.(\ref{n/p}),
in this case the correction to $Q$ enters exponentially to
determine the final neutron-to-proton ratio. However, the actual
computation performed by Grasso and Rubinstein \cite{GraRub95}
showed that the effect on the light element abundances is
sub-dominant whenever the field strength is smaller than  $\simleq
10^{18}~\mathrm{Gauss}$.

\section{The effects on the expansion and cooling rates of the
Universe}\label{sec:rhob}

In the previous section we discussed how the presence of strong
\mfs affects the rates of the weak reactions which are responsible
for the chemical equilibrium of neutrons and protons before BBN.
The knowledge of such rates is, however, not sufficient to predict
the relic abundances of the elements synthesized during BBN. In
fact, the temperature $T_F$ at which $(n/p)_{eq}$ is frozen is
determined by the competition of the weak reaction and the
Universe expansion according to the condition \cite{KolTur}
\begin{equation}
\label{TFdef} \Gamma_{n\,\to\,p}(T_{F}) = H(T_{F})~.
\end{equation}
From this expression it is clear that in order to determine $T_F$
the knowledge of the Universe expansion rate $H(T)$ is also
required.

In the absence of a cosmological term and assuming the effect of
the \mf on the Universe geometry to be negligible, $H$ is
determined by
\be
\label{Hubble} H \equiv \frac {\dot a}{a} = \left({8\pi G
\rho_{tot}\over 3}\right)^{1/2}~.
\ee
 where, according to the
standard notation, $a$ is the scale factor of the Universe, $G$ the
Newton gravitational constant and  $\rho(T)$ is the energy density
of the Universe. In the presence of a magnetic field
\begin{equation}
\rho(T,B) = \rho_{em}(T,B) + \rho_\nu + \rho_B(B)
\end{equation}
where $\rho_{em}(T,B)$ is the energy density of the standard
electromagnetic component (photons + electrons and positrons) of
the heat-bath and $\rho_\nu$ is the energy density of all neutrino
species (the reason why $\rho_{em}$ depends on the \mf strength
will be discussed in the next section). We see that to the
standard components of $\rho$ it adds now the contributions of the
magnetic field energy density
\begin{equation}
\label{rhoB} \rho_B(T) = \frac {B^2(T)}{8\pi}~.
\end{equation}
It is worthwhile to observe that, concerning its direct
contribution to $\rho$, the \mf behaves like any relativistic
component of the heat-bath. In fact, by assuming that the field is
not too tangled on scales smaller than the magnetic dissipation
scale, and that the Universe geometry is not affected by the
magnetic field (see Sec.\ref{sec:anisotropy}), the magnetic flux
conservation during the Universe expansion implies
\be
\label{scalingbis}
 B\propto R^{-2} \propto T^2 \ \  \longrightarrow
\ \ \rho_B(T) \propto T^4~,
 \ee
 which is the same behaviour of the radiation.

In the absence of other effects, the relation (\ref{scalingbis})
would allow to parametrize the effect of the \mf in terms of a
correction to the effective number of massless neutrino species
$\Delta N_\nu^B$ \cite{KerSV95}. Indeed, by comparing the
contribution of $N_\nu$ light ($m_\nu \ll 1~\MeV$) neutrino
species with the energy density of the Universe, which is
\be
\rho_\nu = {{7\pi^2} \over {120}} N_\nu  T_\nu^4~,
 \label{rhonu}
\ee
with (\ref{rhoB}) one gets
\be
\Delta N_{\nu}^B = {15 \over 7\pi^3}b^2,
\ee
 where $b \equiv B/T_\nu^2$.

Before closing this section we have to mention another possible
consequence of the faster Universe expansion induced by the
presence of the magnetic field. The effect is due to shortening of
time between weak reactions freeze-out and breaking of the
deuterium bottleneck. It follows that neutrons have less time to
decay before their confinement into nucleons take place which
turns into a larger abundance of $^4$He. In Ref.\cite{KerSV95} it
was showed that such an effect is generally sub-dominant.

\section{The effect on the electron thermodynamics}\label{sec:bbnthermo}

In the above we discussed how the phase-space of electrons and
positrons is modified by the presence of strong \mfs and how this
effect changes the weak processes rates. The consequences of the
variation of the electron phase-space, however, go well beyond
that effect. Electron and positron thermodynamics functions will
also be affected. In fact, by applying the prescription
(\ref{phase-space}), we find that the number density, the energy
density and the pressure of the electron-positron gas are now
given by
 \bea
 n_e(B) &=& {eB\over (2\pi)^2}\sum_{n = 0}^{\infty}
(2-\delta_{n0}) \int_{-\infty}^{+\infty} f_{FD}(T) dp_z
\label{nume}\\
 \rho_e(B) &=& {eB\over
(2\pi)^2}\sum_{n = 0}^{\infty} (2-\delta_{n0})
\int_{-\infty}^{+\infty} E_n f_{FD}(T) dp_z \label{rhoe}\\
 p_e(B) &=&
{eB\over (2\pi)^2}\sum_{n = 0}^{+\infty} (2-\delta_{n0})
\int_{-\infty}^{+\infty} {E_n^2 - m_e^2 \over 3 E_n} f_{FD}(T)
dp_z \label{Pe}
 \eea
 where
 \be
 f_{FD}(T) \equiv \frac 1 {1+e^{\beta E_n(p_{z})}}
 \ee is
the Fermi-Dirac distribution function, and $E_n(p_z)$ is given by
(\ref{edisp}). As for the case of the weak processes rates, it is
possible to show that Eqs.(\ref{nume}-\ref{Pe}) reduce to the
their well know standard from in the limit $B \rightarrow 0$ (see
e.g. Ref.\cite{KerSV95}).

Numerical computations \cite{GraRub95} show that, for small $T$, $\rho_e$
grows roughly linearly with $B$ when $B > B_c$. This effect is
mainly due to - 1) the reduction, for each Landau level, of the area
occupied by the cyclotron motion of the electron  in plane perpendicular
to the field;\ 2) the growth of the energy
gap among the lowest Landau level and the $n > 0$ levels, which
produces an overpopulation of the lowest Landau level. The first
effect is the dominant one. The net number density and the pressure of
the electron-positron gas follow a similar behaviour. As we
already noted in Sec.\ref{sec:weakrates}, the energy cost of
producing the electron-positron pairs excess cannot be paid by the
\mf which is supposed to be static. Rather, the ``power bill'' is
paid from the heat-bath, or better, from its photon component
\cite{Dittrich81,Miller}. Especially in the context of BBN this
point is quite relevant since the energy transfer from the photons
to the lowest Landau level of the electron-positron gas will
affect the expansion rate of the Universe, its cooling rate and
the effective baryon-to-photon ratio
$\eta$~\cite{GraRub95,GraRub96,KerSV95}. We start discussing the
first two effects.
We observe that the growth of the electron and positron energy
density, due to the presence of the magnetic field, gives rise
to a faster expansion rate of the Universe.
This point was first qualitatively discussed by Matese and
O'Connell \cite{MatOco69} and recently analyzed in more detail by
Grasso and Rubinstein \cite{GraRub95}. The time-temperature
relation will also be modified. The relevance of latter effect has
been first showed by Kernan, Starkman and Vachaspati
\cite{KerSV95} by solving numerically the relation
\be
\label{ttemp} {dT\over dt}= -3 H {\rho_{em} + p_{em} \over
d\rho_{em}/dT}~, \ee where  $\rho_{em} \equiv \rho_e +
\rho_{\gamma}$ and $p_{em} \equiv p_{\gamma} + p_e$ are the energy
density and the pressure of the electromagnetic component of the
primordial heat-bath. In agreement with our previous
considerations, Eq.(\ref{ttemp}) has been obtained by
imposing energy conservation of the electromagnetic component
plasma.

For small values of the ratio $eB/T^2$, the most relevant effect
of the magnetic field enters in the derivative
$d\rho_{em}/dT_\gamma$ that is smaller than the free field value.
This effect can be interpreted as a delay in the electron-positron
annihilation time induced by the magnetic field. This will give
rise to a slower entropy transfer from the electron-positron pairs
to the photons, then to a slower reheating of the heat bath. In
fact, due to the enlarged phase-space of the lowest Landau level
of electrons and positrons, the equilibrium of the process $e^+e^-
\leftrightarrow \gamma$ is shifted towards its left side. Below we
will discuss as this effect has a clear signature on the deuterium
and $^3$He predicted abundances. Another point of interest is that
the delay in the $e^+e^-$ annihilation causes a slight decrease in
the $T_\nu/T$ ratio with respect to the canonical value,
$(4/11)^{1/3}$ \cite{KerSV95}.

The delay in the entropy transfer from the $e^+e^-$ gas to the
heat-bath induces also an increment in the value of
baryon-to-photon ratio $\eta$. In the absence of other effects a
larger value of $\eta$ would induce smaller relic abundances of
Deuterium and $^3$He. This effect, was first predicted in the
Refs.\cite{GraRub95,GraRub96}, Furthermore, it is interesting to observe
that in the case the primordial magnetic field is inhomogeneous
and it is confined in finite volume regions where its strength
exceed the cosmic mean value (e.g. flux tubes), this effect may give
rise to spatial variation in the relic element abundances.

\section{Derivation of the constraints}\label{sec:bbnconst}

In order to account for all the effects discussed in the previous
sections, the use of a numerical analysis is required. Usually,
this is done by modifying properly the famous BBN numerical code
developed by Wagoner and improved by Kawano \cite{bbncode}. After
some discussion around the relative importance of the different
effects, the results of different groups have
converged to a common conclusion: the most relevant effect
of a cosmological magnetic field on BBN is that produced by the
energy density of the field on the Universe expansion rate. This
is in qualitative agreement with the early result of Greenstein
\cite{Greenstein}. From a more quantitative point of view,
however, the effect of the \mf on the electron thermodynamics
cannot be totally neglected. In fact, it was showed in
\cite{GraRub96} that such an effect produces sizable changes in
the relic abundance of $^4$He, Deuterium and $^3$He (see e.g.
Fig.\ref{fig:helium} for the  $^4$He relic abundance prediction)
As a consequence, we think
that the effect of the \mf on the BBN cannot be simply
parameterized in terms of a contribution to the effective number
of neutrino species. Although, in this respect, a different
conclusion was reached in \cite{KerSV95,CheOST96} it should be
noted that, differently from \cite{GraRub96}, in those papers only
approximate expressions for the electron thermodynamic quantities
in the presence of a strong \mf were used. Such an approximation
may be not justified when $eB \simgeq T^2$.

According to the standard procedure, the upper limit on the
strength of the cosmological \mf was obtained in
\cite{GraRub95,GraRub96,CheST94} by comparing the numerically
predicted relic abundance of $^4$He with the observational upper
limit. In \cite{GraRub96} however, the information about Deuterium
and $^3$He was also used. In fact, since the effective value of
$\eta$ is also affected by the \mf, $\eta$ was chosen in the actual
numerical simulation so to saturate the predicted value of
$\mathrm{D} + ^3\mathrm{He}/ \mathrm{H}$ to the observational
upper limit. This choice assured the minimal predicted abundance
of $^4$He for each considered value of $B$.

\begin{figure}[t]
\vskip -1cm
\centerline{\protect \hbox{
\psfig{file=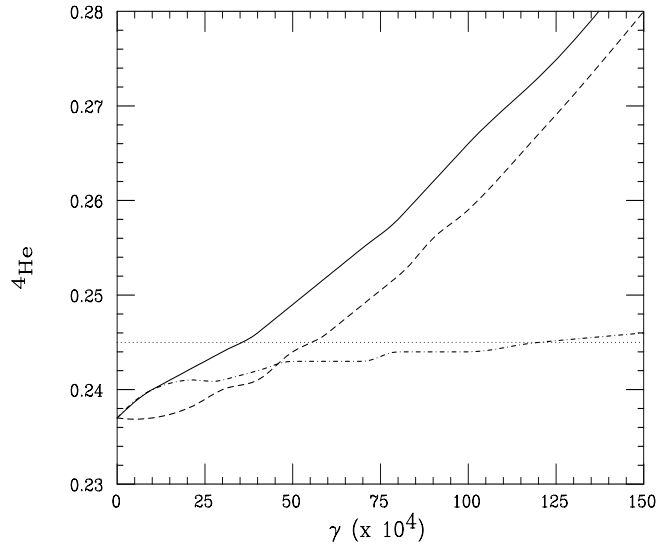,
height=9.cm,
width=10.cm,angle=0}}}
\vskip -1.0cm
\caption{The $^4$He predicted abundance is represented
in function of the parameter $\gamma$, considered at
$T = 10^9~^oK$, in three different cases:
only the effect of the magnetic field energy density
is considered (dashed line); only the effect of
the field on the electron statistics is considered (dotted-dashed
line); both effects are considered (continuous line).
The dotted line represents the observational upper limit.
From Ref.\cite{GraRub96}}
\label{fig:helium}
\end{figure}

By fixing $N_\nu = 3$, requiring $Y_P < 0.245$ and $\mathrm{D} +
^3\mathrm{He}/ \mathrm{H} < 1.04 \times 10^{-4}$ Grasso and
Rubinstein derived the upper limit
\begin{equation}\label{BBNlimit}
  B(T = 10^9~~\mathrm{K}) \simleq 1 \times 10^{13}~\mathrm{G}~.
\end{equation}
Similar results has been obtained by the authors of
Refs.\cite{CheOST96,KerSV96}.

It is useful to remind to the reader under which assumptions the
previous limit has been derived. They are the following.
\begin{enumerate}
\item Universe dynamics was assumed to be well described by
Friedman-Robertson-Walker metric. In other words we
are assuming that the \mf does not lead to a sizeable anisotropic
component in the Universe expansion
 \footnote{BBN in the presence of anisotropic Universe, possibly
 due to a homogeneous cosmic magnetic field, has been considered by
 Thorne \cite{Thorne67}}.
\item The effective number of neutrino species is three. This means, for
example, that if neutrinos are of Dirac type their mass and magnetic moment
are negligible. This in order do not populate right handed degree
of freedom by the magnetic dipole interaction of the neutrinos
with the field.
\item The neutrino chemical potential is negligible.
\item Fundamental physical constants are equal to their present time values
(for a discussion on this issue see Ref.\cite{Lars}).
\end{enumerate}
Some of these assumptions will be relaxed in the following part of
this chapter.

In order to translate our limit (\ref{BBNlimit}) into a bound on the
\mf at the time of galaxy formation some caution is required.
 If we just assume  that the \mf re-scales adiabatically with the Universe
expansion, according to Eq.(\ref{scalingbis}), the BBN limit reads
\begin{equation}\label{BBNlimitnow}
  B_0 \simleq 7 \times 10^{-5}~\mathrm{G}~.
\end{equation}
We should keep in mind, however, that in this case we are neglecting
any possible nonadiabatic evolution of the \mf as that which could
be induced by a non-trivial topology of the field.
Even assuming an adiabatic evolution, we note that the limit
(\ref{BBNlimitnow}) cannot be directly interpreted as a limit on
the progenitor of galactic magnetic fields. The reason for that is that
BBN probes \mfs on scales of the order of the horizon radius at BBN time
(the Hubble comoving radius at BBN time is $\sim 100$ pc)
which are much smaller than  typical protogalaxy sizes ($\sim 1-10$ Mpc).
Therefore, if cosmic magnetic field are tangled on scales smaller than the
protogalactic size, the progenitor \mf has to be interpreted as
a proper average of smaller flux elements.
Clearly, the result of such an average will depend on the
statistical properties of the random magnetic field. If the field vector
were to perform a random walk in 3d volume, the scaling would be
$B(L) \equiv \langle B \rangle_{{\rm rms},~L} \sim N^{-3/2}$ \cite{Hogan},
where $L_0$ is the comoving coherence length of the \mf and $N = L/L_0$
is the number of steps. An argument based on the statistical independence of
conserved flux elements gives   $B(L) \sim
 N^{-1}$ \cite{Vachaspati91}, whereas another argument based on the
statistical independence of the field in neighboring cells predicts
$B(L) \sim N^{-1/2}$ \cite{EnqOle93}.
Adopting a phenomenological point of view, one may just write that
the rms field computed on the scale $L$ at the time $t$ is \cite{EnqRez95}
\begin{equation}
\langle B(L,t) \rangle_{\rm rms} = B_0 \left(\frac {a_0}{a(t)} \right)^2
\left(\frac {L_0}{L} \right)^p~,
\end{equation}
where $p$ is an unknown parameter ($p = 3/2, 1, 1/2$ respectively in the three
cases discussed in the above). The meaning of $B_0$ is now understood
as  $B_0 = \lim_{L \rightarrow \infty} B(L,t_0)$~\footnote{A detailed
discussion about
average procedures of tangled \mfs can be found in Ref.\cite{HinEve97}}.
If, for example, we adopt the value $p = 1$  and assume
$L_0 = 100$ pc, the limit (\ref{BBNlimitnow}) implies
\begin{equation}
\langle B(1~{\rm Mpc}, t_0)\rangle_{\rm rms} \simleq 10^{-9}~{\rm G}~.
\end{equation}
Therefore, although the BBN bound is much more stringent than what is usually
claimed in the literature, it cannot exclude a primordial origin of
galactic \mfs by the adiabatic compression of the field lines.

For the same reasons which we have explained in the above, BBN limits
on primordial \mfs cannot be directly compared with bounds derived
from the analysis of CMBR anisotropies. In fact, unless the
\mfs is uniform through the entire Universe, CMBR offers a probe of \mfs
only on comoving scales which are much larger than the horizon radius at
BBN time.
\vskip0.5cm
We shall now consider how the previous limits changes by
relaxing one of the assumptions under which the constraint
(\ref{BBNlimit}) has been derived, namely that related to the neutrino
chemical potential. The effects of a possible neutrino-antineutrino asymmetry
in this context has been recently considered by
Suh and Mathews \cite{SuhMat}. This issue is interesting
since several recent leptogenesis scenarios predict the formation
of such asymmetry during the radiation era.  It is well know that
even in the absence of a primordial \mf a non-vanishing neutrino
chemical potential can affect the predictions of BBN (see
Ref.\cite{MalMat} and references therein). In fact, a degeneracy
of the electron neutrino changes both the weak reaction rates and
the neutron-to-proton equilibrium ratio, whereas a degeneracy in
any of the neutrino species modifies the expansion rate of the
Universe. Clearly, the presence of any of these effects would
affect the BBN limit on the strength of a primordial \mf. Suh and
Mathews found that if the limit is $B_0 \leq 5.8\times 10^{-7}~$ G
with $\xi_e \equiv \mu_{\nu_e}/T_{\nu_e}= 0$ (in good agreement
with the limit (\ref{BBNlimitnow})), it becomes $B_0 \leq 2.8\times
10^{-6}~~\mathrm{Gauss}$ if $\xi_e \equiv \mu_{\nu_e}/T_{\nu_e}=
0.15$. Therefore, we see that in the presence of
phenomenologically acceptable values of the neutrino chemical
potential the BBN constraint on the magnetic field can be
considerably relaxed.

\section{Neutrino spin-oscillations in the presence of a
\mf}\label{sec:bbnspinosc}
It is  interesting to consider how the limit obtained in the
previous section changes if neutrinos carry non-standard
properties which may change the effective neutrino number during
BBN.  We are especially interested here to the possibility that
neutrinos carry non-vanishing masses and magnetic moments. If this
is the case, the dipole interaction of the neutrinos with the \mf
may give rise to {\it spin-oscillations} of the neutrinos, i.e.
periodic conversion of a helicity state into another. In the case
of Dirac neutrinos, this phenomenon may have crucial consequences
for BBN. In fact, spin-oscillation can populate the right-handed
helicity state of the neutrino which, being practically sterile
(for $m_\nu \ll T$) to weak interactions, would otherwise play no
effective role. By adding a new degree of freedom to the thermal
bath, such an effect may produce dangerous consequences for the
outcome of BBN. This problem was first pointed-out by
Shapiro and Wasserman \cite{ShaWas81} and, independently, by Lynn
\cite{Lynn81} who used the argument to put a constraint on the
product of the \mf with the neutrino magnetic moment.
In both works, however, the important role played by  neutrino refractive
properties determined by the neutrino  collective interaction with
the heat-bath, as well as that played by neutrino scattering
with leptons, were disregarded. A more complete treatment was developed by
Fukugita et al. \cite{FukNRS}. They showed that the conditions under which
the neutrino wrong-helicity state can be effectively populated are the
following:
\begin{enumerate}
 \item the spin-oscillation frequency
 $ \Delta E_{\mathrm{magn}} = 2 \mu_\nu B$ has to exceed the
Universe expansion rate;
 \item since neutrino scattering destroy the phase relationship
between the left-handed and right-handed helicity states, $\Delta
E_{\mathrm{magn}}$ has to be larger than the scattering rate;
 \item since the refractive indices for left-handed and right-handed
states, $n_L$ and $n_R$, are not equal, oscillations can occur only if
\begin{equation}
  \Delta E_{\mathrm{magn}} \simleq \Delta E_{\mathrm{refr}}
\end{equation}
where $\Delta E_{\mathrm{refr}} \equiv (n_L - n_R) E_\nu$.
\end{enumerate}
The BBN is affected only if such conditions are simultaneously
satisfied at some temperature $T_{\mathrm{osc}}$ in the range
$T_{\mathrm{dec}} \simleq T_{\mathrm{osc}} \simleq
T_{\mathrm{QCD}}$ where $T_{\mathrm{dec}} \approx 1~\mathrm{MeV}$
is the neutrino decoupling temperature and $T_{\mathrm{\rm QCD}
}\approx 200~\mathrm{MeV}$. Note that in the case right-handed neutrinos
decouples before the QCD phase transition, the huge amount of entropy which
is expected to be released during this transition would dilute their relative
abundance so to prevent any effect on the BBN.
From the previous considerations the authors of Ref.\cite{FukNRS} derived the
limit
\begin{equation}\label{Raffeltlimit}
  \mu_\nu \simleq
  10^{-16}~\mu_B~\left(\frac{10^{-9}~\mathrm{G}}{B_0}\right)~,
\end{equation}
where $\mu_B$ is the Bohr magneton.
The work of Fukugita et al. has been reconsidered by several
authors. For example, Enqvist, Olesen and Semikoz \cite{EnqOle92},
improved the previous analysis by considering the effect of the
neutrino refractive properties on the left-right transition probability.
Elmfors, Grasso and Raffelt \cite{ElmGR} accounted for the
effect of the \mf on the neutrino refractive properties and used an
improved treatment of neutrino collisions. First of all, Elmfors et al.
noted that by affecting the thermodynamics properties of the
electromagnetic component of the heat-bath (see Sec.\ref{sec:bbnthermo})
a strong \mf changes also the neutrino potentials. This
may have relevant consequences both for neutrino spin-oscillations
and flavour oscillations in a magnetized medium \cite{Grasso98}.
The interplay between spin-oscillations and collisions was then accounted in
Ref.\cite{ElmGR} by means of the following evolution equation \cite{Stodolsky}
\be
 \label{stodolsky}
    \frac {\partial {\bf P}}{\partial t} ={\bf V}\times {\bf P}
    - D {\bf P}_{\rm T}~~,
\ee
where $\bf P$ is the neutrino polarization vector and
 ${\bf P}_{\rm T}$ is its  transverse component respect to the neutrino
direction of motion. $\bf V$ is a vector of effective magnetic
interaction energies which can be decomposed into its transverse
and longitudinal components
 \bea
    {\bf V}_{\rm T}&=&2\mu_\nu {\bf B}_{\rm T}~,\\
    |{\bf V}_{\rm L}|&=& \Delta E_{\rm refr}~,
 \eea
where $\mu_\nu$ is the neutrino magnetic moment and \cite{ElmGR}
\begin{equation}
\label{DErefr}
 \Delta E_{\rm refr}(B) =\frac{8\sqrt2\,G_{\rm F}E}{3}
    \sum_{\ell=e,\mu,\tau} \left(\frac{\rho^L_{\nu_\ell}(B)}{m_Z^2}+
      \frac{\rho_{\ell}}{m_W^2}\right)
\end{equation}
is the left-right neutrino energy difference in the magnetized
medium  \footnote{For a computation of the neutrino refractive
properties in a magnetized medium see also Refs.\cite{SemVal,Smirnov97}.}.
It is worthwhile to note that in Eq.(\ref{DErefr})
the expression (\ref{rhoe}) has to be used for $\rho_e(B)$.
As we wrote above, collisions destroy the phase coherence between the
left-handed  and right-handed component of a neutrino state, which
amounts to a damping of the transverse part ${\bf P}_{\rm T}$ of
the polarization vector. The main contribution to the damping rate
$D$ comes from neutrino elastic and inelastic scattering with
leptons and equals half the total collision rate of the
left-handed component \cite{McKellar}. In the early Universe at $T
\sim 1$ MeV one finds:
\be
     \langle D\rangle=f_{\rm D}\,\frac{7\pi}{16}\,G_{\rm F}^2 T^5~,
\ee
 where $f_{\rm D}$ is a order one numerical factor.
Inserting the previous expressions in  Eq.(\ref{stodolsky}) it is
easy to derive the neutrino depolarization rate $\Gamma_{\rm
depol}$. In the small mixing angle limit,
\begin{equation}
  \tan 2\theta=\frac{V_{\rm T}}{V_{\rm L}}
    =\frac{2\mu B_{\rm T}}{\Delta E_{\rm refr}} \ll 1~,
\end{equation}
one finds
\begin{equation}
\label{depol}
  \Gamma_{\rm depol}\approx
    \frac{(2\mu B_{\rm T})^2\,\langle D\rangle}
    {\langle V_L^2\rangle}
    \approx \frac{f_{\rm D}}{f_{\rm L}^2}\,
    \frac{400\,\alpha^2}{7\pi}\,
    \frac{\mu_\nu^2B_{\rm T}^2}{G_{\rm F}^2 T^5}~,
\end{equation}
where $f_L = 1$ for $\mu$ and $\tau$ neutrinos, while for $e$ neutrinos
$f_L \approx 3.6$.
By requiring this rate to be smaller than the Universe expansion
rate $H(T)$ in the temperature interval $T_{\rm dec} < T < T_{\rm
QCD}$, Elmfors at al. \cite{ElmGR} found the upper limit
\footnote{Note that in Ref. \cite{ElmGR} $B_0$ was defined as the
\mf at BBN time.}
\begin{equation}\label{newRaffeltlimit}
  \mu_\nu \simleq
  7 \times
  10^{-17}~\mu_B~\left(\frac{10^{-9}~\mathrm{G}}{B_0}\right)~,
\end{equation}
which is not too different from the limit (\ref{Raffeltlimit})
previously found by the authors of Ref.\cite{FukNRS}.
Limits on $\mu_\nu$ were also found by the authors of
Refs.\cite{EnqRez95,EnqSem93} who considered the case of
random magnetic fields.

In principle, right-handed neutrinos could also be populated by
direct spin-flip interactions mediated by  virtual photons
produced by scattering on charged particles or by annihilation
processes \cite{Morgan}, as well as by the interaction with small
scale \mfs produced by  thermal fluctuations \cite{ElmERS}. In practice,
however, bounds on $\mu_\nu$ from a possible large scale magnetic
field are found to be more stringent even for very weak magnetic fields.
The most stringent upper limit on  Dirac type neutrino magnetic
moment with mass $m_\nu < 1$ MeV, comes from stellar evolution considerations.
It is $\mu_\nu \simleq 3 \times 10^{-12}\mu_{\rm B}$ \cite{Raffeltbook,
Dipolebounds}. It is interesting
to observe that if one of the neutrinos saturate this limit,
Eq.(\ref{newRaffeltlimit}) implies the following quite stringent
bound on the present time cosmic magnetic field, $B_0 \simleq 10^{-13}$
G.

In the particle physics standard model, neutrinos have no magnetic
dipole moment. However, if the neutrino has a Dirac mass $m_\nu$,
radiative corrections automatically give rise to a finite dipole
moment \cite{Lynn81}
\begin{equation}\label{munuSM}
  \mu_\nu =3.2\times 10^{-19}~\mu_{\rm B}~\left(\frac{m_\nu}
  {1 {\rm eV}}\right)~,
\end{equation}
even without invoking any further extension of the standard model
beside that required to account for the finite neutrino mass. On
the basis of this consideration Enqvist et al. \cite{EnqSem93b}
derived the following upper limit for the
present time local magnetic field
\begin{equation}
  B_0 \simleq \frac{2 \times 10^{-3}~{\mathrm G}}
 {\displaystyle \sum_i \frac{m_{\nu_i}}{1~{\mathrm eV}}}~.
\end{equation}
Clearly, this limit cannot compete with the constraint derived in the previous
section.

Spin oscillations in the presence of twisted primordial magnetic fields
(i.e. \mf with a nonvanishing  helicity, see Sec.\ref{sec:evolution}) have
been considered by Athar \cite{Athar}. Athar showed that in such a situation
the left-right conversion probabilities for neutrino and antineutrinos may be
different. This result may open the interesting possibility that a
neutrino-antineutrino asymmetry may be generated during the big-bang
by a preexisting non trivial topology of a primeval magnetic field.
As we shall see in Sec.\ref{sec:helicity}, the production of a net magnetic
helicity of the Universe is indeed predicted by some models.

It is also interesting to speculate on the effects when the number of
dimensions change, and these are large \cite{Dimo}.
In fact BBN is one of the most serious objections to this idea, together
with the background diffuse gamma radiation \cite{Hall}.
Detailed studies of effects of magnetic fields in these scenarios are not
available yet.


\chapter{Generation of magnetic fields}\label{chap:generation}

\section{Magnetic fields from primordial vorticity}\label{sec:vorticity}

The idea that primordial magnetic fields can be produced by plasma
vortical motion during the radiation era of the early Universe has
been first proposed by Harrison \cite{Harrison70}. Since this
mechanism has been reviewed in several papers (see e.g.
\cite{Kronberg,Sicotte}) we shall not discuss it in detail here.
Harrison's mechanism is based on the consideration that the electron and
ion rotational velocities should decrease differently in the
expanding Universe in the pre-recombination era. The reason is
that Thomson scattering is much more effective for electrons than
ions. Therefore electrons remains tightly coupled for a longer
time to the radiation and behave like relativistic matter whereas ions are
already non-relativistic. It follows that during Universe
expansion angular velocity decreases like $\omega \propto a^{-1}$
for electrons and like $\omega \propto a^{-2}$ for ions, where $a$
is the Universe scale factor. The difference between these two
velocities causes an electromotive force, hence an electric
current which generate a magnetic field. Harrison
\cite{Harrison73} showed that if a primordial turbulence was
present at the recombination this mechanism may lead to present
time intergalactic magnetic fields as large as $10^{-8}$ G on a
scale-length of 1 Mpc.

A problem, however, arises with this scenario.
In fact, it was noted by Rees \cite{Rees} that since rotational,
or vector, density perturbations decay with cosmic expansion,
in order to produce sizeable effects at recombination time
this kind of fluctuations should has been dominant at the radiation-matter
equality time. This seems to be incompatible with the standard scenario
for galaxy formation.

Another related problem is that, in contrast to scalar or tensor perturbations,
rotational perturbations cannot arise from
small deviations from the isotropic Friedmann Universe near the
initial singularity. This is a consequence of the Helmholtz-Kelvin circulation
theorem which states that the circulation around a closed
curve following the motion of matter is conserved.
Such a problem, however,
may be partially circumvented if collisionless matter was present during
the big-bang (e.g. decoupled gravitons after the Planck era).
Rebhan \cite{Rebhan92} showed that in this case the
Helmholtz-Kelvin theorem does not apply and growing modes of
vorticity on superhorizon scale can be obtained. In fact, a non
perfect fluid can support anisotropic pressure which may generate
nonzero vorticity even if it was zero at the singularity. It
follows that the only constraint to the amount of primordial
vorticity comes from the requirement that it does not produce too
large anisotropies in the CMBR. Rebhan showed that this
requirement implies the following upper limit to the strength of a
present time intergalactic magnetic field, with coherence length
$L$, produced by vortical plasma motion
\begin{equation}
B_0(L) < 3 \times 10^{-18}~h^{-2}~L^{-3}_{\rm Mpc}~{\rm G}~.
\end{equation}
Such a field might act as a seed for galactic dynamo.

If primordial vorticity is really not incompatible with standard
cosmology, another interesting possibility is to generate primordial \mfs
arises. It was noted by Vilenkin \cite{Vilenkin78} that, as a consequence of
parity violation in the Weinberg-Salam model of the electroweak interactions,
macroscopic parity-violating currents may develop in a vortical
thermal background.  Vilenkin and Leavy \cite{VilLea} suggested that this
currents may effectively give rise to strong magnetic fields.
It was also recently noted by Brizard, Murayama and Wurtele \cite{Murayama}
that in the presence of vorticity and a neutrino-antineutrino asymmetry,
collective neutrino-plasma interactions
may power astrophysical as well as cosmological magnetic fields.

Clearly, in order to implement these scenarios, a suitable mechanism
to produce the required amount of primordial vorticity has to be
found. Among other exotic possibilities, generation of vorticity
and \mfs  by the anisotropic
collapse of conventional matter into the potential well of
pre-existing dark matter condensations in a texture-seeded
scenario of structure formation \cite{Sicotte},
and from rotating primordial black-holes \cite{VilLea} have been considered
in the literature.

Primordial phase transitions, may provide a more realistic source
of vorticity. The generation of primeval \mfs during some of these
transitions will be the subject of the next sections of this
chapter.

Another interesting possibility which is currently under study
\cite{DolGra}, is that large scale vorticity and \mfs are
generated at neutrino decoupling in the presence of a large
neutrino degeneracy.

\section{Magnetic fields from the quark-hadron phase transition}

It is a prediction of Quantum-Chromo-Dynamics (QCD) that at some very high
temperature and/or density strongly interacting matter undergoes a
deconfinement transition, where quark/gluon degrees of freedom are
``melted-out" from hadrons. In the early Universe, due to the
cosmological expansion, the process proceeds in the opposite
direction starting from a ``quark-gluon plasma" which at some
critical temperature $T_{QCD}$ condenses into colorless hadrons
\cite{Ornella}.
Lattice computations suggest that the QCD phase transition (QCDPT)
is a first-order phase transition taking place at $T_{QCD} \sim 150 \MeV$
\cite{Kajantie}. Typically a first-order phase transition takes
place by bubble nucleation. As the temperature supercools below
$T_{QCD}$, sub-critical bubbles containing the hadronic phase grow
as burning deflagration fronts releasing heat in the quark-gluon
plasma in the form of supersonic shock fronts. When the shock
fronts collide they reheat the plasma up to $T_{QCD}$ stopping
bubble grow. Clearly, up to this time, the transition is an
out-of-equilibrium process. Later on, the growth of newly nucleated
bubbles proceeds in thermal equilibrium giving rise to the,
so-called, coexistence phase. The latent heat released by these
bubbles compensate for the cooling due to the Universe expansion
keeping temperature at $T_{QCD}$. The transition ends when
expansion wins over and the remaining quark-gluon plasma pockets
are hadronized.

The first step of the magnetogenesis scenario at the QCDPT
proposed by Quashnock, Loeb and Spergel \cite{QuaLS89} consists in
the formation of an electric field behind the shock fronts which
precede the expanding bubbles. This is a consequence of the baryon
asymmetry, which was presumably already present and which makes
the baryonic components of the primordial plasma positively
charged. At the same time, the leptonic component must be
negatively charged to guarantee the Universe charge neutrality.
The other crucial ingredient of the mechanism is the difference in
the equation of state of the baryonic and leptonic fluids. As a
consequence, the strong pressure gradient produced by the passage
of the shock wave gives rise to a radial electric field behind the
shock front. Such a generation mechanism is usually known as a
battery. Quashnock et al. gave the following estimate for the
strength of the electric field
\be
eE \simeq 15 \left(\frac {\epsilon}{10\%}\right) \left(\frac
{\delta}{10\%}\right)\left(\frac{kT_{QCD}}{150~\MeV}\right)
\left(\frac{100~{\rm cm}}{\ell}\right)~\frac{\rm keV}{\rm cm}~,
\label{elQCD}
\ee
  where $\epsilon$ represents the ratio of the
energy densities of the two fluids, $\delta \equiv (\ell \Delta p/p)$
is the pressure gradient and $\ell$ is the average distance between
nucleation sites.

Small scale magnetic fields are generated by the electric currents
induced by the electric fields. These fields, however, live on a
very small scale ($\ll \ell$ )and presumably they are rapidly
dissipated. Phenomenologically more interesting fields should be
produced when the shock fronts collide giving rise to turbulence
and vorticity on a scales of order $\ell$. Magnetic fields are
produced on the same scale by the circulation of electric fields of
magnitude given by Eq.(\ref{elQCD}). Then, using standard
electrodynamics, Quashnock et al. found that the magnetic field
produced on the scale $\ell \sim 100$ cm has a magnitude
\be
B_\ell \simeq  v E \simeq 5~{\rm G}~.
\ee
Following the approach which was first developed by Hogan \cite{Hogan},
the \mf on scales $L \gg \ell$ can be estimated by performing a proper
{\it volume average} of the fields produced by a large number of
magnetic dipoles of size $\ell$ randomly oriented
in space. Such an average gives
\begin{equation}
\label{Hogan}
 B_L = B_\ell \left(\frac \ell L \right)^{3/2}~.
\end{equation}
After  the  QCDPT the \mf evolves according to the frozen-in law
(\ref{scaling}). It is straightforward to estimate the \mf strength at
the recombination time on a given scale $L$.  The smaller
conceivable coherence length is given by the dissipation length at that time
which, following the argument already described in Sec.\ref{sec:evolution},
is found to be $L_{diss}(t_{rec}) \simeq 5
\times 10^{10}$ cm, corresponding to 1 A.U. at present time. On
this scale the \mf produced at the QCDPT is $\simeq 2 \times
10^{-17}$ G. This small strength is further dramatically
suppressed if one considers scales of the order of the galactic
size $\sim 10$ kpc. Therefore, it looks quite unplausible that the
\mfs generated by this mechanism could have any phenomenological relevance
even if the galactic dynamo was effective.

According to a more recent paper by Cheng and Olinto
\cite{CheOli94} stronger fields might be produced during the
coexistence phase of the QCDPT. The new point raised by the authors of
Ref.\cite{CheOli94} is that even during such equilibrium phase a  baryon
excess build-up in front of the bubble wall, just as a consequence
of the difference of the baryon masses in the quark and hadron
phases. According to some numerical simulations \cite{Kurky88},
this effect might enhance the baryon density contrast by few orders
of magnitude. Even more relevant is the thickness of the charged
baryonic layer which, being controlled by baryon diffusion, is $\sim
10^7$ fm  rather than the microphysics QCD length scale $\sim 1$
fm. In this scenario  \mfs are generated by the peculiar motion of
the electric dipoles which arises from the convective transfer of
the latent heat released by the expanding bubble walls. The field
strength at the QCDPT time has been estimated by Cheng and Olinto
to be as large as
\begin{equation}
B_{QCD} \simeq 10^8~{\rm G}
\end{equation}
on a maximal coherence length $l_{\rm coh} \simeq H^{-1}_{QCD}$.
 Once again, by assuming frozen-in evolution of the field, one can
determine present time values:
\begin{equation}
 B_0 \simeq 10^{-16}~{\rm G}\qquad l_0 \simeq 1~{\rm pc}~.
\end{equation}
Using Eq.(\ref{Hogan}) Cheng and Olinto found that on the galactic
length scale $B({\rm kpc}) \simeq 10^{-20}$ G which may have same
phenomenological relevance if the galactic dynamo is very
effective.

In a following work Sigl, Olinto and Jedamzik \cite{SiglOL97}
investigated the possible role that hydrodynamic instabilities
produced by the expanding bubble walls may have in generating
strong \mfs. Although is not clear whether these instabilities can
really develop during the QCDPT, Sigl at al. claimed that this
phenomenon is not implausible for a reasonable choice of the QCDPT
parameters. By taking into account the damping due to the finite
viscosity and heat conductivity of the plasma, the authors of
Ref.\cite{SiglOL97} showed that the instability may grow
non-linearly producing turbulence on a scale of the order of the
bubble size at the percolation time. As a consequence, a MHD dynamo
may operate to amplify seed \mfs and equipartition of the magnetic
field energy with the kinetic energy may be reached. If this is
the case, magnetic fields of the order of $10^{-20}$ G may be
obtained at the present time on a very large scale $\sim 10$
Mpc. Larger fields may be obtained by accounting for an inverse cascade
(see Sec.\ref{sec:evolution}).  In the most optimistic situation in
which the \mf was produced having maximal helicity at the QCDPT and
equipartion between magnetic and thermal energy was realized
at that time, Eqs.(\ref{helLscalT},\ref{helBscalT}) for the time evolution
of the rms field strength and coherence length apply.
By substituting in such equations the initial scale $l(T_i)
\sim r_H(T_{QCD}) \sim 30$ km, and $B_{rm rms}(T_i) \sim 10^{17}$ G,
one finds the present time values
\begin{equation}
 B_{\rm rms}(T_0) \sim 10^{-9}~~{\rm G} \qquad L_{\rm coh} \sim
 100~~{\rm kpc}~.
\end{equation}
Remarkably, we see that in this optimistic case no dynamo amplification is
required to explain galactic, and probably also cluster, magnetic fields.

\section{Magnetic fields from the electroweak phase
transition}\label{sec:EWPT}

\subsection{Magnetic fields from a turbulent charge flow}
Some of the ingredients which may give rise to magnetogenesis at
the QCDPT may also be found at the electroweak phase transition
(EWPT). As for the case of the QCDPT, magnetogenesis at the weak
scale seems to require a first order transition. Although
recent lattice computations performed in the framework of the
standard electroweak theory \cite{Kajantie96} give a strong
evidence against a first order transition, this remains a viable
possibility if supersymmetric extension of the standard model are
considered \cite{MSSM}. It is noticeable that a first order EWPT is also
required for the successful realization of the electroweak
baryogenesis scenario \cite{toni}. Indeed, as we shall see in the rest of
this review, this is only one among several common aspects of baryogenesis and
magnetogenesis.

According to  Baym, B\"odeker and McLerran \cite{McLerran} strong
magnetic fields can be generated by a first order EWPT
\footnote{At the time the paper by Baym et al. was written, a
first order EWPT was thought to be compatible with the standard
model. Therefore all computation in \cite{McLerran} were done in
the framework of that model.} via a dynamo mechanism. In this
scenario seed fields are provided by random \mf fluctuations which
are always present on a scale of the order of a thermal wavelength
\footnote{It is worthwhile to observe here that thermal
fluctuation in a dissipative plasma can actually produce
stochastic magnetic fields on a scale larger than the thermal
wavelength \cite{Tajima,LemoineD}}. The amplification of such seed
fields proceeds as follows. When the Universe supercooled below
the critical temperature ($T_c \sim 100$ GeV) the Higgs field
locally tunneled from the unbroken SU$(2)\times{\rm U}(1)_Y$ phase
to the broken $U(1)_{em}$  phase. The tunneling gave rise to the
formation of broken phase bubbles which then expanded by
converting the false vacuum energy into kinetic energy. Although
the bubble wall velocity is model dependent, one can find that for
a wide range of the standard model parameters the expansion is
subsonic (deflagration) which give rise to a supersonic shock wave
ahead of the burning front. As the shock fronts collided
turbulence should have formed in the cone associated with the
bubble intersection. The Reynold number for the collision of two
bubbles is
\begin{eqnarray}
   Re \sim \frac{v_{\rm fluid} R_{\rm bubble}}{\lambda},
\end{eqnarray}
where $v_{\rm fluid}\sim v_{\rm wall} \sim 10^{-1}$ is the typical fluid
velocity, $R_{\rm bubble}$ is the size of a bubble at the
collision time and  $\lambda$ is the scattering length of excitations in the
electroweak plasma.
The typical size of a bubble after the phase transition is completed is in
the range
\begin{eqnarray}
R_{\rm bubble} \sim f_{\rm b} H^{-1}_{\rm ew}
\label{bubsize}
\end{eqnarray}
where
\begin{eqnarray}
H^{-1}_{\rm ew} \sim \frac{m_{\rm Pl}}{g_*^{1/2}T_{\rm c}^2}\sim 10~{\rm cm}
\end{eqnarray}
is the size of the event horizon at the electroweak scale, $m_{\rm Pl} $
is the Planck mass, $g_*\sim 10^2$ is the number of massless degrees of
freedom in the matter, and the fractional size $f_{\rm b}$ is $\sim
10^{-2}-10^{-3}$.  The typical
scattering length $\lambda$ of excitations in the plasma is of order
\begin{eqnarray}
   \lambda \sim \frac{1}{T g_{\rm ew}\alpha_{\rm w}^2|\ln\alpha_{\rm w}|},
\end{eqnarray}
where $\alpha_{\rm w}$ is the fine structure constant at the
electroweak scale, and $g_{\rm ew}\sim g_*$ is the number of degrees of
freedom that scatter by electroweak processes.
By substituting these expressions Baym et al. found
\begin{eqnarray}
   Re \sim 10^{-3} \frac{m_{\rm Pl}}{T_{\rm c}}\alpha_{\rm w}^2
         |\ln\alpha_{\rm w}| \sim 10^{12}~.
\end{eqnarray}
Such a huge Reynolds number means that turbulence fully develops
at all scales smaller than  $R_{\rm bubble}$. Since conductivity
is expected to be quite large at that time \cite{conductivity}, \mfs
followed the fluid motion so that a strong magnetic turbulence
should also have been produced. In such a situation it is known that
the kinetic energy of the turbulent flow is equipartitioned with
the magnetic field energy. Therefore
\begin{eqnarray}
   B^{2}(R_{\rm bubble}) \sim \epsilon(T_{\rm c}) v_{\rm fluid}^2,
\label{equipartition}
\end{eqnarray}
where $\epsilon(T_{\rm c})\sim g_* T_{\rm c}^4$ is the energy density
of the electroweak plasma.

In order to estimate the \mf strength on scale larger than $R_{\rm
bubble}$, Baym et al. treated the large scale field as a
superposition of the field of dipoles with size $R_{\rm bubble}$.
This is similar to what done by other authors
\cite{Hogan,CheOli94} (see what we wrote above for the QCDPT)
apart for the fact that Baym et al. used a continuum approximation for
the distribution of dipoles rather than to assume a random walk of
the field lines. The density $\nu^i({\bf r})$ of dipoles pointing
in the $i$-th direction was assumed to be  Gaussianly distributed.
This implies the following correlation functions for the density
of dipoles
\begin{equation}
    \left\langle\nu^i({\bf r}) \nu^j(0)\right\rangle =
                  \kappa \delta^{ij} \delta^{(3)} ({\bf r})
\end{equation}
 and for the magnetic field
\begin{equation}
   \left\langle {\bf B}({\bf r})\cdot{\bf B}({\bf 0})\right\rangle
   \sim e^2\kappa \int d^3 r_{\rm d}\frac1{\mid{\bf r}-{\bf r}_{\rm
   d}\mid^3} \frac1{\mid{\bf r}_{\rm d}\mid^3}~.
\end{equation}
The logarithmic divergence of the integral in these regions
is cut off by the size of the typical dipole, $f_{\rm b}H^{-1}_{\rm ew}$, so
that for $r \gg f_{\rm b}H^{-1}_{\rm ew}$,
\begin{equation}
   \left\langle {\bf B}({\bf r})\cdot{\bf B}({\bf 0})\right\rangle
   \sim \frac{e^2\kappa}{r^3}\ln\left(\frac{H_{\rm ew} r}{f_{\rm b}}\right).
\end{equation}
By using this expression and the equipartition relation (\ref{equipartition})
one finds that the strength of $B^2$ measured by averaging on a size scale
$R$ is
\begin{equation}
   \label{Bew}
   \left\langle B^2 \right\rangle_R\sim v_{\rm fluid}^2 g_* T_{\rm c}^4
    \left( \frac{f_{\rm b}}{H_{\rm ew} R }\right)^3
   \ln^2 \left(\frac{H_{\rm ew}R}{f_{\rm b} }\right).
\end{equation}
This result can be better expressed in terms of the ratio $r$ of
$\left\langle B^2\right\rangle$ to the energy $\rho_\gamma$ in photons
which is a constant during Universe expansion in the absence of flux diffusion.
From Eqs.(\ref{Bew})one gets
\begin{equation}
         r_R \sim v_{\rm fluid}^2 f_{\rm b}^3
         \left(\lambda_{\rm ew} \over R \right)^3
         \ln^2\left(\frac{R}{f_{\rm b} \lambda_{\rm ew}}\right),
\end{equation}
where $\lambda_{\rm ew}$ is the Hubble radius at the electroweak
phase transition ($\sim 1$ cm) times the scale factor, $T_{\rm c}/T_\gamma$,
where $T_\gamma$ is the photon temperature.

From the previous results the authors of Ref.\cite{McLerran} estimated
the average magnetic field strength at the present time. This is
\begin{equation}
B(l_{\rm diff}) \sim 10^{-7}~-~10^{-9}~{\rm G},
\end{equation}
where $l_{\rm diff} \sim 10$ AU is present time diffusion length,
and
\begin{equation}
B(l_{\rm gal}) \sim 10^{-17}~-~10^{-20}~{\rm G},
\end{equation}
on the galactic scale $l_{\rm gal} \sim 10^9$ AU.
\vskip 1.cm
\noindent

\subsection{Magnetic fields from Higgs field equilibration}
In the previous section we have seen that, concerning the generation of
magnetic fields, the QCDPT and the EWPT share several common aspects.
However, there is one important aspect which makes the EWPT much
more interesting than the QCDPT. In fact, at the electroweak
scale the electromagnetic field is directly influenced by the
dynamics of the Higgs field which drives the EWPT.

To start with we remind that, as a consequence
of the Weinberg-Salam theory,  before the EWPT is not even
possible to define the electromagnetic field, and that this
operation remains highly non-trivial until the transition is
completed. In a sense, we can say that the electromagnetic field
was ``born" during the EWPT. The main problem in the definition
of the electromagnetic field at the weak scale is the breaking of
the translational invariance: the Higgs field module and its
$SU(2)$ and $U_Y(1)$ phases take different values in different
positions. This is either a consequence of the presence of thermal
fluctuations, which close to $T_c$ are locally able to
break/restore  the $SU(2)\times U_Y(1)$ symmetry or
of the presence of large stable domains, or bubbles, where the broken
symmetry has settled.

The first generalized definition of the electromagnetic field in
the presence of a non-trivial Higgs background was given by
t'Hooft \cite{tHooft} in the seminal paper where he introduced
magnetic monopoles in a $SO(3)$ Georgi-Glashow model. t'Hooft
definition is the following
\begin{equation}\label{Hooft}
{\cal F}^{{\rm em}}_{\mu\nu} \equiv {\hat \phi}^a G^{a}_{\mu\nu} +
  {g}^{-1} \epsilon^{abc} {\hat{\phi}}^{a}
(D_\mu{\hat{\phi}})^{b} (D_\nu {\hat{\phi}})^{c}~.
\end{equation}
In the above $G^{a}_{\mu\nu} \equiv \partial W^a_\mu -
\partial W^a_\nu$, where
\begin{equation}\label{phia}
  {\hat{\phi}}^{a} \equiv
\frac{\Phi^{\dag}\tau^{a}\Phi}{\Phi^{\dag}\Phi}
\end{equation}
($\tau^{a}$ are the Pauli matrices) is a unit isovector which
defines the ''direction" of the Higgs field in the $SO(3)$
isospace (which coincides with $SU(2)$) and
$(D_\mu{\hat{\phi}})^{a} =
\partial_\mu\hat{\phi}^a + g \epsilon^{abc} W_\mu^b\hat{\phi}^c$,
where $W_\mu^b$ are the gauge fields components in the adjoint
representation. The nice features of the definition (\ref{Hooft})
are that it is gauge-invariant and it reduces to the standard
definition of the electromagnetic field tensor if a gauge rotation
can be performed so to have ${\hat \phi}^a = - \delta^{a3}$
(unitary gauge). In some models, like that considered by t'Hooft, a
topological obstruction may prevent this operation to be possible
everywhere. In this case singular points (monopoles) or lines
(strings) where $\phi^a = 0$ appear which become the source of
magnetic fields. t'Hooft result provides an existence proof of \mfs
produced by non-trivial vacuum configurations.

The  Weinberg-Salam theory, which is based on the  $SU(2)\times
U_Y(1)$ group representation, does not predict topologically stable
field configurations. We will see, however, that vacuum
non-topological configurations possibly produced during the EWPT
can still be the source of magnetic fields.

A possible generalization of the definition (\ref{Hooft}) for the
Weinberg-Salam model was given by Vachaspati \cite{Vachaspati91}.
It is
\begin{eqnarray}\label{HooftSM}
{F^{{\rm em}}_{\mu\nu}} &\equiv& - \sin\theta_W{\hat \phi}^a
F^{a}_{\mu\nu}  + \cos\theta_W F^{Y}_{\mu\nu}\nonumber\\ &-& i\frac
{\sin\theta_W}{g}\frac{2}{\Phi^{\dag}\Phi} \left[
\left(D_{\mu}{\Phi}\right)^{\dag} D_{\nu}{\Phi} -
\left(D_{\nu}{\Phi}\right)^{\dag} D_{\mu}{\Phi} \right],
\end{eqnarray}
where $D_\mu = \partial_\mu - i\frac{g}{2} \tau^a W_\mu^a -
i\frac{g'}{2}  Y_\mu$.

This expression was used by Vachaspati to argue that magnetic
fields should have been produced during the EWPT. Synthetically,
Vachaspati argument is the following. It is known that well below
the EWPT critical temperature $T_c$ the minimum energy state of
the Universe corresponds to a spatially homogeneous vacuum in
which $\Phi$ is covariantly constant, i.e. $D_{\nu}{\Phi} =
D_\mu{\hat{\phi}}^{a} = 0$. However, during the EWPT, and
immediately after it, thermal fluctuations give rise to a finite
correlation length $\xi \sim (eT_c)^{-1}$. Therefore, there are
spatial variations both in the Higgs field module $\vert \Phi
\vert $ and in its $SU(2)$ and $U(1)_Y$ phases which take random
values in uncorrelated regions {\footnote{ Vachaspati
\cite{Vachaspati91} did also consider Higgs field gradients
produced by the presence of the cosmological horizon. However,
since the Hubble radius at the EWPT is of the order of 1 cm
whereas $\xi \sim (eT_c)^{-1} \sim 10^{-16}$ cm, it is easy to
realize that \mfs possibly produced by the presence of the
cosmological horizon are phenomenologically irrelevant.}}. It was
noted by Davidson \cite{Davidson} that gradients in the radial
part of the Higgs field cannot contribute to the production of
\mfs as this component is electrically neutral. While this
consideration is certainly correct, it does not imply the failure
of Vachaspati argument. In fact, the role played by
the spatial variations of the $SU(2)$ and $U(1)_Y$ ``phases" of the the
Higgs field cannot be disregarded.
It is worthwhile to observe that gradients of these phases are not a
mere gauge artifact as they correspond to a nonvanishing kinetic
term in the Lagrangian. Of course one can always rotate Higgs
fields phases into gauge boson degrees of freedom (see below) but
this operation does not change $F^{{\rm em}}_{\mu\nu}$ which is a
gauge-invariant quantity. The contribution to the electromagnetic
field produced by gradients of ${\hat \phi}^a$ can be readily
determined by writing the Maxwell equations in the presence of an
inhomogeneous Higgs background \cite{GraRio}
\begin{eqnarray}
\label{MaxwellSM} \partial^\mu F^{\rm{em}}_{\mu\nu} =
&-& \sin\theta_W \left\{ D^\mu {\hat \phi}^a F^a_{\mu\nu}
\right.\nonumber \\
 &+& \left. \frac{i}{g}\partial^\mu
\left[\frac{4}{\Phi^{\dag}\Phi}\left(
\left(D_{\mu}{\Phi}\right)^{\dag}D_{\nu}{\Phi} -
D_{\mu}{\Phi}\left(D_{\nu}{\Phi}\right)^{\dag}
\right) \right] \right\} ~.
\end{eqnarray}
Even neglecting the second term on the righthand side of
Eq.(\ref{MaxwellSM}), which depends on the definition of
$F^{\rm{em}}_{\mu\nu}$ in a Higgs inhomogeneous background (see
below), it is evident
that a nonvanishing contribution to the electric 4-current arises
from the covariant derivative of ${\hat \phi}^a$. The physical
meaning of this contribution may look more clear to the reader if
we write Eq.(\ref{MaxwellSM}) in the unitary gauge
\begin{eqnarray}
\partial^{\mu} F^{\rm{em}}_{\mu\nu} &=&
+ie\left[ W^{\mu \dag}\left(D_\nu W_\mu\right) -
W^{\mu}\left(D_\nu W_\mu\right)^{\dag} \right]\nonumber\\ &-&
ie\left[ W^{\mu \dag}\left(D_\mu W_{\nu}\right) -
W^{\mu}\left(D_\mu W_\nu\right)^{\dag} \right]\label{MaxwellUG} \\
&-& ie \partial^\mu\left(W_\mu^{\dag} W_\nu - W_\mu W_\nu^{\dag}
\right)~. \nonumber
\end{eqnarray}
Not surprisingly, we see that the electric currents produced by
Higgs field equilibration after the EWPT are nothing but $W$ boson
currents.

Since, on dimensional grounds, $D_{\nu}{\Phi} \sim v/\xi$ where
$v$ is the Higgs field vacuum expectation value,  Vachaspati
concluded that magnetic fields (electric fields were supposed to
be screened by the plasma) should have been produced at the EWPT
with strength
\begin{equation}
  B \sim \sin\theta_W g T_c^2 \approx 10^{23}~~{\rm G}~.
\end{equation}
Of course these fields live on a very small scale of the order of
$\xi$ and in order to determine fields on a larger scale
Vachaspati claimed that a suitable average has to be performed
(see return on this issue below in this section).

Before discussing averages, however, let us try to understand
better the nature of the magnetic fields which may have been
produced by the Vachaspati mechanism. We notice that Vachaspati's
derivation does not seem to invoke any out-of-equilibrium process
and indeed the reader may wonder what is the role played by the
phase transition in the magnetogenesis. Moreover, magnetic fields
are produced anyway on a scale $(eT)^{-1}$ by thermal fluctuations
of the gauge fields so that it is unclear what is the difference
between \mfs produced by the Higgs fields equilibration and these
more conventional fields. In our opinion, although Vachaspati's
argument is basically correct its formulation was probably
oversimplified. Indeed, several works showed that in order to
reach a complete understanding of this physical effect a more
careful study of the dynamics of the phase transition is
called for. We shall now review these works starting from the case
of a first order phase transition.
\vskip0.5cm \noindent
\underline{The case of a first order EWPT}
\vskip0.5cm
Before discussing the $SU(2)\times U(1)$
case we cannot overlook some important work which was previously done
about phase equilibration during bubble collision in the framework
of more simple models. In the context of a $U(1)$ Abelian gauge
symmetry, Kibble and Vilenkin \cite{KibVil} showed that the
process of phase equilibration during  bubble collisions
give rise to relevant physical effects. The main tool developed by Kibble
and Vilenkin to investigate this kind of processes is the, so-called, {\it
gauge-invariant phase difference} defined by
\begin{equation}\label{gaugeinvphase}
  \Delta\theta = \int_A^B dx^k~D_k\theta~,
\end{equation}
where $\theta$ is the $U(1)$ Higgs field phase and $D_\mu \theta
\equiv \partial_\mu \theta + e A_\mu$ is the phase covariant
derivative. $A$ and $B$ are  points taken in the bubble interiors
and $k = 1,2,3$. $\Delta\theta$ obeys the Klein-Gordon equation
\begin{equation}\label{KG}
  (\partial^2 + m^2)\Delta\theta = 0
\end{equation}
where $m = ev$ is the gauge boson mass. Kibble and Vilenkin
assumed that during the collision the radial mode of the Higgs
field is strongly damped so that it rapidly settles to its
expectation value $v$ everywhere. One can choose a frame of
reference in which the bubbles are nucleated simultaneously with
centers at $(t,x,y,z)=(0,0,0,\pm R_c)$. In this frame, the bubbles
have equal initial radius $R_i=R_0$. Their first collision occurs
at $(t_c,0,0,0)$ when their radii are $R_c$ and $t_c=\sqrt{R_c^2 -
R_0^2}$. Given the symmetry of the problem about the axis joining
the nucleation centers ($z$-axis), the most natural gauge is the
axial gauge. In this gauge
\begin{equation} \label{axial}
 \theta(x)=\theta(\tau,z), \quad
 A^{\alpha}(x)=x^\alpha a(\tau,z)~,
\end{equation}
where $\alpha = 0,1,2 $ and $\tau^2 = t^2-x^2-y^2 $.
 The condition $\theta_a(\tau, 0)= 0 $ fixes the gauge completely.
At the point of first contact $z=0$, $\tau = t_c$ the Higgs field
phase was assumed to change from $\theta_0$ to $-\theta_0$ going
from a bubble into the other. This constitutes the initial
condition of the problem. The following evolution of $\theta$ is
determined by the Maxwell equation
\begin{equation}\label{MaxwellKV}
  \partial^\nu F_{\mu\nu} = j_\mu = - ev^2~D_\mu \theta
\end{equation}
and the Klein-Gordon equation which splits into
\begin{eqnarray}
\label{thetalineq}
\partial_\tau^2\theta_a + \frac{2}{\tau} \partial_\tau\theta -
\partial_z^2\theta +m^2\theta &=& 0~,\\
\label{alineq}
\partial_\tau^2 a + \frac{4}{\tau} \partial_\tau a -
\partial_z^2 a +m^2 a &=& 0~.
\end{eqnarray}
The solution of the linearized equations (\ref{thetalineq}) and
(\ref{alineq}) for $\tau > t_c$ then becomes
\begin{eqnarray}
\label{Kibbletheta} \theta_a(\tau,z) &=& \frac{\theta_0
t_c}{\pi\tau}\int_{-\infty}^\infty \frac{dk}{k}\sin k
z\left(\cos\omega(\tau-t_c) + \frac{1}{\omega t_c}
\sin\omega(\tau- t_c)\right)~,\\
  \label{Kibblea} a(\tau,z) &=&
\frac{\theta_0 m^2 t_c}{\pi e\tau^3} \int_{-\infty}^\infty
\frac{dk}{k}\sin k z\left[ - \frac{\tau- t_c}{\omega^2
t_c}\cos\omega(\tau- t_c)\right.\nonumber\\  &&
\left. + \left(\frac{\tau}{\omega}+\frac{1}{\omega^3
t_c}\right) \sin\omega(\tau- t_c)\right]~,
\end{eqnarray}
where $\omega^2=k^2+m^2$. The gauge-invariant phase difference is
deduced by the asymptotic behavior at $z \rightarrow \pm \infty$
\begin{eqnarray}\label{gipd}
  \Delta \theta &=& \theta_a(t,0,0,+\infty) -
  \theta_a(t,0,0,-\infty)\\
   &=& \frac{2\theta_0
t_c}{t} \left( \cos m(t - t_c) + \frac{1}{mt_c}\sin m(t -
t_c)\right)~.\nonumber
\end{eqnarray}
Thus, phase equilibration occurs with a time scale $t_c$
determined by the bubble size, with superimposed oscillations with
frequency given by the gauge-field mass. As we see from
Eq.(\ref{Kibblea}) phase oscillations come together with
oscillations of the gauge field. It follows from
Eq.(\ref{MaxwellKV}) that these oscillations give rise to an
``electric" current. This current will source an
''electromagnetic" field strength $F_{\mu\nu}$ {\footnote{It is
understood that since the toy model considered by Kibble e
Vilenkin is not $SU(2)\times U(1)_Y$, $F_{\mu\nu}$ is not the
physical electromagnetic field strength.}}. Because of the symmetry
of the problem the only nonvanishing component of $F_{\mu\nu}$ is
\begin{equation}\label{fmunuKV}
 F^{\alpha z} = x^\alpha \partial_z a(\tau,z)~.
\end{equation}
Therefore, we have an azimuthal \mf
$B^\varphi=F^{z\rho}=\rho\partial_z a$ and a longitudinal electric
field $E^z=F^{0z}= -t\partial_z a = -(t/\rho) B^\varphi(\tau,z)$,
where we have used cylindrical coordinates $(\rho, \phi)$.
We see that phase equilibration during bubble collision
indeed produces some real physical effects.

Kibble and Vilenkin did also consider the role of electric
dissipation. They showed that a finite value of the electric
conductivity $\sigma$ gives rise to a damping in the ''electric"
current which turns into a damping for the phase equilibration.
They found
\begin{equation}\label{smallsi}
\Delta \theta(t) = 2\theta_0 e^{-\sigma t/2}  \left( \cos m t +
\frac{\sigma}{2m}\sin m t\right)
\end{equation}
for small values of $\sigma$, and
\begin{equation}\label{largesi}
\Delta \theta(t) = 2\theta_0 \exp\left(-m^2t/\sigma\right)
\end{equation}
in the opposite case. The dissipation time scale is typically much
smaller than the time which is required for two colliding bubble
to merge completely. Therefore the gauge-invariant phase
difference settles rapidly to zero in the overlapping region of
the two bubbles and in its neighborhood. It is interesting to
compute the line integral of $D_k\theta$ over the path ABCD
represented in the Fig.\ref{KibVil-fig}.
\begin{figure}[htb]
\begin{center}
\begin{picture}(350,150)(0,0)
\CArc(100,75)(70,20,340)
\CArc(230,75)(70,0,160)
\CArc(230,75)(70,200,360)
\ArrowLine(130,75)(200,75)
\ArrowLine(200,75)(200,125)
\ArrowLine(200,125)(130,125)
\ArrowLine(130,125)(130,75)
\Text(130,75)[rt]{$A$}
\Text(200,75)[lt]{$B$}
\Text(202,125)[lb]{$C$}
\Text(130,125)[rb]{$D$}
\end{picture}
\end{center}
\vspace{0.5cm}
\caption{ Two colliding bubbles are depicted. The gauge invariant
phase difference is computed along the path ABCD. (From Ref.\cite{KibVil}).}
\label{KibVil-fig}
\end{figure}
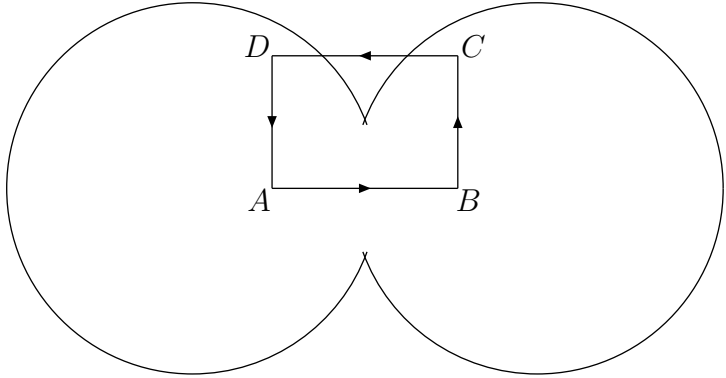
%
From the previous considerations
it follows that $\Delta \theta_{AB} = 0$, $\Delta \theta_{AD} =
\Delta \theta_{BC} = 0$ and $\Delta \theta_{DC} = 2\theta_0$. It
is understood that in order for the integral to be meaningful, the
vacuum expectation value of the Higgs field has to remain nonzero
in the collision region and around it, so that the phase $\theta$
remains well defined and interpolates smoothly between its values
inside the bubbles. Under these hypothesis we have
\begin{equation}\label{KVloop}
  \oint_{ABCD} D_k\theta dx^k = 2\theta_0~.
\end{equation}
The physical meaning of this quantity is recognizable at a glance in
the unitary gauge, in which each $\Delta \theta$ is given by a
line integral of the vector potential $\bf A$. We see that the
gauge-invariant phase difference computed along the loop is
nothing but the magnetic flux trough the loop itself
\begin{equation}\label{KVflux}
  \phi(B) = \oint_{ABCD} A_k\theta dx^k =
  \frac 1 e \oint_{ABCD} D_k\theta dx^k = \frac {2\theta_0}{e}~.
\end{equation}
In other words, phase equilibration give rise to a ring of
magnetic flux near the circle on which bubble walls intersect. If
the initial phase difference between the two bubbles is $2\pi$,
the total flux trapped in the ring is exactly one flux quantum,
$2\pi/e$.

Kibble and Vilenkin did also consider the case in which three
bubbles collide. They argued that in this case the formation of a
string, in which interior symmetry is restored, is possible.
Whether or not this happens is determined by the net phase
variation along a closed path going through the three bubbles. The
string forms if this quantity is larger than $2\pi$. According to
Kibble and Vilenkin strings cannot be produced by two bubble
collisions because, for energetic reasons, the system will tend to
choose the shorter of the two paths between the bubble phases so
that a phase displacement $\geq 2\pi$ can never be obtained. This
argument, which was first used by Kibble \cite{Kibble} for the
study of defect formation, is often called the ``geodesic rule''.

The work of Kibble and Vilenkin was reconsidered by Copeland and
Saffin \cite{CopSaf96} and more recently by  Copeland, Saffin and
T\"ornkvist \cite{CopST} who showed that during bubble collision
the dynamics of the radial mode of the Higgs field cannot really
be disregarded. In fact, violent fluctuations in the modulus of
the Higgs field take place and cause symmetries to be restored
locally, allowing the phase to ``slip'' by an integer multiple of
$2\pi$ violating the geodesic rule. Therefore strings, which
carry a magnetic flux, can be produced also by the collision of
only two bubbles. Saffin and Copeland \cite{SafCop97} went a step
further by considering phase equilibration in the $SU(2)\times
U(1)$ case, namely the electroweak case. They showed that for some
particular initial conditions the $SU(2)\times U(1)$ Lagrangian is
equivalent to a $U(1)$ Lagrangian so that part of Kibble and
Vilenkin \cite{KibVil} considerations can be applied. The
violation of the geodesic rule allows the formation of vortex
configurations of the gauge fields. Saffin and Copeland argued
that these configurations are related to the Nielsen-Olesen
vortices \cite{NieOle73}. Indeed, it is know that such a kind of
non-perturbative solutions are allowed by the Weinberg-Salam model
\cite{Nambu77} (for a comprehensive review on electroweak strings
see Ref.\cite{AchVac}). Although electroweak string are not
topologically stable, numerical simulations performed in
Ref.\cite{SafCop97} show that in presence of small perturbations the
vortices survives on times comparable to the time required for bubble
to merge completely.

The generation of \mfs in the $SU(2)\times U(1)_Y$ case was not
considered in the work by Saffin and Copeland. This issue was the
subject of a following paper by Grasso and Riotto \cite{GraRio}.
The authors of Ref.\cite{GraRio} studied the dynamics of the gauge fields
starting from the following initial Higgs field configuration
\be
\label{phiin} \Phi_{\rm in}({x}) = \frac{1}{\sqrt{2}}\left(
\begin{array}{c} {0}\\ {\rho({\bf x})} \end{array}\right) +
\frac{1}{\sqrt{2}}\:\exp\left(-i \frac {\theta_0}{2} n^a\tau^a\right)
\left(
\begin{array}{c} {0}\\ {\rho({\bf x} - {\bf b})e^{i\varphi_0}}
\end{array}\right)
\ee
 which represents the superposition of the Higgs fields of two
 bubbles which are separated by a distance $b$.
In the above $n^a$ is a unit vector in the $SU(2)$ isospace and
$\tau^a$ are the Pauli matrices. The phases and the orientation of
the Higgs field were chosen to be uniform across any single
bubble. It was assumed that Eq.(\ref{phiin}) holds until the two
bubble collide ($t = 0$). Since $n^a\tau^a$ is the only
Lie-algebra direction which is involved before the collision, one
can write the initial Higgs field configuration in the form
\cite{SafCop97}
\begin{equation}\label{phiin2}
  \Phi_{{\rm in}}({\bf x}) = \frac{1}{\sqrt{2}} {\exp}\left(
  -i \frac {{ \theta}({\bf x})}{2} n^a\tau^a\right)
  \left( \begin{array}{c} 0\\ \rho({\bf x}) e^{i\varphi({\bf x})}~.
\end{array}\right)
\end{equation}
In order to disentangle the peculiar role played by the Higgs
field phases,  the initial gauge fields $W_\mu^a$  and their
derivatives were assumed to be zero at $t = 0$. This condition is
of course gauge dependent and should be interpreted as a gauge
choice. It is convenient to write the equation of motion for the
gauge fields in the adjoint representation. For the $SU(2)$ gauge
fields we have
\begin{equation}\label{SU2eq}
  D^\mu F^a_{\mu\nu} = g |\rho|^2~ \epsilon^{abc} D_\nu{\hat \phi}^b
{\hat \phi}^c
\end{equation}
where the isovector ${\hat \phi}^a$ has been defined in
Eq.(\ref{phia}). Under the assumptions mentioned in the above, at
$t = 0$, this equation reads
\begin{equation}
\label{Weqin}
\partial^\mu F^{a}_{\mu\nu} = - g|\rho|^2 \partial_{\nu}\theta(x)
\left(n^a - n^c {\hat \phi}^a{\hat \phi}^c\right)~.
\end{equation}
In general, the unit isovector ${\hat \phi}^a$ can be decomposed
into
\begin{equation}\label{phidecomp}
 {\hat \phi} = \cos\theta~{\hat \phi}_0 + \sin\theta~{\hat
n}\times {\hat \phi}_0 + 2 \sin^2{\frac \theta 2}~\left({{\hat
n}}\cdot {\hat \phi}_0\right)~{{\hat n}}
\end{equation}
where ${\hat \phi}^T_0 \equiv - (0,0,1)$. It is straightforward to
verify that in the unitary gauge, ${\hat \phi}$ reduces to  ${\hat
\phi}_0$. The relevant point in Eq.(\ref{phiin2}) is that the
versor ${\hat n}$, about which it is performed the $SU(2)$ gauge
rotation, does not depend on the space coordinates. Therefore,
without loosing generality, we have the freedom to choose ${\hat
n}$ to be everywhere perpendicular to ${\hat {\bf \phi}}_0$. In
other words, ${\hat {\bf \phi}}$ can be everywhere obtained by
rotating ${\hat \phi}_0$ by an angle $\theta$ in the plane
identified by $\hat{n}$ and ${\hat {\phi}}_0$. Formally, ${\hat
\Phi} = \cos\theta {\hat \phi}_0 + \sin\theta~{\hat n}\times {\hat
\phi}_0$, which clearly describes a simple  $U(1)$ transformation.
In fact, since it is evident that the  condition $\hat{n}\perp
{\bf {\hat \phi}}_0$ also implies $\hat{n}\perp {\bf {\hat
\phi}}$, the equation of motion (\ref{Weqin}) becomes
\begin{equation}
\label{Weqin2}
\partial^\mu F^a_{\mu\nu} = - g|\rho|^2 \partial_{\nu}\theta(x) n^a~.
\end{equation}
As expected, we see that only the gauge field component along the
direction $\hat{n}$, namely $A_\mu=n^aW^a_\mu$, that has  some
initial dynamics which is created by a nonvanishing gradient of
the phase between the two domains. When we generalize this result
to the full $SU(2)\times U(1)_Y$ gauge structure, an extra
generator, namely the hypercharge, comes-in. Therefore in this
case is not any more possible to choose an arbitrary direction for
the unit vector $\hat{n}$ since different orientations of the unit
vector ${\hat n}$ with respect to ${\hat \phi}_0$ correspond to
different physical situations. We can still consider the case in
which $\hat{n}$ is parallel to  $\hat{\phi}_0$ but we should take
in mind that this is not the only possibility. In this case we
have
\begin{eqnarray}
\partial^\mu F^3_{\mu\nu}&=&  \frac{g}{2} \rho^2(x)\left(\partial_\nu\theta+
\partial_\nu\varphi\right), \label{F3}\\
\partial^\mu F^Y_{\mu\nu} &=& - \frac{g'}{2} \rho^2(x) \left(\partial_\nu
\theta+\partial_\nu\varphi\right) \label{FY}
\end{eqnarray}
where $g$ and $g'$ are respectively the $SU(2)$ and $U(1)_Y$
gauge coupling constants. It is noticeable that in this case the charged gauge
fields are not excited by the phase gradients at the time when bubble first
collide. We can combine Eqs.(\ref{F3}) and (\ref{FY}) to obtain the
equation of motion for the $Z$-boson field
\begin{equation}
\label{Zeq0}
\partial^\mu F^Z_{\mu\nu} =  \frac{\sqrt{g^2 + g^{'2}}}{2}\rho^2(x)
\left(\partial_\nu\theta+ \partial_\nu\varphi\right).
\end{equation}
This equation tells us that a gradient in the phases of the Higgs
field gives rise to a nontrivial dynamics of the $Z$-field with an
effective gauge coupling constant $\sqrt{g^2 + g^{'2}}$. We see
that the equilibration of the phase $(\theta+\varphi)$ can be now
treated in analogy to the $U(1)$ toy model studied by Kibble and
Vilenkin \cite{KibVil}, the role of the $U(1)$ "electromagnetic"
field being now played by the $Z$-field. However, differently from
Ref.\cite{KibVil} the authors of Ref.\cite{GraRio} left the Higgs
field modulus free to change in space. Therefore, the equation of
motion of $\rho(x)$ has to be added to (\ref{Zeq0}). Assuming the
charged gauge field does not evolve significantly, the complete set
of equations of motion that we can write at finite, though small,
time after the bubbles first contact, is
\begin{eqnarray}
\label{nieole}
&&\partial^\mu F^{Z}_{\mu\nu} =  \frac{g}{2\cos\theta_W}\rho^2(x)
\left(\partial_\nu\varphi + \frac{g}{2\cos\theta_W} Z_\nu\right),\nonumber\\
&&d^\mu d_\mu\left(\rho(x) {\rm e}^{i\frac{\varphi}{2}}\right) +
2\lambda \left(\rho^2(x) -
\frac{1}{2}\eta^2\right)\rho(x){\rm e}^{i\frac{\varphi}{2}}=0,
\end{eqnarray}
 where $d_\mu = \partial_\mu + i\frac{g}{2\cos\theta_W} Z_\mu$, $\eta$ is the
vacuum expectation value of $\Phi$ and $\lambda$ is the quartic coupling.
Note that, in analogy with \cite{KibVil}, a  gauge invariant phase difference
can be introduced by making use of the covariant derivative $d_\mu$.
Equations (\ref{nieole}) are the Nielsen-Olesen equations of
motion  \cite{NieOle73}. Their solution describes
a $Z$-vortex where $\rho=0$ at its core \cite{Vachaspati93}.
The geometry of the  problem
implies that the vortex is closed, forming a ring which axis coincide
with the conjunction of bubble centers. This result provides further support
to the possibility that electroweak strings are produced during the EWPT.

In principle, in order to determine the magnetic field produced during the
process that we illustrated in the above,  we need a gauge-invariant
definition of the electromagnetic field strength in the
presence of the non-trivial Higgs background.
We know however that such definition is not unique \cite{Coleman}.
For example, the authors of Ref.\cite{GraRio} used the definition
given in Eq.(\ref{HooftSM}) to find that the electric current is
\begin{equation}
\partial^\mu F^{\rm em}_{\mu\nu}= 2\tan\theta_W \partial^\mu\left(
{Z}_\mu\partial_\nu \ln{\rho(x)} -  {Z}_\nu\partial_\mu \ln{\rho(x)} \right)
\end{equation}
whereas other authors \cite{Ola98}, using the definition
\begin{equation}\label{oladef}
{\cal F}^{{\rm em}}_{\mu\nu} \equiv - \sin\theta_W{\hat \phi}^a
F^{a}_{\mu\nu}  +
\cos\theta_W F^{Y}_{\mu\nu} +
\frac {\sin\theta_W}{g}  \epsilon^{abc}
{\hat{\phi}}^{a} (D_\mu{\hat{\phi}})^{b} (D_\nu {\hat{\phi}})^{c}~,
\end{equation}
found no electric current, hence no magnetic field, at all.
We have to observe, however, that the choice between these, as others,
gauge invariant definitions is more a matter of taste than physics.
Different definitions just give the same name to different
combinations of the gauge fields. The important requirement
which acceptable definitions of the electromagnetic field have to fulfill
is that they have to reproduce the standard definition in the broken phase
with a uniform Higgs background.
This requirement is fulfilled by both the definitions used in the
Refs.\cite{GraRio} and \cite{Ola98}.
In our opinion, it is not really meaningful to ask what
is the electromagnetic field inside, or very close to, the electroweak
strings. The physically relevant question is what are the electromagnetic
relics of the electroweak strings once the EWPT is concluded.

One important point to take in mind is that electroweak strings are
not topologically stable (see \cite{AchVac} and references therein)
and that, for the physical value of the Weinberg angle, they rapidly decay
after their formation. Depending on the nature of the decay process two
scenarios are possible.
According to Vachaspati \cite{Vachaspati94} long strings should decay in short
segments  of length $\sim m_W^{-1}$.
Since the $Z$-string carry a flux of $Z$-magnetic flux in its interior
\be
\phi_Z = {{4 \pi } \over {\alpha}} =
          {{4\pi} \over e} \sin\theta_W~ \cos\theta_W ~.
\ee
and the $Z$ gauge field is a linear superposition of the
$W^3$ and $Y$ fields  then, when the string terminates,
the $Y$ flux cannot terminate because it is a $U(1)$ gauge field and the
$Y$ magnetic field is divergenceless. Therefore some field must continue
even beyond the end of the string. This has to be the massless field
of the theory, that is, the electromagnetic field. In some sense,
a finite segment of $Z$-string terminates on magnetic monopoles
\cite{Nambu77}.
The magnetic flux emanating from a monopole is:
\be
\phi_{B} = {{4\pi} \over {\alpha}} \tan\theta_W
                   = {{4\pi} \over e} \sin^2 \theta_W~.
\ee
This flux may remain frozen-in the surrounding plasma and become a seed
for cosmological magnetic fields.

Another possibility is that $Z$-strings decay by the formation of
a $W$-condensate in their cores. In fact, it was showed  by
Perkins \cite{Perkins} that while electroweak symmetry restoration
in the core of the string reduces  $m_W$, the magnetic field via
its coupling to the anomalous magnetic moment of the $W$-field,
causes, for $eB > m_W^2$, the formation of a condensate of the
$W$-fields. Such a process is based on the Ambj\/orn-Olesen
instability which will be discussed in some detail in the
Chap.\ref{chap:stability} of this review. As noted in \cite{GraRio}
the presence of an
inhomogeneous $W$-condensate produced by string decay gives rise
to electric currents which may sustain magnetic fields even after
the $Z$ string has disappeared. The formation of a $W$-condensate
by strong \mfs at the EWPT time, was also considered by Olesen
\cite{Olesen92}.

We can now wonder what is the predicted strength of the magnetic
fields at the end of the EWPT. An attempt to answer to this question
has been done by Ahonen and Enqvist \cite{AhoEnq97} (see also
Ref.\cite{Enqvist98})  where the formation of
ring-like magnetic fields in collisions of bubbles of broken phase
in an Abelian Higgs model were inspected. Under the assumption
that magnetic fields are generated by a process that resembles the
Kibble and Vilenkin \cite{KibVil} mechanism, it was concluded that
a  magnetic field of the order of  $B  \simeq 2 \times 10^{20}$ G
with a coherence length of about $10^2~\rm{GeV}^{-1}$ may be
originated. Assuming turbulent enhancement the authors of
Ref.\cite{AhoEnq97} of the field by inverse cascade \cite{BraEO},
a root-mean-square value of the magnetic field $B_{{\rm rms}}
\simeq 10^{-21}$ G  on a comoving scale of $10$ Mpc might be
present today. Although our previous considerations give some
partial support to the scenario advocated in \cite{AhoEnq97} we
have to stress, however, that only in some restricted  cases it is
possible to reduce the dynamics of the system to the dynamics of a
simple $U(1)$ Abelian group. Furthermore, once $Z$-vortices are
formed the non-Abelian nature of the electroweak theory shows due
to the back-reaction of the magnetic field on the charged gauge
bosons and it is not evident  that  the same numerical values
obtained in \cite{AhoEnq97} will be obtained in the case of the
EWPT.

However the most serious problem with the kind of scenario
discussed in this section comes form the fact that, within the
framework of the standard model, a first order EWPT seems to be
incompatible with the Higgs mass experimental lower limit
\cite{Kajantie96}. Although some parameter choice of the minimal
supersymmetric standard model (MSSM) may still allow a first order transition
\cite{MSSM}, which may give rise to \mfs in a way similar to that
discussed in the above, we think it is worthwhile to keep an open
mind and consider what may happen in the case of a second order
transition or even in the case of a cross over.
\vskip0.5cm \noindent
\underline{The case of a second order EWPT}
\vskip0.5cm
As we discussed in the first part of this section, magnetic fields
generation by Higgs field equilibration share several common
aspects with the formation of topological defects in the early
Universe. This analogy holds, and it is even more evident, in the
case of a second order transition. The theory of defect formation
during a second order phase transition was developed in a seminal
paper by Kibble \cite{Kibble}. We shortly review some relevant
aspects of the Kibble mechanism. We start from the Universe being
in the unbroken phase of a given symmetry group $G$. As the
Universe cools and approach the critical temperature $T_c$
protodomains are formed by thermal fluctuations where the vacuum
is in one of the degenerate, classically equivalent, broken symmetry
vacuum states. Let $M$ be the manifold of the broken
symmetry degenerate vacua. The protodomains size is determined by
the Higgs field correlation function. Protodomains become stable
to thermal fluctuations when their free energy becomes larger than
the temperature. The temperature at which this happens is usually named
Ginsburg temperature $T_G$. Below $T_G$ stable domains are formed which, in
the case of a topologically nontrivial manifold $M$, give rise to
defect production. Rather, if $M$ is topologically trivial, phase
equilibration will continue until the Higgs field is uniform
everywhere. This is the case of the Weinberg-Salam model, as well
as of its minimal supersymmetrical extension.

Higgs phase equilibration, which occurs when stable domains merge,
gives rise to magnetic fields in a way similar to that
described by Vachaspati \cite{Vachaspati91} (see the beginning of
this section). One should keep in mind, however, that as a matter of
principle, the domain size, which determine the Higgs field gradient,
is different from the correlation length at the critical
temperature \cite{GraRio}. At the time when stable domains form,
their size is given by the correlation length in the broken phase
at the Ginsburg temperature. This temperature was computed, in the case
of the EWPT, by  the authors of Ref.\cite{GraRio} by comparing the expansion
rate of the Universe with the nucleation rate per unit volume of sub-critical
bubbles of symmetric phase (with size equal to the correlation length in
the broken phase) given by
\begin{equation}
\Gamma_{{\rm ub}}=\frac{1}{\ell_{{\rm b}}^4}\:{\rm e}^{-S_3^{{\rm
ub}}/T},
\end{equation}
where $\ell_{{\rm b}}$ is the correlation length in the broken
phase. $S_3^{{\rm ub}}$ is the high temperature limit of the
Euclidean action (see e.g. Ref.\cite{Enqvist91}). It was shown
that for the EWPT the Ginsburg temperature is  very close to the
critical temperature, $T_G=T_c$ within a few percent. The
corresponding size of a broken phase domain is determined by the
correlation length in the broken phase at $T = T_G$
\begin{equation}
\frac{1}{\ell(T_G)^2_{{\rm
b}}}=V^{\prime\prime}\left(\langle\phi(T_G)\rangle,T_G\right)
\end{equation}
where $V(\phi,T)$ is the effective Higgs potential. $\ell
(T_G)^2_{{\rm b}}$ is weakly dependent on $M_H$, $\ell_{{\rm
b}}(T_G)\simeq 11/T_G$ for $M_H=100$ GeV and $\ell_{{\rm
b}}(T_G)\simeq 10/T_G$ for $M_H=200$ GeV. Using this result and
Eq.(\ref{HooftSM}) the authors of Ref.\cite{GraRio} estimated the
magnetic field strength at the end of the EWPT to be of order of
\begin{equation}
\label{Bell}
 B_\ell \sim 4e^{-1} \sin^2\theta_W \ell_{{\rm
b}}^2(T_G) \sim 10^{22} ~\rm{G}~,
\end{equation}
on a length scale $\ell_{{\rm b}}(T_G)$.

Although it was showed by Martin and Davis \cite{MarDav95} that
\mfs produced on such a scale may be stable against thermal
fluctuations,  it is clear that  \mfs of phenomenological interest
live on scales much larger than $\ell_{{\rm b}}(T_G)$. Therefore,
some kind of average is required. We are ready to return to the
discussion of the Vachaspati mechanism for \mf generation
\cite{Vachaspati91}. Let us suppose we are interested in the \mf
on a scale $L = N \ell$. Vachaspati argued that, since the Higgs
field is uncorrelated on scales larger than $\ell$, its gradient
executes a random walk as we move along a line crossing $N$
domains. Therefore, the average of the gradient $D_\mu \Phi$ over
this path should scale as $\sqrt N$. Since the \mf is proportional
to the product of two covariant derivatives, see Eq.(\ref{HooftSM}),
Vachaspati concluded that it scales as $1/N$. This conclusion,
however, overlooks the difference between $\langle D_\mu \Phi^\dag
\rangle \langle D_\mu \Phi \rangle$ and $\langle D_\mu \Phi^\dag
D_\mu \Phi \rangle$. This point was noticed by Enqvist and Olesen
\cite{EnqOle93} (see also Ref.\cite{HinEve97}) who produced a
different estimate for the average \mf, $\langle B \rangle_{{\rm
rms}, ~L} \equiv B(L) \sim B_\ell/\sqrt N$. Neglecting the possible
role of the magnetic helicity (see the next section) and of
possible related effects, e.g. inverse cascade, and using
Eq.(\ref{Bell}), the line-averaged field today on a scale $L \sim
1$ Mpc ($N \sim 10^{25}$) is found to be of the order $B_0(1 {\rm
Mpc}) \sim 10^{-21}$ G.
 \vskip0.5cm
Another important point of this kind of scenario (for the reasons
which will become clear in the next section) is
that it naturally gives rise to a nonvanishing vorticity. This
point can be understood by the analogy with the process which lead
to the formation of superfluid circulation in a Bose-Einstein
fluid which is rapidly taken below the critical point by a
pressure quench \cite{Zurek}. Consider a circular closed path
through the superfluid of length  $C = 2\pi R$. This path will
cross $N \simeq C/\ell$ domains, where $\ell$ is the
characteristic size of a single domain. Assuming that the phase
$\theta$ of the condensate wave function is uncorrelated in each
of the $N$ domains (random-walk hypothesis) the typical mismatch
of $\theta$ is given by:
\begin{equation} \label{dtheta} \delta\theta = \int_C {\bf
\nabla}\theta\cdot {\bf ds} \sim \sqrt{N}
\end{equation} where ${\bf \nabla}\theta$ is the phase gradient
across two adjacent domains and ${\bf ds}$ is the line element
along the circumference. It is well known (see e.g.
\cite{FetWal}) that from the Schr\"odinger equation it follows
that the velocity of a superfluid is given by the gradient of the
phase trough the relation ${\bf v}_s = (\hbar/m) {\bf
\nabla}\theta $, therefore  (\ref{dtheta}) implies
\begin{equation}
v_s = (\hbar/m) \frac 1 d \frac {1}{\sqrt{N}}~.
\end{equation}
It was argued by Zurek \cite{Zurek} that this phenomenon can
effectively simulate the formation of defects in the early
Universe. As we discussed in the previous section, although the
standard model does not allows topological defects, embedded
defects, namely electroweak strings, may be produced through a
similar mechanism. Indeed a close analogy  was showed to exist
\cite{VacVol} between the EWPT and the ${}^3He$ superfluid
transition where formation of vortices is experimentally observed.
This hypothesis received further support by some recent lattice
simulations which showed evidence for the formation of a cluster
of Z-strings just above the cross-over temperature
\cite{Chernodub} in the case of a $3D$ $SU(2)$ Higgs model.
Electroweak strings should lead to the generation of magnetic
fields in the same way we discussed in the case of a first order
EWPT. Unfortunately, to estimate the strength of the \mf produced
by this mechanism requires the knowledge of the string density
and net helicity which, so far, are rather unknown quantities.

\section{Magnetic helicity and electroweak baryogenesis}\label{sec:helicity}

As we discussed in the introduction of our report, the cosmological magnetic
flux is a nearly conserved quantity  due to the high conductivity of the
Universe. In this section we will focus on another quantity which, for the
same reason, is approximatively conserved during most of the Universe
evolution. This is the so called {\it magnetic helicity} defined by
\begin{equation}
\label{helicity}
 {\cal  H} \equiv \int d^3x {\bf A}\cdot {\bf B}
\end{equation}
where ${\bf A}$ is the electromagnetic vector potential and ${\bf B}$
is the magnetic field.
In the presence of a small value of the electric conductivity $\sigma$ the
time evolution of $\cal H$ is given by
\begin{equation}
\frac{d{\cal H}}{dt} = - \frac{1}{\sigma} \int d^3x~ {\bf B}\cdot
{\bf \nabla}\times{\bf B}~.
\end{equation}

Besides for the fact to be a nearly conserved quantity, the
magnetic helicity is a very interesting quantity for a number of
different reasons. The main among these reasons are:
\begin{itemize}
\item In a field theory language $\cal H$ coincides with the Chern-Simon
number which is know to be related to the topological properties of the
gauge fields.
\item  Since $\cal H$ is a P (parity) and CP-odd function, the observation of
a nonvanishing net value of this quantity would be a manifestation of a
macroscopic  violation of both these symmetries.
\item It is know from magneto-hydro-dynamics that the presence of magnetic
helicity can lead to the amplification of magnetic fields and contribute to
their self-organization into a large scale ordered configuration
(see Sec.\ref{sec:evolution}). The same phenomenon could take place at
a cosmological level.
\end{itemize}

In the last few years, several authors proposed mechanisms for the
production of magnetic helicity in the early Universe starting from particle
physics processes. Cornwall \cite{Cornwall} suggested that magnetic helicity
was initially stored in the Universe under the form of baryons (B) and leptons
(L) numbers possibly generated by some GUT scale baryogenesis mechanism.
He assumed that an order one fraction of the total classically conserved $B+L$
charge was dissipated by anomalous processes at the EW scale and showed
that a small fraction of this dissipated charge, of the order of
$n_{B+L} T^{-3}$, may have been converted into a magnetic helicity of the
order of
\begin{equation}
{\cal H} \sim \alpha^{-1}(N_B + N_L) \simeq 10^{66}~~{\rm erg~cm}~.
\end{equation}
Another possibility is that before symmetry breaking of
a non-Abelian gauge symmetry vacuum configurations existed which carried
nonvanishing winding number. It was shown by Jackiw and Pi \cite{Jackiw}
that after symmetry breaking, one direction in isospin space is identified
with electromagnetism, and the projection of the vacuum configuration
becomes a \mf with non vanishing helicity.

A different mechanism was proposed by Joyce and Shaposhnikov
\cite{JoySha}. In this case it was assumed that some excess of
right-handed electrons over left-handed positrons was produced by
some means ({\it e.g.} from some GUT scale leptogenesis) above a
temperature $T_R$. At temperatures higher than $T_R$ perturbative
processes which changes electron chirality are out of thermal
equilibrium ($T_R \sim 3$ TeV in the SM \cite{CamDEO}). Therefore,
a chemical potential for right electrons $\mu_R$ can be introduced
above $T_R$. On the other hand, the corresponding charge is not
conserved because of the Abelian anomaly, which
gives
\begin{equation}
\label{Ranomaly}
\partial_\mu j_R^\mu = - \frac{g'^2y_R^2}{64 \pi^2}
~f_{\mu\nu}{\tilde f}^{\mu\nu}~.
\end{equation}
In the above, $f_{\mu\nu}$ and ${\tilde f}^{\mu\nu}$ are,
respectively, the $U(1)_Y$ hypercharge field strength and its
dual, $g' $ is the associated gauge coupling constant, and $y_R =
- 2$ is the hypercharge of the right electron. As it is well
known, Eq.(\ref{Ranomaly})  relates the variation in the number of
the right handed electrons $N_R$ to the variation of the
topological properties (Chern-Simon number) of the hypercharge
field configuration. By rewriting this expression in terms of the
hypermagnetic ${\vec B}_Y$ and of the hyperelectric ${\vec E}_Y$
fields,
\begin{equation}
\label{Ranomalybis}
\partial_\mu j_R^\mu = - \frac{g'^2}{4 \pi^2}~{\bf B}_Y\cdot {\bf E}_Y~,
\end{equation}
it is  evident the relation of $j_R^\mu$ with the hypermagnetic
helicity. It is worthwhile to observe that only the hypermagnetic
helicity is coupled to the fermion number by the chiral anomaly
whereas such a coupling is absent for the Maxwell magnetic
helicity because of the vector-like coupling of the electromagnetic
field to fermions. From Eq.(\ref{Ranomaly}) it follows that the
variation in $N_R$,  is related to the variation in the
Chern-Simon number,
\begin{equation}
N_{CS}=-\frac{g'^2}{32 \pi^2}\int d^3{\bf x} \epsilon_{ijk}
f_{ij}b_k,
\label{integratedanomaly}
\end{equation}
by  $\Delta N_R = \frac{1}{2} y_R^2\Delta N_{CS}$. In the above
$b_k$ represents the hypercharge field potential. The energy
density sitting in right electrons is of order $\mu_R^2T^2$ and
their number density of order $\mu_R T^2$. Such fermionic number
can be reprocessed into hypermagnetic helicity of order $g'^2 k
b^2$, with energy of order $k^2 b^2$, where $k$ is the momentum of
the classical hypercharge field and $b$ is its amplitude.
Therefore, at $ b> T/g'^2$ it is energetically convenient for the
system to produce hypermagnetic helicity by ``eating-up''
fermions. It was showed by Joyce and Shaposhnikov \cite{JoySha},
and in more detail by Giovannini and Shaposhnikov \cite{GioSha},
that such a phenomenon corresponds to a {\it magnetic dynamo
instability}. In fact, by adding the anomaly term to the Maxwell
like equations for the hyperelectric and hypermagnetic fields
these authors were able to write the anomalous
magneto-hydrodynamical (AMHD) equations for the electroweak plasma
in the expanding Universe, including the following generalized
hypermagnetic diffusivity equation
\begin{equation}
\label{adynamo}
\frac{\partial{\bf B}_Y}{\partial\tau} = - \frac{4a\alpha' }{\pi\sigma}
~{\bf \nabla}\times(\mu_R {\bf B}_Y) +
~{\bf \nabla}\times({\bf v}\times {\bf B}_Y) +
\frac 1 \sigma \nabla^2 {\bf B}_Y~.
\end{equation}
In the above $a$ is the Universe scale factor, $\alpha' \equiv
g'^2/4\pi$, $\tau = \int a^{-1}(t) dt$ is the conformal time, and
$\sigma$ the electric conductivity of the electroweak plasma
\cite{conductivity}. By comparing Eq.(\ref{adynamo}) with the
usual magnetic diffusivity equation ( see e.g. Ref.\cite{Jackson})
one sees that the first term on the r.h.s. of Eq.(\ref{adynamo})
corresponds to the so-called dynamo term which is know to be
related to the vorticity in the plasma flux \cite{Parker}. We see
that the fermion asymmetry, by providing a macroscopic parity
violation, plays a similar role of that played by the fluid
vorticity in the dynamo amplification of magnetic fields. In the
scenario advocated in Ref.\cite{JoySha,GioSha}, however, it is not
clear what are the seed fields from which the dynamo amplification
start from. Clearly, no amplification takes place if the initial
value of the hypermagnetic field vanishes. Perhaps, seeds field
may have been provided by thermal fluctuations or from a previous
phase transition although this is a matter of speculation.
According to Joyce and Shaposhnikov  \cite{JoySha}, assuming that
a large right electron asymmetry $\mu_R/T \sim 10^{-2}$ was
present when $T = T_R$, magnetic field of strength $B \sim
10^{22}$ G may have survived until the EWPT time with typical
inhomogeneity scales $\sim 10^6/T$.
 \vskip0.5cm
 Another interesting point raised by Giovannini and Shaposhnikov
\cite{GioSha} is that the Abelian anomaly may also process a
preexisting hypermagnetic helicity into fermions. In this sense
the presence of tangled magnetic fields in the early Universe may
provide a new leptogenesis scenario.

Indeed, assuming that a primordial hypermagnetic field ${\vec
B}_Y$ was present before the EWPT with some non-trivial topology
(i.e. $\langle {\bf B}_Y \cdot {\bf \nabla}\times {\bf B}_Y
\rangle \neq 0 $) the kinetic equation of right electrons for $T >
T_c$ is
\begin{equation}
\frac{\partial}{\partial t}\left(\frac{\mu_R}{T}\right)=
-\frac{g'^2}{4\pi^2 \sigma a T^3} \frac{783}{88}
{\bf B}_{Y}\cdot {\bf \nabla}\times {\bf B}_{Y} - (\Gamma +
\Gamma_{np} ) \frac{\mu_R}{T}, \label{muR}
\end{equation}
where
\begin{equation}
\Gamma_{np} = \frac{783}{22} \frac{\alpha'^2}{\sigma a \pi^2}
\frac{|{\bf B}_{Y}|^2}{T^2}
\end{equation}
is the rate of the $N_R$ nonconserving anomalous processes whereas
$\Gamma$ is the rate of the perturbative ones. In the case
$\Gamma_{np}> \Gamma$, as a consequence of Eq.(\ref{muR}), one
finds
\begin{equation}
n_{R} \simeq -\frac{88 \pi^2}{783 g'^2}\left( \frac{ {\bf B}_{Y}\cdot
{\bf \nabla}\times {\bf B}_{Y} }
{ |{\bf B}_{Y}|^2  }\right) + O\left(\frac{\Gamma}{\Gamma_{np}}\right)~.
\label{nR}
\end{equation}
Below the critical temperature the hypermagnetic fields are
converted into ordinary Maxwell \mfs. Similarly to the usual EW
baryogenesis scenario, the fermion number asymmetry produced by
the Abelian anomaly may survive the sphaleron wash-out only if the
EWPT is strongly first order, which we know to be incompatible
with the standard model in the absence of primordial magnetic
fields. However, Giovannini and Shaposhinikov argued that this
argument might not apply in the presence of strong \mfs (we shall
discuss this issue in Chap.\ref{chap:stability}).  If this is the case a
baryon asymmetry compatible with the observations might have been
generated at the EW scale. Another prediction of this scenario is
the production of strong density fluctuations at the BBN time which
may affect the primordial synthesis of light elements
\cite{GioSha97}.

Primordial magnetic fields and the primordial magnetic helicity
may also have been produced by the interaction of the hypercharge
component of the electromagnetic field with a cosmic pseudoscalar
field condensate which provides the required macroscopic parity
violation. This idea was first sketched  by Turner and Widrow
\cite{Widrow} in the framework of an inflationary model of the
Universe which we shall discuss in more details in the next
section.  Turner and Widrow assumed the pseudoscalar field to be
the axion, a particle which existence is invoked for the solution
of the strong CP problem (for a review see Ref. \cite{axion}).
Although the axion is supposed to be electrically neutral it
couples to electromagnetic field by means of the anomaly. Indeed,
the effective Lagrangian for axion electrodynamics is
\begin{equation}\label{axionL}
  {\cal L} = - \frac 1 2 \partial_\mu \theta  \partial^\mu \theta
  - \frac 1 4 F_{\mu\nu}F^{\mu\nu}
  + g_a \theta F_{\mu\nu}{\tilde F}^{\mu\nu}~,
\end{equation}
where $g_a$ is a coupling constant of order $\alpha$, the vacuum
angle $\theta = \phi_a/f_a$, $\phi_a$ is the axion field,  $f_a$
is the Peccei-Quinn symmetry breaking scale (see
Ref.\cite{axion}), $F_{\mu\nu}$ is the electromagnetic field
strength and ${\tilde F}^{\mu\nu}$ is its dual. Since the axion
field, as any other scalar field, is not conformally invariant
(see the next section), it will be amplified during the
inflationary expansion of the Universe starting from quantum
fluctuations, giving rise to $\langle \theta^2 \rangle \sim
(H_0/f_a)^2$, which can act as a source term for the
electromagnetic field \footnote{The careful reader may wonder
what is the fate of axions in the presence of cosmic magnetic fields.
Interestingly, it was showed by Ahonen, Enqvist and Raffelt \cite{EnqRaf}
that although oscillating cosmic axions drives an oscillating electric
field, the ensuing dissipation of axions is found to be inversely proportional
to the plasma conductivity and is, therefore, negligible.}.

Carroll and Field \cite{CarFie91} reconsider in more details
the idea of Turner and Widrow and found that the evolution of a Fourier
mode of the magnetic field with wave number $k$ is governed by the
equation
\begin{equation}\label{Fmodes}
  \frac {d^2 F_\pm}{d\tau^2} + \left( k^2 \pm g_a k
  \frac{d\phi_a}{d\tau} \right) F_\pm = 0
\end{equation}
where $F_\pm = a^2(B_x \pm iB_y)$ are the Fourier modes
corresponding to different circular polarizations, and $\tau$ is
the conformal time. One or both polarization modes will be
unstable for $k < g_a \vert d\phi_a/d\tau\vert$, whereas both
polarization modes can becomes unstable to exponential growth if
$\phi$ is oscillating. In this case it seems as if a  quite strong \mf
could be produced during inflation. However, such a conclusion was
recently criticized by Giovannini \cite{Giovannini99} who noted
that above the EWPT temperature QCD sphalerons \cite{QCDsphal} are
in thermal equilibrium which can effectively damp axion
oscillations. In fact, because of the presence of QCD sphaleron
the axion equation of motion becomes
\begin{equation}\label{axionsphal}
  {\ddot \phi}_a + \left( 3H + \gamma \right) {\dot \phi}_a = 0
\end{equation}
where \cite{QCDsphal}
\begin{equation}\label{qcdspahl}
  \gamma = \frac{\Gamma_{sphal}}{f_a^2 T} \simeq
  \frac {\alpha_s^4 T^3}{f_a^2}
\end{equation}
(where $\alpha_s = g_s^2/4\pi$). Giovannini founds that sphaleron
induced damping dominates over damping produced by the expansion
of the Universe if $f_a > 10^9$ GeV. Since astrophysical and
cosmological bounds \cite{Raffeltbook} leave open a window
$10^{10}~{\rm GeV} < f_a < 10^{12}~{\rm GeV}$, it follows that no
magnetic fields amplification was possible until QCD sphaleron
went out of thermal equilibrium. A very tiny magnetic helicity
production from axion oscillations may occur at lower
temperatures. In fact, Giovannini \cite{Giovannini99} estimated
that in the temperature range $1~\GeV
> T > 10\MeV$ a magnetic helicity of the order of
\begin{equation}
  \frac {\langle {\bf B}\cdot{\bf \nabla}\times{\bf B} \rangle}
  {\sigma \rho_c} \sim 10^{-22}
\end{equation}
may be generated, which is probably too small to have any
phenomenological relevance.

Generation of \mfs from  coherent oscillations or roll-down of
pseudoscalar fields different from the axion has been also
considered in the literature. It is interesting that pseudoscalar
fields with an axion-like anomalous coupling to the
electromagnetic field appear in several possible extension of the
SM. Typically these fields have only perturbative derivative
interactions and therefore vanishing potential at high
temperatures, and acquire a potential at lower temperatures
through non-perturbative interactions. The generic form of this
potential is $V(\phi) = V_0^4~A(\phi/f)$, where $A$ is a bounded
function of the ''Peccei-Quinn" scale $f$ which in this case can
be as large as the Planck scale. The pseudoscalar mass $m \sim
V_0^2/f$ may range from a few eV to $10^{12}$ GeV.
 The amplification of magnetic fields proceeds in a way
quite similar to that discussed for axions. The evolution equation
of the electromagnetic Fourier modes were derived by Brustein and
Oaknin \cite{BruOak99} who added  to Eq.(\ref{Fmodes}) the effect
of a finite electric conductivity, finding
\begin{equation}\label{Fmodesbis}
  \frac {d^2 F_\pm}{d\tau^2} + \sigma \frac{\partial F_\pm}{d\tau}
  + \left( k^2 \pm g k \frac{d\phi}{d\tau} \right) F_\pm = 0~.
\end{equation}
If the pseudoscalar field is oscillating, the field velocity
$d\phi/d\tau$ changes sign periodically and both polarization
modes are amplified, each during a different semi-cycle. Each mode
is amplified during one part of the cycle and damped during the
other part of the cycle. Net amplification results when $g k
d\phi/d\tau > \sigma^2$ and the total amplification is exponential
in the number of cycles. Taking into account that $\sigma \sim 10
T$ before the EWPT \cite{conductivity}, Brustein and Oaknin found
that a huge amplification can be obtained from pseudoscalar field
oscillations at $T \sim 1$ TeV, for a scalar mass $m$ of few TeV
and $g f \sim 10$.  This is a particularly interesting range of
parameters as in this case the pseudoscalar field dynamics could
be associated with breaking of supersymmetry  at the TeV scale.
Since only a limited range of Fourier modes are amplified, with
$k$ not too different from $T$, and modes with $k/T < \tau_{EW}
\sigma \ll 1$ are rapidly dissipated, amplified fields may survive
until the EWPT only if amplification occurred just before the
transition. If this is a natural assumption may be a matter of
discussion. However, it was pointed-out by Brustein and Oaknin
\cite{BruOak99} that if, depending on the form of the pseudoscalar
field potential, this field rolls instead of oscillating, then the
hypermagnetic fields would survive until EWPT even if the
amplification takes place at higher temperatures before the
transition. In this case there could be interesting consequences
for the EW baryogenesis. This is so, because the pseudoscalar
field may carry a considerable amount of helicity. According to what
discussed in the above, this
number will be released in the form of fermions and baryons if the
EWPT is strongly first order. Brustein and Oaknin argued that this
mechanism could naturally generate the observed BAU.

Clearly, a more serious problem of this mechanism is the same of
other EW baryogenesis scenarios, namely to have a strongly first
order EWPT. An interesting possibility which was proposed by two
different groups \cite{GioSha,ElmEK} is that strong magnetic
fields may enhance the strength of the EWPT  (see Chap.\ref{chap:stability}).
Unfortunately, detailed lattice computations \cite{KajLPRS} showed that this is
not the case. Furthermore, in a recent work by Comelli et al.
\cite{labanda} it was shown that strong \mfs also increase the
rate of EW sphalerons so that the preservation of the baryon
asymmetry calls for a much stronger phase transition than required
in the absence of a magnetic field. The authors of Ref.\cite{labanda} showed
that this effect overwhelms the gain in the phase transition
strength (see Sec.\ref{sec:sphalerons}). Therefore, the only way for the kind
of EW baryogenesis mechanism discussed in the above to work is to invoke for
extensions of the standard model which allow for a strong first
order EWPT \cite{MSSM}.
 \vskip0.5cm\noindent
 {\bf Electroweak strings}, which we introduced in the
previous section, may also carry a net hypermagnetic helicity and
act as a source of the observed BAU. One of the interesting
features of this objects is that they should have been formed
during the EWPT even if this transition is second order or just a
cross over \cite{Chernodub}. It is known that electroweak strings
can carry magnetic helicity, hence a baryon number, which is
related to the twisting and linking  of the string gauge fields
\cite{AchVac,VacFie}. Several authors (see {\it e.g.}
Ref.\cite{BraDav93}) tried to construct a viable EW baryogenesis
scenario based on these embedded defects. Many of these models,
however, run into the same problems of more conventional EW
baryogenesis scenarios. An interesting attempt was done by
Barriola \cite{Barriola} who invoked the presence of a primordial
\mf at the EWPT time. Barriola observed that the production and
the following decay of electroweak strings give rise, in the
presence of an external magnetic field, to a variation in the
baryon number $\Delta B$  in the time interval $dt$, given by
\begin{equation}\label{barriola}
  \Delta B = \frac{N_f}{32 \pi^2} \int dt \int d^3x~ \alpha^2 \cos 2\theta_W
  {\vec E}_Z\cdot {\vec B}_Z + \frac{\alpha^2}{2} \sin^2 \theta_W
  \left( {\vec E}\cdot {\vec B}_Z + {\vec E}_Z\cdot {\vec B}
  \right)~,
\end{equation}
where ${\vec E}_Z$ and  ${\vec B}_Z$ are the Z-electric and
Z-magnetic fields of the strings whereas ${\vec E}$ and  ${\vec
B}$ are the corresponding standard Maxwell fields. The first term
in the r.h.s. of (\ref{barriola}) represents the change in the
helicity of the string network, and the second and third terms
come form the coupling of the string fields with the external
Maxwell fields. Whereas, the first term  averages to zero over a
large number of strings, the other terms may not. Clearly, some
bias is required to select a direction in the baryon number
violation. In the specific model considered by Barriola this is
obtained by a CP violation coming from the extension of the Higgs
sector of the SM, and from the dynamics of strings which are
supposed to collapse along their axis.  It was concluded by
Barriola that such a mechanism could account for the BAU.
Unfortunately, it is quite unclear if the invoked out-of-equilibrium
mechanism  based on the string collapse could indeed be effective
enough to avoid the sphaleron wash-out. We think, however, that
this possibility deserves further study.

\section{Magnetic fields from inflation}\label{sec:inflation}

As noted by Turner and Widrow \cite{Widrow} inflation (for a
comprehensive introduction to inflation see Ref.\cite{KolTur})
provides four important ingredients for the production of primeval
\mfs.
\begin{itemize}
\item Inflation naturally produces effects on very large scales,
larger than the Hubble horizon, starting from microphysical
processes operating on a causally connected volume. If
electromagnetic quantum fluctuations are amplified during
inflation they could appear today as large-scale static magnetic
fields (electric field should be screened by the high conductivity
plasma).

\item Inflation also provides the dynamical means to amplify these
long-wavelength waves. If the conformal invariance of the
electromagnetic field is broken in some way (see below) magnetic
fields could be excited during the de Sitter Universe expansion.
This phenomenon is analogous to particle production occurring  in
a rapidly changing space-time metric.

\item During inflation (and perhaps during most of reheating) the Universe
is not a good conductor so that magnetic flux is not conserved and
the ratio $r$ of the magnetic field with the radiation energy
densities can increase.

\item Classical fluctuations with wavelength $\lambda \simleq
H^{-1}$ of massless, minimally coupled fields can grow {\it
super-adiabatically}, i.e. their energy density decrease only as
$\sim a^{-2}$ rather than $a^{-4}$.
\end{itemize}

The main obstacle on the way of this nice scenario is given by the
fact that in a conformally flat metric, like the Robertson-Walker
usually considered, the background gravitational field does not
produce particles if the underlying theory is conformally
invariant \cite{Parker68}. This is the case for photons since the
classical electrodynamics is conformally invariant in the limit of
vanishing fermion masses. Several ways out this obstacle have been
proposed. Turner and Widrow \cite{Widrow} considered  three
possibilities. The first is to break explicitly conformal
invariance by introducing a gravitational coupling, like $R A_\mu
A^\mu$ or $R_{\mu\nu} A^\mu A^\mu$, where $R$ is the curvature
scalar, $R_{\mu\nu}$ is the Ricci tensor, and $A^\mu$ is the
electromagnetic field. These terms breaks gauge invariance and
give the photons an effective, time-dependent mass. In fact, one
of the most severe constraints to this scenario come from the
experimental upper limit to the photon mass, which today is
$m_\gamma < 2 \times 10^{-16}$ eV \cite{pdg}. Turner and Widrow
showed that for some suitable (though theoretically unmotivated)
choice of the parameters, such a mechanism may give rise to
galactic \mfs even without invoking the galactic dynamo.
We leave to the reader to judge if such a
booty deserve the abandonment of the theoretical prejudice in
favor of gauge invariance. A different model invoking a
spontaneous breaking of gauge symmetry of electromagnetism,
implying nonconservation of the electric charge, in the early
stage of the evolution of the Universe has been proposed by Dolgov
and Silk \cite{DolSil}.

The breaking of the conformal invariance may also be produced by
terms of the form $R_{\mu\nu\lambda\kappa}
F^{\mu\nu}F^{\lambda\kappa}/m^2$ or $R F^{\mu\nu}F_{\mu\nu}$,
where $m$ is some mass scale required by dimensional
considerations. Such terms arise due to one-loop vacuum
polarization effects in curved space-time, and they have the
virtue of being gauge invariant. Unfortunately, Turner and Widrow
showed that they may account only for a far too small contribution
to primordial magnetic fields.  The third way to break conformal invariance
discussed by Turner and Widrow invoke a coupling of the photon to
a charged field which is not conformally coupled or the anomalous
coupling to a pseudoscalar. This mechanism was already illustrated
in the previous section.

The anomaly can give rise to breaking of the conformal invariance
also in a different way. The kind of anomaly we are now discussing
about is the conformal anomaly, which is related to the triangle
diagram connecting two photons to a graviton. It is known (for a
review see Ref.\cite{BirDav}) that this kind of diagrams breaks
conformal invariance by producing a nonvanishing trace of the
energy-momentum tensor
\begin{equation}\label{traceT}
  T_\mu^\mu = \frac{\alpha \beta}{8\pi} F^a_{\mu\nu}F^{a~\mu\nu}~,
\end{equation}
where $\alpha$ is the fine-structure constant of the theory based
on the $SU(N)$ gauge-symmetry with $N_f$ fermion families, and
\begin{equation}\label{beta}
 \beta = \frac{11}{3} N - \frac{2}{3} N_f~.
\end{equation}
Dolgov \cite{Dolgov93} pointed-out that such an effect may lead to
strong electromagnetic fields amplification during inflation.
In fact, Maxwell equations are modified by the anomaly in the
following way
\begin{equation}\label{anomMax}
  \partial_\mu F_\nu^\mu + \beta \frac{\partial_\mu}{a} F_\nu^\mu
  = 0
\end{equation}
which, in the Fourier space, gives rise to the equation
\begin{equation}\label{Amodes}
  A''  + \beta \frac{a'}{a} A' + k^2 A = 0~,
\end{equation}
where $A$ is the amplitude of the vector potential, and a prime
stands for a derivation respect to the conformal time $\tau$. At
the inflationary stage, when $a'/a = - 1/\tau$ Dolgov found a
solution of (\ref{Amodes}) growing like $\displaystyle \left(
\frac H k \right)^{\beta/2}$. Since $k^{-1}$ grows well above the
Hubble radius during the de Sitter phase, a huge amplification can
be obtained if $\beta > 0$. Dolgov showed that for $\beta \sim 1$
the \mf generated during the inflationary stage can be large
enough to give rise to the observed fields in galaxies even
without a dynamo amplification. Unfortunately, such a large value
of $\beta$ seems to be unrealistic ($\beta \approx 0.06$ for SU(5)
with three charged fermions). The conclusion is that galactic
magnetic fields might be produced by this mechanism only invoking
a group larger than $SU(5)$ with a large number of fermion
families, and certainly no without the help of dynamo
amplification.

As we discussed in the above conformal invariance of the electromagnetic
field is generally
spoiled whenever the electromagnetic field is coupled to a scalar
field. Ratra \cite{Ratra} suggested that a coupling of the form
$e^{\kappa \phi} F^{\mu\nu}F_{\mu\nu}$, where $\kappa$ is a
arbitrary parameter, may lead to a huge amplification of
electromagnetic quantum fluctuations into large scale magnetic
fields during inflation. Such a coupling is produced in some
peculiar models of inflation with an exponential inflaton
potential \cite{RatPeb}. It should be noted by the reader that the
scalar field $\phi$ coincide here with the inflaton field. According to
Ratra, present time intergalactic magnetic fields as large as
$10^{-9}$ G may be produced by this mechanism which would not
require any dynamo amplification to account for the observed
galactic fields. Unfortunately, depending on the parameter of the
underling model, the predicted field could also be as low as
$10^{-65}$ G !

A slightly more predictive, and perhaps theoretically better
motivated, model has been proposed independently by Lemoine and
Lemoine \cite{Lemoinebrothers} , and Gasperini, Giovannini and
Veneziano \cite{Veneziano}, which is based on superstring cosmology
\cite{stringcosmo,stringcosmo2}.
This model is based on the consideration that
in string theory the electromagnetic fields is coupled not only to
the metric ($g_{\mu\nu}$), but also to the dilaton field $\phi$.
In the low energy limit of the theory, and after dimensional
reduction from 10 to 4 space-time dimensions, such a coupling
takes the form
\begin{equation}\label{stringcoup}
  \sqrt{- g} e^{-\phi} F^{\mu\nu}F_{\mu\nu}
\end{equation}
which breaks conformal invariance of the electromagnetic field
and coincides with the coupling
considered by Ratra \cite{Ratra} if $\kappa = -1$. Ratra, however,
assumed inflation to be driven by the scalar field potential,
which is not the case in string cosmology. In fact, typical
dilaton potentials are much too steep to produce the required
slow-roll of the inflaton (= dilaton) field. According to string
cosmologists, this problem can be solved by assuming inflation to
be driven by the kinetic part of the dilaton field, i.e. from
$\phi'$ \cite{stringcosmo}. In such a scenario the Universe
evolves from a flat, cold, and weakly coupled ($\phi = - \infty$)
initial unstable vacuum state toward a curved, dilaton-driven,
strong coupling regime. During this period, called {\it
pre-big-bang} phase, the scale factor and of the dilaton evolve as
\begin{equation}
  a(\tau) \sim (-\tau)^\beta, \quad \phi(\tau) \sim \kappa \ln a,
  \quad \tau < - \tau_1~,
\end{equation}
with $\beta > 1$ and $\kappa < 0$. At $\tau > - \tau_1$  it begins
the standard FRW phase with a radiation dominated Universe. In the
presence of the non-trivial dilaton background the modified
Maxwell equations takes the form \cite{Lemoinebrothers}
\begin{equation}\label{Maxwellstring}
  \nabla_\mu F^{\mu\nu} -  \nabla_\mu \phi~F^{\mu\nu} = 0~.
\end{equation}
 Electromagnetic field amplification from quantum fluctuations
take place during the {\it pre-big-bang} phase when $\phi' =
\delta/ \tau$, where $\delta = \beta \kappa$. By following the
evolution of the electromagnetic field modes from $t = - \infty$
to now, Lemoine and Lemoine estimated that, in the most simple
model of dilaton-driven inflation (with $V(\phi) = \rho = p = 0$)
a very tiny magnetic field is predicted today
\begin{equation}\label{lemoineB}
 \langle B^2 \rangle^{1/2} \sim 10^{-62}~
 \left(\frac {L}{1 {\rm Mpc}}\right)^{-2.1}
 \left(\frac {H_1} {M_{Pl}} \right)^{0.07}~{\rm G}~,
\end{equation}
where $H_1 = 1/\tau_1$, which is far too small to account for
galactic fields.

Gasperini et al. \cite{Veneziano} reached a different conclusion,
claiming that magnetic fields as large those required to explain
galactic fields without dynamo amplification may be produced on
the protogalactic scale. The reason for such a different result is
that they assumed a new phase to exist between the
dilaton-dominated phase and the FRW phase during which dilaton
potential is nonvanishing. The new phase, called the {\it string
phase}, should start when the string length scale $\lambda_S$
becomes comparable to the horizon size at the conformal time
$\tau_S$ \cite{BruVen}. Unfortunately, the duration of such a
phase is quite unknown, which makes the model not very predictive.

Recently, several papers have been published (see e.g.
Refs.\cite{Calzetta, GioSha00, Davis00}) which proposed the
generation of \mfs by fluctuations of scalar (or pseudo-scalar)
fields which were amplified during, or at the end, of inflation.
In some of those papers \cite{Calzetta, Davis00} the authors claim
that \mfs as strong as those required to initiate a successful
galactic dynamo may be produced. This conclusion, however, is
based on incorrect assumptions. According to Giovannini and
Shaposhnikov \cite{GioSha00,GioShacom}, the main problem resides
in the approximate treatment of dissipative effects adopted in
Refs.\cite{Calzetta,Davis00}. Whereas in Ref.\cite{Davis00}
dissipation of electric fields was totally neglected, in
Ref.\cite{Calzetta} an incorrect dependence of the electric
conductivity on the temperature was used. Adopting the correct
expression for the conductivity (see e.g.
Refs.\cite{conductivity}) the authours of Ref.\cite{GioSha00}
found that inflation produces fields which are too small to seed
galactic dynamo.

\section{Magnetic fields from cosmic strings}\label{sec:defects}

At the beginning of this chapter we shortly discussed the
Harrison-Rees mechanism for vortical production of \mfs in the
primordial plasma. In this section we will shortly review as
cosmic string may implement this mechanism by providing a
vorticity source on  scales comparable to galactic sizes. Cosmic
strings are one-dimensional topological defects which are supposed
to have been formed during some primordial phase transition
through the Kibble mechanism \cite{Kibble}, which we already
discussed in Sec.\ref{sec:EWPT} (for a review on cosmic strings
see Ref. \cite{Vilenkin}). The idea that cosmic string may produce
plasma vorticity and \mfs was first proposed by Vachaspati and
Vilenkin \cite{VacVil}. In this scenario vorticity is generated in
the wakes of fast moving cosmic strings after structure formation
begins. Differently from Harrison's \cite{Harrison70} scenario
(see Sec.\ref{sec:vorticity}), in this case vorticity does not
decay with Universe expansion since the vortical eddies are
gravitationally bounded to the string. Even if the mechanism we
are considering is supposed to take place after recombination, a
sufficient amount of ionization should be produced by the violent
turbulent motion so that the Harrison-Rees \cite{Rees} mechanism
can still operate. The scale of coherence of the generated \mfs is
set by the scale of wiggles of the string and, for wakes created
at recombination time it can be up to 100 kpc. The predicted field
strength is of the order of $\sim 10^{-18}$ G, that could enough
to seed the galactic dynamo. The main problem with this scenario
is that is not clear whether stable vortical motion can be really
generated by the chaotic motion of the string wiggles.

An alternative mechanism has been proposed by Avelino and Shellard
\cite{AveShe}. In their model, vorticity is generated not by the
wiggles but by the strings themselves which, because of the finite
dynamical friction, drag matter behind them inducing circular
motions over inter-string scales. Unfortunately, the magnetic field
strength predicted by this model at the present time is very weak
$\sim 10^{-23}$ G, which can only marginally seed the galactic
dynamo.

Larger fields may be produced if cosmic strings are
superconducting. String superconductivity was first conceived by
Witten \cite{Witten85}. The charge carriers on these strings can
be either fermions or bosons, and the critical currents can be as
large as $10^{20}$ Ampere. If primordial magnetic fields
pre-exist, or formed together with, cosmic string they may play a
role charging up superconducting string loops and delaying their
collapse \cite{BrandDMT92}. Otherwise, superconducting cosmic
strings can themselves give rise to \mfs in a way similar to that
proposed by Avelino and Shellard. An important difference,
however, arises with respect to the non-superconducting case
discussed in Ref.\cite{AveShe}. As shown by Dimopoulos and Davis
\cite{DimDav98}, superconducting strings networks may be more
tangled and slower than conventional cosmic strings because of the
strong current which increases dynamical friction. It was shown by
Dimopoulos \cite{Dimopoulos98} that if the string velocity is
small enough the gravitational influence of the string on the
surrounding plasma becomes a relevant effect. As a consequence,
plasma is dragged by the string acquiring substantial momentum.
Such a momentum may induce turbulence which could generate
magnetic fields on scale of the order of the inter-string
distance. Such a distance is smaller than the conventional cosmic
string distance. Quite strong \mfs may be produced by this
mechanism. The only known constraint come from the requirement
that the string network does not produce too large temperature
anisotropies in the CMBR. By imposing this constraint, Dimopoulos
estimated that present time magnetic fields as large as $10^{-19}$
G with a coherence scale of $\sim 1$ Mpc may be produced.

In a recent paper by Voloshin it was claimed
that the generation of large scale magnetic fields by domain walls is not
possible \cite{Voloshin}.



\chapter{Particles and their couplings in the presence of strong magnetic
fields}\label{chap:stability}

In the previous chapter we have seen that, very strong \mfs could
have been produced in the early Universe.
Here we investigate the effects of such strong fields
on bound states of quarks and on condensates created by spontaneous
symmetry breaking. For example, we have already seen in
Chap.\ref{chap:bbn} that strong \mfs can affect masses and decay
rates of charged particles and modify the rate of weak processes.
As we already discussed in Sec.\ref{sec:evolution} another
crucial issue concerns the stability of strong magnetic fields.
QED allows the existence of arbitrary large magnetic fields
provided matter constituents have spin less than $\frac{1}{2}$
\cite{ItzZub}.

This is so because the Lorentz force cannot perform any work on
charged particles so that real particle-antiparticles free pairs
cannot be produced. In Sec.\ref{sec:weakrates} we have already seen
that quantum corrections do not spoil this classical argument. We
have also seen that although pair production can be catalyzed by
strong \mfs at finite temperature and density, in this case it is
the heat-bath that pays for the energy cost of the effect.
This situation changes when, in the presence of
very strong fields, QCD and electroweak corrections cannot be
disregarded. We shall see that the QCD and electroweak field
allow for the formation of condensates of charged particles with no
energy cost. This may lead to screening of magnetic, or hypermagnetic,
fields resembling the Meissner effect in superconductors.

\section{Low Lying States for Particles in Uniform Magnetic Fields}

Following Refs.\cite{Bander1,Bander2} in this section we will
consider questions related to the mass shifts and decay of
bound states of quarks. Mass shifts occur both due to the effect
that magnetic fields have on the strong binding forces, and due to
the direct interactions of charged spinning particles with
external fields.  The modifications of the strong forces are such
as to close the gap between the proton and neutron masses and
ultimately make the proton heavier. A delicate interplay between
the anomalous magnetic moments of the proton and neutron drives
the mass shifts due to the direct interactions in the same
direction. For $B>1.5\times 10^{18}$~G the neutron becomes stable
and as the field is increased past $2.7\times 10^{18}$~G the
proton becomes unstable to decay into a neutron, positron and
neutrino.
 \vskip0.5cm
 The quantum mechanics of a Dirac particle with no
anomalous magnetic moment in a uniform external magnetic field is
straightforward. We shall present the results for the case where
particles do have such anomalous moments. In reality, in fields so
strong that the mass shifts induced by such fields are of the
order of the mass itself one cannot define a magnetic moment as
the energies are no longer linear in the external field. Schwinger
\cite{Schwinger} calculated the self energy of an electron in an
external field and we shall use here his results. We
cannot follow this procedure for the proton or  neutron as we do
not have a good field theory calculation of the magnetic moments
of these particles, even for small magnetic fields; all we have at
hand is a phenomenological anomalous magnetic moment. However, for
fields that change the energies of these particles by only a few
percent, we will consider these as point particles with the given
anomalous moments. In the Sec.\ref{sec:limitations} we will discuss
possible limitations of this approach.

\subsection{Protons in an External Field}

The Dirac Hamiltonian for a
proton with a uniform external magnetic field  ${\bf B} $ is
\begin{equation}
H=\mbox{\boldmath $\alpha$}\cdot\left ({\bf p}-e{\bf
A(r)}\right )+\beta M_p -
 {e\over {2M_p}}
\left ({g_p\over 2}-1\right )\beta{\bf\Sigma\cdot B}\, .\label{protonham}
\end{equation}
The vector potential ${\bf A(r)}$ is related to the \mf by ${\bf
A(r)}={1\over 2}{\bf r\times B}$ and $g_p=5.58$ is the proton's
Land\'{e} g factor. We first solve this equation for the case
where the momentum along the magnetic field direction is zero and
then boost along that direction till we obtain the desired
momentum. For ${\bf B}$ along the ${\bf {z}}$ direction and
$p_z=0$ the energy levels are \cite{ItzZub}
\begin{equation}
E_{n,m,s}=\left [2eB(n+{1\over 2})-eBs+{M_p}^2 \right ]^{1\over 2}-
{e\over {2M_p}}\left ({g_p\over 2}-1\right )Bs\, .\label{Landauener}
\end{equation}
In the above, $n$ denotes the Landau level, $m$ the orbital angular
momentum about the magnetic field direction and $s=\pm 1$ indicates
whether the spin is along or opposed to that direction; the levels are
degenerate in $m$. $n=0$ and $s=+1$ yield the lowest energy
\begin{equation}
E={\tilde M}_p=M_p-{e\over {2M_p}}\left ({g_p\over 2}-1\right )B\, .
\label{protmass}
\end{equation}
As we shall be interested in these states only we will drop the $n$
and $s$ quantum numbers. The Dirac wave function for this
state is
\begin{equation}
\psi_{m,p_z=0}({\bf r})=\left ( \begin{array}{c} 1 \\0 \\0 \\0
\end{array}\right ) \phi_m({x,y})\, ,
\end{equation}
$\phi_m$'s are the standard wave functions of the lowest Landau
level;
\begin{equation}
\phi_m(x,y)={{\left [{1\over 2}|eB|\right ]^{{m+1}\over 2}}\over
{\sqrt{\pi m!}}}[x+iy]^m\exp\left [-{1\over 4}|eB|(x^2+y^2)\right ]\, .
\label{Landauwavefunction}
\end{equation}
Boosting to a finite value of $p_z$ is straightforward; we obtain
\begin{equation}
E_m(p_z)=\sqrt{{p_z}^2+{\tilde M}^2}\, ,
\end{equation}
with a wave function
\begin{equation}
\psi_{m,p_z}({\bf r})=\left ( \begin{array}{c} \cosh \theta \\0
\\ \sinh \theta \\0  \end{array}\right ) {{e^{ip_zz}}\over\sqrt{2\pi}}
\phi_m({x,y})\, ,\label{protonwavefunction}
\end{equation}
where $2\theta$, the rapidity, is obtained from $\tanh
2\theta=p_z/E_m(p_z)$.

In the non-relativistic limit the energy becomes
\begin{equation}
E_m(p_z)={\tilde M}+{{{p_z}^2}\over {2{\tilde M}}} \label{nrpe}
\end{equation}
and the wave function reduces to
\begin{equation}
\psi_{m,p_z}({\bf r})=\left ( \begin{array}{c} 1 \\0
\\ 0 \\0  \end{array}\right ) {{e^{ip_zz}}\over\sqrt{2\pi}}
\phi_m({x,y})\, . \label{nrpwf}
\end{equation}

\subsection{Neutrons in an External Field}

For a neutron the Dirac Hamiltonian is somewhat simpler
\begin{equation}
H=\mbox{\boldmath $\alpha$}\cdot{\bf p}+\beta M_n - {e\over {2M_n}}
\left ({g_n\over 2}\right )\beta{\bf\Sigma\cdot B}\, .\label{neutronham}
\end{equation}
with $g_n=-3.82$. Again for $p_z=0$ the states of lowest energy, the
ones we shall be interested in, have energies
\begin{equation}
E({\bf p_{\perp}}, p_z=0)={e\over {2M_n}}\left ({g_n\over 2}\right
)B+\sqrt{{\bf p_{\perp}}^2+{M_n}^2}
\end{equation}
Boosting to a finite $p_z$ we obtain
\begin{equation}
E({\bf p})=\sqrt{{E({\bf p_{\perp}}, p_z=0)}^2+{p_z}^2}\, .
\label{neutronener}
\end{equation}
The wave functions corresponding to this energy are
\begin{equation}
\psi_{\bf p}({\bf r})= {e^{i\mbox{\boldmath $p$}\cdot\mbox{\boldmath $r$}}
\over {(2\pi)^{3\over 2}}}u({\bf p},s=-1)\, ,
\end{equation}
where $u({\bf p},s=-1)$ is the standard spinor for a particle with
momentum ${\bf p}$, energy $\sqrt{{\bf p}^2+{M_n}^2}$ (not the energy
of Eq.~\ (\ref{neutronener})) and spin down.

In the non-relativistic limit
\begin{equation}
E({\bf p})=M_n+{e\over {2M_n}}\left ({g_n\over 2}\right )B+
{{{\bf p}^2}\over {2M_n}} \label{nrne}
\end{equation}
and the wave functions are
\begin{equation}
\psi_{\bf p}({\bf r})= \left ( \begin{array}{c} 0 \\1
\\ 0 \\0  \end{array}\right )
{e^{i\mbox{\boldmath $p$}\cdot\mbox{\boldmath $r$}}
\over {(2\pi)^{3\over 2}}}\, .\label{nrnwf}
\end{equation}

\subsection{Electrons in an External Field}

We might be tempted to use, for the electron, the formalism
used for the proton with the Land\'{e} factor replaced by $g_e=2+
\alpha/\pi$. However as we shall see for magnetic fields sufficiently
strong as to make the proton heavier than the neutron the change in
energy of the electron would appear to be larger than the mass of the
electron itself.
The point particle formalism breaks down and we have solve QED, to one
loop, in a strong magnetic field; fortunately this problem was treated
by Schwinger{\cite{Schwinger}}. The energy of an electron with $p_z=0$,
spin up and in the lowest Landau level is
\begin{equation}
E_{m,p_z=0}=M_e\left [1+{\alpha\over {2\pi}} \ln \left ( {{2eB}\over
{{M_e}^2}} \right )\right ]\, .\label{electronener}
\end{equation}
For field strengths of subsequent interest this correction is negligible;
the energy of an electron in the lowest Landau level, with spin
down and a momentum of $p_z$ is
\begin{equation}
E_{m,p_z}=\sqrt{{p_z}^2+{M_e}^2}\, .
\end{equation}
and with wave function similar to those of the proton
\begin{equation}
\psi_{m,p_z}({\bf r})=\left ( \begin{array}{c} 0\\ \cosh \theta \\0
\\ \sinh \theta  \end{array}\right ) {{e^{ip_zz}}\over\sqrt{2\pi}}
{\phi_m^*}({x,y})\, ,
\end{equation}
where the boost rapidity, $2\theta$, is defined below
Eq. (\ref{protonwavefunction}) while the Landau level wave function is
defined in Eq. (\ref{Landauwavefunction}). The reason the complex
conjugate wave function appears is that the electron charge is
opposite to that of the proton.

\subsection{Effects of Magnetic Fields on Strong Forces}

We must be sure that shifts due to changes of color strong forces will not
shift states in the opposite direction. The best method to study masses of
QCD bound states is the use of sum rules \cite{Reinders}. This method uses
the SVZ \cite{SVZ} generalized short distance expansion that includes not
only perturbative pieces, but also higher dimensional operators like the
chiral and gluon condensates reflecting the non Abelian nature of the
vacuum. Fortunately the proton has a simple structure \cite{Reinders} which
reflects the fact that if chiral symmetry is restored the proton and
neutron masses vanish.
\begin{equation}
M_{p,n}=3a\, {\langle q\bar{q}\rangle}^{\frac{1}{3}}+
\mbox{\rm small\ corrections}\, \label{nucleoncond}
\end{equation}
where $a$ is a constant. Meson mass terms are more involved; for example
the $\rho$ mass is
\begin{equation}
M_{\rho}=b\, (\mbox{\rm perturbative\ terms})+
c\,\langle G_{\mu\nu}G^{\mu\nu}\rangle +
d\, {\langle q\bar{q}\rangle}\, .
\end{equation}
$b$, $c$ and $d$ are constants of comparable magnitude \cite{Reinders}.
As we shall show it is only the change of ${\langle q\bar{q}\rangle}$ due
to external magnetic fields that may be obtained in a reliable manner.

In the presence of external fields we expect the chiral condensates for
quarks of different charges to vary and Eq.~(\ref{nucleoncond}) becomes
\begin{eqnarray}
M_p^3 &=&a\, \left(2 \langle u\bar{u}\rangle +\langle
d\bar{d}\rangle\right)\, ,\nonumber\\
M_n^3 &=&a\, \left(2 \langle d\bar{d}\rangle+\langle u\bar{u}\rangle\right)\, .
\label{nucleoncond2}
\end{eqnarray}
To first order in condensate changes we find
\begin{eqnarray}
\delta M_p&=&\frac{M_p}{9} \left( 2\frac{\delta \langle u\bar{u}\rangle}
{\langle u\bar{u}\rangle}+\frac{\delta \langle
d\bar{d}\rangle}{\langle d\bar{d}\rangle}
\right)\, , \nonumber\\
\delta M_n&=&\frac{M_n}{9}  \left( 2\frac{\delta \langle d\bar{d}\rangle}
{\langle d\bar{d}\rangle}+\frac{\delta \langle u\bar{u}\rangle}{\langle u\bar{u}
   \rangle}
\right )\, .
\end{eqnarray}
Combining
\begin{equation}
\delta M_p -\delta M_n=\frac{M}{9} \left (\frac{\delta \langle u\bar{u}\rangle}
{\langle u\bar{u}\rangle}-\frac{\delta \langle d\bar{d}\rangle}
{\langle d\bar{d}\rangle}\right )\, .
\end{equation}
A simple method for studying the behavior of the chiral condensates in
the presence of external constant fields is through the use of
the Nambu-Jona-Lasinio model \cite{NJL}. This has been done by Klevansky
and Lemmer \cite{Klevansky} and a fit to their results is
\begin{equation}
\langle q\bar{q}\rangle{(B)}=\langle q\bar{q}\rangle{(0)}\left[
1+\left(\frac{e_qB}{\Lambda^2}\right)^2\right]^{\frac{1}{2}}\, ,
\label{KL1}
\end{equation}
with $\Lambda= 270$ MeV and $e_q$ the charge on the quark. To lowest order
we find
\begin{equation}
\frac{\delta \langle q\bar{q}\rangle}{\langle q\bar{q}\rangle}=
\frac{1}{2}\left(\frac{e_qB}{\Lambda^2}\right)^2\, .\label{KL2}
\end{equation}
and
\begin{equation}
\delta M_p -\delta M_n = \frac{M}{54}\left(\frac{eB}{\Lambda^2}\right
)^2\, ;\label{qcddiff}
\end{equation}
As in the previous section, these corrections are such as drive the proton
energy up faster than that of the neutron. One can understand the sign of
this effect; the radius of a quark-antiquark pair will decrease with
increasing magnetic field thus making the condensate larger. As the $u$
quark has twice the charge of the $d$ quark, its condensate will grow
faster and as there are more $u$ quarks in the proton than in the neutron
its mass will increase faster.

The fact that our estimate of the sign of the neutron-proton mass
difference is the same as that due to electromagnetic effects is
crucial. QCD sum rules and our method of evaluating the chiral
condensates are both crude and the magnitude of the mass difference
is uncertain. Had the sign of the hadronic correction been opposite,
cancellations could have occurred and the argument for a narrowing of
the mass difference and ultimate reversal could not have been made.
We are quite sure of the sign of the chiral change. The magnetic field
acts in the naive way in the spin of the scalar bound state in the condensate
so it is sure that goes as calculated \cite{Nambu}.

\subsection{Proton Life Time}

First we study the decay kinematics.
\vskip .1cm
Combining Eq.~(\ref{protmass}), Eq.~(\ref{nrne}) and Eq.~(\ref{qcddiff})
we find the proton-neutron energy difference as a function of the
applied magnetic field,
\begin{equation}
\Delta (B)=\left(-1.3+0.38B_{14}+0.11B_{14}^2\right)\ \mbox{\rm MeV}\, .
\label{Delta}
\end{equation}
$B_{14}$ is the strength of the magnetic field in units of $10^{14}$ T
($1~{\rm T} = 10^4\ $G).
The neutron becomes stable for $B>1.5\times 10^{14}$ T and the proton
becomes unstable to $\beta$ decay for $B>2.7\times 10^{14}$~T.
We shall now turn to a calculation of
the life time of the proton for fields satisfying the last inequality.

With the wave functions of the various particles in the
magnetic fields we may define field operators for these particles. For the
proton and electron we shall restrict the summation over states to the
lowest Landau levels with spin up, down respectively; for
magnetic fields of interest the other states will not contribute to the
calculation of decay properties. For the same reason, the neutron
field will be restricted to spin down only. The proton and neutron
kinematics will be taken as non-relativistic.
\begin{eqnarray}
\Psi_p({\bf r})=\sum_m\int dp_z\left [a_m(p_z)\left ( \begin{array}{c} 1 \\0
\\ 0 \\0  \end{array}\right ) {{e^{ip_zz}}\over\sqrt{2\pi}}
\phi_m({x,y})\right. \nonumber\\
 \left.
+{b^{\dag}}_m(p_z)\left ( \begin{array}{c} 0 \\0
\\ 1 \\0  \end{array}\right ){{e^{-ip_zz}}\over\sqrt{2\pi}}
\phi_m({x,y})\right ]\, ,
\label{protonfield}
\end{eqnarray}
with $\phi_m({x,y})$ defined in Eq.~\ (\ref{Landauwavefunction})
and the energy,
$E_m(p_z)$ in Eq.~\ (\ref{nrpe}). $a_m(p_z)$ is the annihilation operator
for a proton with momentum $p_z{\bf z}$ and angular momentum $m$;
$b_m(p_z)$ is the same for the negative energy states. For the
neutron the field is
\begin{equation} \Psi_n({\bf r})=\int d^3p \left
[a({\bf p})\left ( \begin{array}{c} 0 \\1  \\ 0 \\0  \end{array}\right )
{e^{i\mbox{\boldmath $p$}\cdot\mbox{\boldmath $r$}} \over {(2\pi)^{3\over
2}}} +b^{\dag}({\bf p})\left ( \begin{array}{c} 0 \\0  \\ 0 \\1
\end{array}\right )  {e^{-i\mbox{\boldmath $p$}\cdot\mbox{\boldmath $r$}}
\over {(2\pi)^{3\over 2}}}\right ]\, ,\label{neutron field}
\end{equation}
with an obvious definition of the annihilation operators.
For the electron we use fully relativistic kinematics and the field is
\begin{eqnarray}
\Psi_e({\bf r})=\sum_m\int dp_z \sqrt{{M_e}\over E}\left [a_m(p_z)
\left ( \begin{array}{c} 0\\ \cosh \theta \\0
\\ \sinh \theta  \end{array}\right ) {{e^{ip_zz}}\over\sqrt{2\pi}}
{\phi_m^*}({x,y})\right. \nonumber\\ \left.
+b^{\dag}_m(p_z)
\left ( \begin{array}{c} 0\\ \cosh \theta \\0
\\ \sinh \theta  \end{array}\right ) {{e^{-ip_zz}}\over\sqrt{2\pi}}
{\phi_m^*}({x,y})\right ]\, .\label{electron field}
\end{eqnarray}
\subsection{Decay Rates and Spectrum}
The part of the weak Hamiltonian responsible for the decay
$p\rightarrow n+e^{+}+\nu_e$ is
\begin{equation}
H={{G_F}\over {\sqrt{2}}}\int d^3x \bar{\Psi}_n\gamma_{\mu}(1+\gamma_5)\Psi_p
\bar{\Psi}_{\nu}\gamma^{\mu}(1+\gamma_5)\Psi_e\, .
\end{equation}
For non-relativistic heavy particles the matrix element of this Hamiltonian
between a proton with quantum numbers $p_z=0$ , $m=m_i$, a neutron
with momentum
${\bf p}_n$, a neutrino with momentum  ${\bf p}_{\nu}$ and an electron in state
$m=m_f$ and with ${p_{z,e}}$ is
\begin{eqnarray}
\langle H\rangle ={{2G_F}\over {(2\pi)^3}}\left( {{E_e+{p_{z,e}}}\over
{E_e-{p_{z,e}}}}\right )^{1\over 4}\sin (\theta_{\nu}/2)
\sqrt{M_e\over {E_e}}\delta (p_{z,e}+p_{z,\nu}+p_{z,n})\nonumber\\
\int dx\, dy\, {\phi_{m_f}^*}({x,y}){\phi_{m_i}}({x,y})\exp
[-i(\mbox{\boldmath  $p_{\perp,n}+p_{\perp,\nu}$})\cdot\mbox{\boldmath
$r_{\perp}$}]\, ;\label{matelem}
\end{eqnarray}
$\theta_{\nu}$ is the azimuthal angle of the neutrino. The integral in
the above
expression can be evaluated in a multipole expansion. Note that the natural
extent of the integral in the transverse direction is $1/\sqrt{eB}$ whereas the
neutron momenta are, from Eq.~(\ref{Delta}), of the order of
$\sqrt{eB(0.12+0.04B_{14})}$; thus setting the exponential term in
this
integral equal to one
will yield a good estimate for the rate and spectrum of this decay.
The positron spectrum is given by
\begin{equation}
{{d\Gamma}\over {dp_{z,e}}}={4\over 3}{{G_F^2 M_p}\over {(2\pi)^6}}
{{E_e+p_{z,e}}\over {E_e}}(\Delta-E_e)^3\, ;
\end{equation}
where $\Delta$ is defined in Eq.~(\ref{Delta}). For $\Delta\gg M_e$
the total rate is easily obtained
\begin{equation}
\Gamma={2\over 3}{{G_F^2 M_p}\over {(2\pi)^6}}\Delta^4\, .
\end{equation}
For $B=5\times 10^{14}$ T, the lifetime is $\tau=6$ s.

\subsection{Caveats and Limitations}\label{sec:limitations}

For general magnetic fields we expect the masses of particles to be
nonlinear functions of these fields. Such an expression has been
obtained, to order $\alpha$, for the electron\cite{Schwinger}. For
small fields this reduces to a power series, up to logarithmic terms,
in $B/B_c$, where $B_c$ is some scale. For the electron $B_c=m_e^2/e$.
For the hadronic case the value of $B_0$ is uncertain.
$B_c=M_p^2/e=1.7\times 10^{16}$~T is probably too large and $M_p$
should be replaced by a quark constituent mass and $e$ be $e_q$; in
that case $B_c=(2-4)\times 10^{15}$~T, depending on the quark type.
This is also the range of values of $\Lambda^2/e_q$ in
Eq.~(\ref{KL1}).  The effects we have studied need fields around a few
$\times 10^{14}$~T or an order of magnitude smaller than the lowest
candidate for $B_c$. Eq.~(\ref{Delta}) bay be viewed as a power series
expansion up to terms of order $(B/B_c)^2$; as the coefficient of the
quadratic term was obtained from a fit to a numerical solution,
logarithms of $B/B_c$ may be hidden in the coefficient.  As in
Ref.~\cite{Schwinger}, even powers will be spin independent and the
odd ones will be linear in the spin direction and may be viewed as
field dependent corrections to the magnetic moment. We cannot prove,
but only hope, that the coefficient of the $(B/B_c)^3$ term, the first
correction to the magnetic moment is not unusually large; should it
turn out to be big and of opposite sign to the linear and quadratic
terms, the conclusions of this analysis would be invalidated. These
arguments, probably, apply best to the field dependence of the
magnetic moments of the quarks rather than the total moment of the
baryons. We may ask what is the effect on these magnetic moments due
to changes in the ``orbital'' part of the quark wave functions. To
first order we expect no effect as all the quarks are in S states and
there is no orbital contribution to the total moment. The next order
perturbation correction will be {\it down} by $(r_b/r_c)^4$ compared to
the leading effect; $r_b$ is a hadronic radius and $r_c$ is the quarks
cyclotron radius. This again contributes to the $(B/B_c)^3$ term in
the expansion for the energy of a baryon.

Another limitation is due to the results of Ref.~\cite{Bander2}
where it is shown that fields of the order of a few $\times
10^{14}$~T are screened by a changes in chiral condensates. In
fact, as the chiral condensate will, in large fields, point in the
charged $\pi$ direction, the baryonic states will not have a
definite charge.  Whether the proton-neutron reversal takes place
for fields below those that are screened by chiral condensates or
vice versa is a subtle question; the approximations used in this
paper and in  Ref.~\cite{Bander2} are not reliable to give an
unambiguous answer. The treatment of the effects of magnetic
fields on the strong force contributions to the baryon masses
relies on the Nambu-Jona-Lasinio model and in Ref.\cite{Bander2}
the variation of $f_{\pi}$ with magnetic field was not taken into
account. It is clear from this discussion and the one from the
previous paragraph that we cannot push the results of this
calculation past few $\times 10^{14}$~T.
 \vskip0.5cm \noindent
Experimental Consequences
 \vskip0.5cm \noindent
The mass evolution of protons, neutrons and electrons
in magnetic fields and due to electromagnetism alone, will force a
proton to decay in a very intense field. Including the effects of chiral
condensates diminishes the field even further. Qualitatively, it is clear
that the effect enhances the electromagnetic contribution but its exact
value depends on the model. This points to a novel astrophysical mechanism
for creation of extra galactic positrons.

\subsection{Proton neutron mass difference by a lattice calculation}

One can take another approach, in principle exact, to calculate the mass
difference using lattice gauge theory.
We introduce the \mf in the lattice by multiplying a link $U_\mu(x)$ with a
phase $U_\mu^B(x)$.
Fixing the field,for convenience, in the direction z we then set:
\begin{equation}
U_x^B(x) = exp(-ieBa^2yL_x)~~{\rm for}~~ x= L_x-1;=~~ {\rm otherwise}
\end{equation}
\begin{equation}
U_y^B(x) =\exp(ieBa^2(x-x_0)
\end{equation}
where $x_0$ is an offset for the magnetic field.

Consequently the plaquette in the x-y plane is
\begin{equation}
P^B(x) =\exp(ieBa^2(-L_xL_y+1))~~{\rm for}~~ x=L_x-1, y=L_y-1
\end{equation}
\begin{equation}
P^B(x) = exp(ieBa^2)~~{\rm otherwise}
\end{equation}
The magnetic field is homogeneous only if $B$ is is quantized as
$a^2eB=2\pi n(L_xL_y)^{-1}$
This is a troublesome condition since the field is very large for
reasonable lattice size.
In the simulations we ignored this condition inducing some inhomogeneity.
The results are very preliminary: at this stage it is difficult to find
these effects with present lattice technology, but in principle it is possible.
For details of the calculations see \cite{Solomon}.
This method is still not efficient given the present state of the art in
simulations.

\section{Screening of Very Intense Magnetic Fields by chiral
symmetry breaking}\label{sec:screening}

Now we discuss another interesting phenomenon if very strong fields
could be created.

In very intense magnetic fields, $B > 1.5\times 10^{18}$ G, the
breaking of the strong interaction $SU(2)\times SU(2)$ symmetry
arranges itself so that instead of the neutral $\sigma$ field acquiring
a vacuum expectation value it is the charged $\pi$ field that does and
the magnetic field is screened.

In the previous section we discussed that fields with complicated
interactions of non-electromagnetic origin can induce various
instabilities in the presence of very intense magnetic fields. By very
intense we mean $10^{18}$ G to $10^{24}$ G. Fields with anomalous
magnetic moments \cite{AmbOle} or fields coupled by transition
moments \cite{Bander1} may induce vacuum instabilities.
The usual breaking of the strong interactions,
chiral symmetry $(\chi SM)$, is incompatible with very intense magnetic fields.
Using the standard $SU(2)\times SU(2)$ chiral $\sigma$ model we show
that magnetic fields $B\ge B_c$ with $B_c={\sqrt 2}m_{\pi}f_{\pi}$ are
screened; $f_{\pi}=132$ MeV is the pion decay constant and $m_{\pi}$ is
the mass of the charged pions. This result is opposite to what
occurs in a superconductor; in that case it is weak fields that are
screened and large ones penetrate and destroy the superconducting state.

As the magnetic fields are going to be screened we must be very careful in
how we specify an external field. One way would be to give $f_{\pi}$ a
spatial dependence and take it to vanish outside some large region of
space. In the region that $f_\pi$ vanishes we could specify the external
field and see how it behaves in that part of space where chiral symmetry
is broken. This is the procedure used in studying the behavior of fields
inside superconductors. In the present situation we find this division
artificial and, instead of specifying the magnetic fields, we shall
specify the external currents. Specifically we will look, at first, at the
electromagnetic field coupled to the charged part of the $\sigma$ model
and to the current $I$ in a long straight wire. From this result it will
be easy to deduce the behavior in other current configurations. We will
discuss a solenoidal current configuration towards the end of this work.

The Hamiltonian density for this problem is
\begin{eqnarray}
H=&&{1\over 2}\mbox{\boldmath$\nabla$}\sigma\cdot \mbox{\boldmath
$\nabla$}\sigma + {1\over 2}\mbox{\boldmath $\nabla$}\pi_0\cdot
\mbox{\boldmath $\nabla$}\pi_0+ (\mbox{\boldmath $\nabla$}+e{\bf
A}){\pi}^{\dag}\cdot(\mbox{\boldmath $\nabla$}-e{\bf A}){\pi}\nonumber\\
+&&g(\sigma^2+{\bf\pi}\cdot{\bf\pi}-f^2_{\pi})^2+m_{\pi}^2(f_{\pi}-
\sigma) +
{1\over 2}(\mbox{\boldmath $\nabla\times A$})^2 -
 {\bf j}\cdot {\bf A}\, ;
\end{eqnarray}
$\bf j$ is the external current.
We have used cylindrical coordinates with \mbox{\boldmath$\rho$}
the two dimensional vector normal to the $z$ direction. We will study
this problem in the limit of very large $g$, where the radial degree
of freedom of the chiral field is frozen out and we may write
\begin{eqnarray}
\sigma&&=f_{\pi}\cos\chi\, ,\nonumber\\
\pi_0&&=f_{\pi}\sin\chi\cos\theta\, ,\nonumber\\
\pi_x&&=f_{\pi}\sin\chi\sin\theta\cos\phi\, ,\nonumber\\
\pi_y&&=f_{\pi}\sin\chi\sin\theta\sin\phi\, .\label{angvar}
\end{eqnarray}
In terms of these variables the Hamiltonian density becomes
\begin{eqnarray}
H=&&{{f^2_{\pi}}\over 2}(\mbox{\boldmath$\nabla$}\chi)^2+
{{f^2_{\pi}}\over 2}\sin^2\chi(\mbox{\boldmath$\nabla$}\theta)^2+
{{f^2_{\pi}}\over2}
\sin^2\chi\sin^2\theta(\mbox{\boldmath$\nabla$}\phi- e{\bf A})^2
\nonumber\\
+&&m^2_{\pi}f^2_{\pi}(1-\cos\chi)+
{1\over 2}(\mbox{\boldmath $\nabla\times A$})^2
-{\bf j}\cdot {\bf A}\, .\label{hamdens}
\end{eqnarray}
The angular field $\phi$ can be eliminated by a gauge transformation.
For a current along a long wire we have
\begin{equation}
{\bf j}=I\delta(\mbox{\boldmath $\rho$}){\bf z}\, ;
\end{equation}
The vector potential will point along the $z$ direction, ${\bf
A}=A{\bf z}$ and the
fields will depend on the radial coordinate only. The equations of
motion become
\begin{eqnarray}
-\mbox{\boldmath$\nabla$}^2\chi+
\sin\chi\cos\chi(\mbox{\boldmath$\nabla$}\theta)^2+
e^2\sin\chi\cos\chi\sin^2\theta A^2+m^2_{\pi}\sin\chi=&&0\, ,\nonumber\\
\mbox{\boldmath$\nabla$}(\sin^2\chi\mbox{\boldmath$\nabla$}\theta)
+e^2\sin^2\chi
\sin\theta\cos\theta A^2=&&0\, ,\nonumber\\
-\mbox{\boldmath$\nabla$}^2A+e^2f^2_{\pi}\sin^2\chi\sin^2\theta A
-I\delta(\mbox{\boldmath $\rho$})=&&0\, .\label{eqmot}
\end{eqnarray}
In the absence of the chiral field the last of the Eqs.\
(\ref{eqmot}) gives the classical vector potential due to a long wire
\begin{equation} A={I\over 2\pi}\ln{\rho\over a}\, ,
\end{equation}
with $a$ an ultraviolet cutoff. The energy per unit length in the
$z$ direction associated with this configuration is
\begin{equation}
E={I^2\over 4\pi}\ln{R\over a}\, ,\label{classener}
\end{equation}
where $R$ is the transverse extent of space (an infrared cutoff).

Before discussing the solutions of (\ref{eqmot}) it is instructive to
look at the case where there is no explicit chiral symmetry breaking,
$m_{\pi}=0$. The solution that eliminates the infrared divergence in
Eq.\ (\ref{classener}) is $\chi=\theta=\pi /2$ and $A$ satisfying
\begin{equation}
-\mbox{\boldmath$\nabla$}^2A+e^2f^2_{\pi}A
-I\delta(\mbox{\boldmath $\rho$})=0\, .
\end{equation}
For any current the field $A$ is damped for distances $\rho >
1/ef_{\pi}$ and there is no infrared divergence in the energy. (Aside
from the fact that chiral symmetry is broken explicitly, the reason
the above discussion is only of pedagogical value is that the
coupling of the pions to the quantized electromagnetic field does break
the $SU(2)\times SU(2)$ symmetry into $SU(2)\times U(1)$ and the charged
pions get a light mass, $m_{\pi}\sim 35$ MeV \cite{pdg}, even in the
otherwise chiral symmetry limit.)

The term in Eq.\ (\ref{hamdens}) responsible for the pion mass prevents
us from setting $\chi=\pi/2$ everywhere; the energy density would behave
as $\pi f^2_{\pi}m^2_{\pi}R^2$, an infrared divergence worse than that
due to the wire with no chiral field present. We expect that $\chi$
will vary from $\pi/2$ to $0$ as $\rho$ increases and that
asymptotically we will recover classical electrodynamics. Although we
cannot obtain a closed solution to Eqs.\  (\ref{eqmot}), if the
transition between $\chi=\pi/2$ and $\chi=0$ occurs at large $\rho$,
we can find an approximate solution. The approximation consists of
neglecting the $(\mbox {\boldmath$\nabla$}\chi)^2$ term in Eq.\
(\ref{hamdens}); we shall return to this shortly. The solution of
these approximate equations of motion is
\begin{eqnarray}
\chi&&=\left\{\begin{array}{ll}
              {\pi\over 2}\ \   & \mbox{for $\rho < \rho_0$}\\
              0     &   \mbox{for $\rho > \rho_0$}\, ,
             \end{array} \right. \nonumber\\
\theta&&={\pi\over 2}\, , \nonumber\\
A&&=\left\{\begin{array}{ll}
     -{I\over 2\pi}\left[ K_0(ef_{\pi}\rho)-
     {{I_0(ef_{\pi}\rho)K_0(ef_{\pi}\rho_0)}/
     I_0(ef_{\pi}\rho_0)}\right]\ \  & \mbox{for $\rho <\rho_0$}\\
   {I\over 2\pi}\ln{\rho\over\rho_0}  & \mbox{for $\rho >\rho_0$}\, ;
    \end{array} \right.
\end{eqnarray}
$\rho_0$ is a parameter to be
determined by minimizing the energy density of Eq.\  (\ref{hamdens}).
Note that for $\rho > \rho_0$ the vector potential as well as the
field return to values these would have in the absence of any chiral
fields and that for $\rho < \rho_0$ the magnetic field decreases
exponentially as $B\sim \exp (-ef_{\pi}\rho)$. The physical picture is
that, as in a superconductor, near $\rho=0$ a cylindrical current sheet
is set up that opposes the current in the wire and there is a return
current near $\rho=\rho_0$; Amp\`{e}re's law insures that the field at
large distances is as discussed above.
The energy density for the above configuration, neglecting
the spatial variation of $\chi$, is
\begin{equation}
H=-{I^2\over 4\pi}\left [{K_0(ef_{\pi}\rho_0)\over
I_0(ef_{\pi}\rho_0)}+\ln(ef_{\pi}\rho_0)\right ] +
\pi m^2_\pi f^2_\pi \rho^2 +\cdots\, ,\label{apprhamdens}
\end{equation}
where the dots represent infrared and ultraviolet regulated terms
which are, however, independent of $\rho_0$. For $\rho_0>1/ef_\pi$
the term involving the Bessel functions may be neglected and
minimizing the rest with respect to $\rho_0$ yields
\begin{equation}
\rho_0={I\over 2{\sqrt 2}\pi m_\pi f_\pi}\, .\label{rho0}
\end{equation}
This is the main result of this work.

We still have to discuss the validity of the two approximations we
have made. The neglect of the Bessel functions in Eq.\
(\ref{apprhamdens}) is valid for $ef_{\pi}\rho_0 > 1$ which in
turn provides a condition on the current $I$, $eI/m_{\pi} > 2{\sqrt
2}\pi$ or more generally
\begin{equation}
I/m_{\pi} >> 1\, . \label{condition}
\end{equation}
The same condition permits us to
neglect the spatial variation of $\chi$ around $\rho=\rho_0$. Let
$\chi$ vary from $\pi/2$ to $0$ in the region $\rho-d/2$ to
$\rho+d/2$, with $1/d$ of the order of $f_{\pi}$ or $m_{\pi}$. The
contribution of the variation of $\chi$ to the energy density is
$\Delta H=\pi^3f^2_{\pi}\rho_0 d$. Eq.\  (\ref{condition}) insures that
$\Delta H$ is smaller than the other terms in Eq.\
(\ref{apprhamdens}).

Eq.(\ref{rho0}) has a very straightforward explanation. It results
from a competition of the magnetic energy density ${1\over 2}B^2$
and the energy density of the pion mass term $m^2_\pi f^2_\pi
(1-\cos\chi)$.  The magnetic field due to the current $I$ is
$B={I/ 2\pi\rho}$ and the transition occurs at $B=B_c$, with
$B_c={\sqrt 2}m_\pi f_\pi$. The reader may worry that the magnetic
fields very close to such thin wires are so large as to invalidate
completely the use of the chiral model as a low energy effective
QCD theory. In order to avoid this problem we may consider the
field due to a solenoid of radius $R$. The field is zero outside
the solenoid, ${\bf B}=B(\rho){\bf z}$ inside with $B(R)=B_0$. At
no point does the field become unboundedly large. Using the same
approximations as previously we obtain the following solutions of
the equations of motion (for $B_0\ge B_c$)
\begin{eqnarray}
\chi&&=\left\{\begin{array}{ll}
              0  &  \mbox{for $\rho > R$}\\
              {\pi\over 2}\ \   & \mbox{for $R > \rho > \rho_0$}\\
              0     &   \mbox{for $\rho < \rho_0$}\, ,
             \end{array} \right. \nonumber\\
\theta&&={\pi\over 2}\, , \nonumber\\
B&&=\left\{\begin{array}{ll}
      0 & \mbox{for $\rho > R$}\\
   a_1 K_0(ef_{\pi}\rho)+a_2 I_0(ef_{\pi}\rho)
\ \    &  \mbox{for $R > \rho > \rho_0$}\\
   B_c  &  \mbox{for $\rho <\rho_0$}\, ;
    \end{array} \right.
\end{eqnarray}
Continuity of the vector potential determines the coefficients $a_1$ and
$a_2$,
\begin{eqnarray}
a_1&=&\left [B_0I_1(ef_{\pi}\rho_0)-B_cI_1(ef_{\pi}R)\right ]/
D(R, \rho_0)\, ,\nonumber\\
a_2&=&\left [B_0K_1(ef_{\pi}R)-B_cK_1(ef_{\pi}\rho_0)\right ]/
D(R, \rho_0)\, ;
\end{eqnarray}
$D(R, \rho_0)=K_1(ef_{\pi}R)I_1(ef_{\pi}\rho_0)-K_1(ef_{\pi}\rho_0)
  I_1(ef_{\pi}R)$ and $\rho_0$ is determined, once more, by minimizing the
energy. For $(R,\ \rho_0) > 1/ef_{\pi}$
\begin{equation}
\rho_0=
    R-{1\over {ef_{\pi}}}\ln\left (B_0/B_c\right )\, .
\end{equation}
For $B_0\le B_c$, $\chi=0$ everywhere and $B(\rho)=B_0$ in the interior
of the solenoid. Thus, for any current configuration, the chiral fields
will adjust themselves to screen out fields larger than $B_c$. Topological
excitations may occur in the form of magnetic vortices; the angular field
$\phi$ of Eq.\ (\ref{angvar}) will wind around a quantized flux tube of
radius $1/ef_\pi$ \cite{FetWal}.

\section{The effect of strong \mfs on the electroweak vacuum}\label{sec:EW}

It was pointed out by Ambj\/orn and Olesen \cite{AmbOle} (see also
Ref.\cite{Skalozub85}) that the Weinberg-Salam model of
electroweak interactions shows an instability at $B \simeq
10^{24}~$ Gauss. The nature of such instability can be understood
by looking at the expression of the energy of a particle with
electric charge $e$, and spin $\bf s$, moving in homogeneous
magnetic field $\bf B$ directed along the $z$-axis. As we already
discussed in Sec.\ref{sec:weakrates}, above a critical field $B_c =
m^2/e$ particle energy is discretized into Landau levels
\be
E_n^2 = k_z^2 + (2n + 1)e|{\bf B}| - 2e{\bf B}\cdot {\bf s} +
m^2~.
\ee
 We observe that energy of scalar ($s = 0$) and spinor
($s_z = \pm 1/2$) is always positive, and indeed no instability
arise in QED (it is possible to verify that quantum one-loop
corrections do not spoil this conclusion). In the case of vector
particles ($s_z = 0, \pm 1$), however, the lowest energy level ($n
= 0,~k_z = 0,~s_z = +1$) becomes imaginary for $B
> B_c$, which could be the signal of vacuum instability. The
persistence of imaginary values of the one loop corrected lowest
level energy \cite{AmbOle} seems to confirm the physical reality
of the instability.

As it is well known the Weinberg-Salam model contains some charged
vector fields, namely the $W^\pm$ gauge bosons. The coupling of
the $W_\mu$ field to an external electromagnetic field
$A^{ext}_\mu$ is given by
\begin{equation}
{\cal L}_{int} = -\frac{1}{4}\vert F^{ext}_{\mu \nu}\vert ^2 -
-\frac{1}{2}\vert D_\mu W_\nu - D_\nu W_\mu \vert^2
- m_W^2 W^\dag_\mu W^\mu - ie F^{ext}_{\mu \nu}W^\mu W^\nu
\end{equation}
with
\begin{equation}
D_\mu = \partial_\mu - ie A^{ext}_\mu~.
\end{equation}
The important term in the previous expression is the ``anomalous'' magnetic
moment term  $ie F^{ext}_{\mu \nu}W^\mu W^\nu$, which arises because the
non-Abelian nature of the $SU(2)$ component of the Weinberg-Salam model gauge
group structure. Due to this term the mass eigenvalues of
the $W$ Lagrangian becomes
\be
m^2 = m_W^2 \pm eB~.
\ee
As expected from the considerations in the above, a tachyonic mode appears
for $B > B_c$. The corresponding eigenvector for zero kinetic energy
is determined by solving the equation of motions
\begin{equation}
D_iW_j - D_jW_i = 0\quad i,~j = x, y~,
\end{equation}
where $W_{1,2} = W_x \pm iW_y$. Ambjorn and Olesen argued that a
suitable solution of this equation is
\begin{equation}
\vert W(x, y)\vert = \displaystyle e^{-\frac 1 4 m_W (x^2 + y^2)}~,
\end{equation}
corresponding to a vortex configuration where $W$-fields wind
around the $z$-axis. This configuration corresponds to the
Nielsen-Olesen vortex solution \cite{NieOle73}. A similar
phenomenon should also take place for $Z$ bosons. Given the
linearity of the equations of motion it is natural to assume that
a superpositions of vortices is formed above the critical field.
This effect resemble the behaviour of a type-II superconductor in
the presence of a critical field magnetic field. In that case
$U(1)$ symmetry is locally broken by the formations of a lattice
of Abrikosov vortices in the Cooper-pairs condensate through which
the \mf can flow. In the electroweak case this situation is
reversed, with the formation of a $W$ condensate along the
vortices. Concerning the back-reaction of the $W$ condensate on
the magnetic field, an interesting effect arises. By writing the
electric current induced by the $W$ fields
\begin{equation}
j_\mu(W) = 2ie \left( W^\dag D_\mu W - W D_\mu W^\dag \right)~,
\end{equation}
Ambj\/orn and Olesen noticed that its sign is opposite to the current
induced by the Cooper pairs in a type-II superconductor, which is responsible
for the Meissner \mf screening effect. Therefore, they concluded that the
$W$-condensate  induce {\it anti-screening} of the external magnetic field.

Although the Higgs field $\Phi$ does not couple directly to the
electromagnetic field (this is different from the case of a
superconductor where the Cooper-pairs condensate couples directly
to $A^{ext}_\mu$), it does  through the action of the $W$
condensate. This can be seen by considering the Higgs, $W$
potential in the presence of the magnetic fields:
\begin{equation}
V(\phi, W) = 2\left( eB - m_W^2\right)\vert W \vert^2 +
g^2 \phi^2 \vert W \vert^2 - 2 \lambda \phi_0^2 \phi^2 +
2 g^2 |W|^4 + \lambda \left( \phi_+^4 \phi_0^4 \right)~.
\end{equation}
In the above $\phi_0$ and $\phi_+$ are respectively the Higgs
field vev and charged component, $g$ is the SU(2) coupling
constant, and $\lambda$ is the Higgs the self-interaction coupling
constant. We see that the $W$-condensate influences the the Higgs
field at  classical level due to the $\phi^2 \vert W \vert^2$
term. It is straightforward  to verify that if $eB < m_W^2 = \frac
1 2 g^2 \phi_0^2$ the minimum of $V(\phi, W)$ sits in the standard
field value $\phi = \phi_0$ with no $W$ condensate. Otherwise a
$W$ condensate is energetically favoured with the minimum of the
potential sitting in
\begin{equation}
\phi_{min}^2  = \phi_0^2~\frac{m_H^2 - eB}{m_H^2 - m_W^2}~
\end{equation}
where
\begin{equation}
m_H^2 \equiv 4\lambda \phi_0^2,\qquad m_W^2 \equiv \frac 1 2 g^2 \phi_0^2~.
\end{equation}
We see that the Higgs expectation value will vanish as the average
\mf strength approaches zero, provided the Higgs mass is larger than the
$W$ mass. This seems to suggest that a $W$-condensate should exist for
\begin{equation}
m_W^2 < eB < m_H^2~,
\end{equation}
and that the $SU(2) \times U_Y(1)$ symmetry is restored above
$H^{(2)}_c \equiv m_H^2/e$. Thus, anti-screening  should produce
restoration of the electroweak symmetry in the core of $W$
vortices. If $m_H < m_W$ the electroweak vacuum is expected to
behave like a type I superconductor with the formation of
homogeneous $W$-condensate above the critical magnetic field. The
previous qualitative conclusion have been confirmed by analytical
and numerical computations performed for $m_H = m_W$ in
Ref.\cite{AmbOle}, and for arbitrary Higgs mass in
Refs.\cite{Skalozub87,McDTor}.

A different scenario seems, however, to arise if thermal
corrections are taken into account. Indeed, recent finite
temperature lattice computations \cite{KajLPRS} showed no evidence
of the Ambjorn and Olesen phase. According to Skalozub and Demchik
\cite{SkaDem} such a behaviour may be explained by properly
accounting the contribution of Higgs and gauge bosons daisy
diagrams to the effective finite temperature potential.

In conclusion, it is quite uncertain if the Ambj\/orn and Olesen
phenomenon was really possible in the early Universe.

\subsection{The electroweak phase transition in a magnetic field}

We shall now consider the possible effects of strong \mfs on the
electroweak phase transition (EWPT).
As it is well known, the properties of the EWPT are determined by
the Higgs field effective potential. In the framework of the
minimal standard model (MSM), taking into account radiative
corrections from all the known particles and for finite temperature effects,
one obtains that
\be\label{veff}
  V_{\rm eff}(\phi , T) \simeq
-\frac{1}{2}(\mu^2 - \alpha T^2)\phi^2
  -  T \delta \phi^3 + \frac{1}{4}(\lambda - \delta\lambda_T )\phi^4~.
\ee
where $\phi$ is the radial component of the Higgs field and $T$ is the
temperature (for the definitions of the coefficients see {\it e.g.}
Ref.\cite{ElmEK}).

A strong hypermagnetic field can produce corrections to the
effective potential as it affects the charge particles propagators
(see below). There is, however, a more direct and simpler effect of
magnetic and hypermagnetic fields on the EWPT which was recently
pointed-out by Giovannini and Shaposhnikov \cite{GioSha} and by
Elmfors, Enqvist and Kainulainen \cite{ElmEK}. The authors of
Refs.\cite{GioSha,ElmEK}
noticed that hypermagnetic fields affect the  Gibbs free  energy
(in practice the pressure) difference between the broken and the
unbroken phase, hence the strength of the transition. The effect
can be understood by the analogy with the Meissner effect, {\it
i.e.} the expulsion of the magnetic field from superconductors as
consequence of photon getting an effective mass inside the
specimen. In our case, it is the $Z$--component of the hypercharge
$U(1)_Y$ magnetic field which is expelled from the broken phase.
This is just because $Z$--bosons are massive in that phase. Such a
process has a cost in terms of free energy. Since in the broken
phase the hypercharge field decomposes into
\begin{equation}\label{Ay}
  A^Y_\mu = \cos\theta_w A_\mu - \sin\theta_w Z_\mu~,
\end{equation}
we see that the Gibbs free energy in the broken and unbroken
phases are
\bea
\label{GbGu}
        G_b&=&V(\phi)-\frac 1 2 \cos^2\theta_w (B^{ext}_Y)^2~,\\
        G_u&=&V(0)-\frac 1 2 (B^{ext}_Y)^2~.
\eea
where $B^{ext}_Y$ is the external hypermagnetic field.
 In other words, compared to the case in which no magnetic field is
present, the energy barrier between unbroken and broken phase,
hence the strength of the transition, is enhanced by the quantity
$\displaystyle \frac{1}{2}\sin^2\theta_w (B^{ext}_{Y})^2$.
According to the authors of Refs.\cite{GioSha,ElmEK} this effect
can have important consequence for baryogenesis.

In any scenario of baryogenesis it is crucial to know at which
epoch do the sphaleronic transitions, which violate the sum ($B +
L$) of the baryon and lepton numbers, fall out of thermal
equilibrium. Generally this happens at temperatures below
$\bar{T}$ such that \cite{Shap86}
\be
\frac{E(\bar{T})}{\bar{T}} \ge A\,\,,
\label{washout}
\ee
where $E(T)$ is the sphaleron energy at the temperature $T$ and
$ A \simeq 35 - 45 $, depending on the poorly known prefactor of the
sphaleron rate.
In the case of baryogenesis at the electroweak scale one requires the
sphalerons to drop out of thermal equilibrium soon after the electroweak
phase transition. It follows that the requirement $\bar{T}=T_c$, where $T_c$
is the critical temperature, turns eq. (\ref{washout}) into a lower bound
on the Higgs vacuum expectation value (VEV),
\be
\frac{v(T_c)}{T_c} \geq 1\,. \label{first}
 \ee
As we already discussed, it is by now agreed \cite{Kajantie96} that the
standard model (SM) does not have a phase transition strong enough
as to fulfill Eq.(\ref{first}), whereas there is still some room
left in the parameter space of the minimal supersymmetric standard
model (MSSM) \cite{MSSM}.

The interesting observation made in Refs.\cite{GioSha,ElmEK} is that
a  magnetic field for the hypercharge $U(1)_Y$ present for $T>T_c$
may help to fulfill Eq.(\ref{first}). In fact,  it follows from the
Eqs.(\ref{GbGu}), that in presence of the magnetic field the critical
temperature is defined by the expression
\be
\label{Tcrit}
        V(0,T_c)- V(\phi,T_c) =
         \frac{1}{2}\sin^2\theta_w (B^{ext}_{Y}(T_c))^2~.
\ee This expression implies a smaller value of $T_c$ than that it
would take in the absence of the magnetic field, hence a larger
value of the ratio (\ref{first}).

Two major problems, however, bar the way of this intriguing
scenario. The first problem is that by affecting fermion, Higgs
and gauge field propagators, the hypermagnetic field changes the
electroweak effective potential in a nontrivial way. Two different
approaches have been used to estimate the relevance of this kind of
effects based either on lattice simulations \cite{Kajantie96} or
analytical computations \cite{SkaDem}. Both approaches agreed in
the conclusion that for a Higgs field mass compatible with the
experimental constraints ($m_H > 75~$ GeV), and for field
strengths $B,~B_Y \simleq 10^{23}~$ G, the standard model EWPT is
second order or a cross-over. Although this negative result could,
perhaps, be overcome by adopting a supersymmetrical extension of
the standard model (see e.g. Ref.\cite{MSSM}), a second, and more
serious problem arises by considering the effect of the \mf on
the anomalous processes (sphalerons) which are responsible for
lepton and baryon violation at the weak scale. This effect will be
the subject of the next section.

\subsection{Sphalerons in strong magnetic fields}\label{sec:sphalerons}

The sphaleron, is a static and unstable solution of the
field equations of the electroweak model, corresponding to the top
of the energy barrier between two topologically distinct vacua \cite{KliMan}.
In the limit of vanishing Weinberg angle, $\theta_w \to 0$, the
sphaleron is a spherically symmetric, hedgehog-like configuration
of $SU(2)$ gauge and Higgs fields. No direct coupling of the sphaleron
to a magnetic field is present in this case. As $\theta_w$ is turned on, the
$U_Y(1)$ field is
excited and the spherical symmetry is reduced to an axial symmetry.
A very good approximation to the exact solution is
obtained using the Ansatz by Klinkhamer and Laterveer \cite{KliLat},
which requires four scalar functions of $r$ only,
 \bea &&\displaystyle
g^\prime a_i \, dx^i = (1 - f_0(\xi))\, F_3 \,,\nonumber\\ &&
\displaystyle gW^a_i \sigma^a \, dx^i = (1-f(\xi)) (F_1 \sigma^1 +F_2 \sigma^2)
+ (1-f_3(\xi)) F_3 \sigma^3 \,,\nonumber\\
&&{\bf \Phi} = \frac{v}{\sqrt{2}} \left(
\begin{array}{c}
\displaystyle 0 \\
\displaystyle h(\xi)
\end{array}
\right)\,\,, \label{ansatz}
 \eea
 where $g$ and $g^\prime$ are the
$SU(2)_L$ and $U(1)_Y$ gauge couplings, $v$ is the Higgs VEV such
that $M_W= g v/2$, $M_h = \sqrt{2 \lambda} v$, $\xi=gvr$,
$\sigma^a$ ($a= 1,2,3$) are the Pauli matrices, and the $F_a$'s
are 1-forms defined in Ref. \cite{KliMan}. The boundary conditions for
the four scalar functions are \bea
 f(\xi)\,,\,f_3(\xi)\,,\,h(\xi)
\to 0\ \ \ \ f_0(\xi)\to 1\;\; \qquad &&{\mathrm for}\,\;\, \xi\to
0 \nonumber \\
 f(\xi)\,,\,f_3(\xi)\,,\,h(\xi)\,,\,f_0(\xi) \to 1
\qquad &&{\mathrm for}\,\;\, \xi\to \infty\,.\label{bcinf}
\eea

 It is known \cite{KliMan,KliLat} that for $\theta_w \neq 0$ the sphaleron
has some interesting electromagnetic properties. In fact,
differently from the pure $SU(2)$ case, in the physical case a
nonvanishing hypercharge current $J_i$ comes-in. At the first
order in $\theta_w$, $J_i$ takes the form
\be
J_i^{(1)} = -  \half g^\prime v^2 \frac{h^2(\xi) [1 -
f(\xi)]}{r^2}\, \epsilon_{3ij} x_j \;, \label{cur_0} \ee where $h$
and $f$ are the solutions in the $\theta_w \to 0$ limit, giving
for the dipole moment
\be
\mu^{(1)} = \frac{2 \pi}{3}  \frac{g^\prime}{g^3 v} \int_0^\infty d\xi
\xi^2 h^2(\xi) [1-f(\xi)]\;.
\label{mu1}
\ee
The reader should note that the dipole moment is a true electromagnetic one
 because in the broken phase only the electromagnetic component of
the hypercharge field survives at long distances.
\begin{figure}[ht]
\vskip -1.cm
\centerline{\protect \hbox{
\epsfig{file=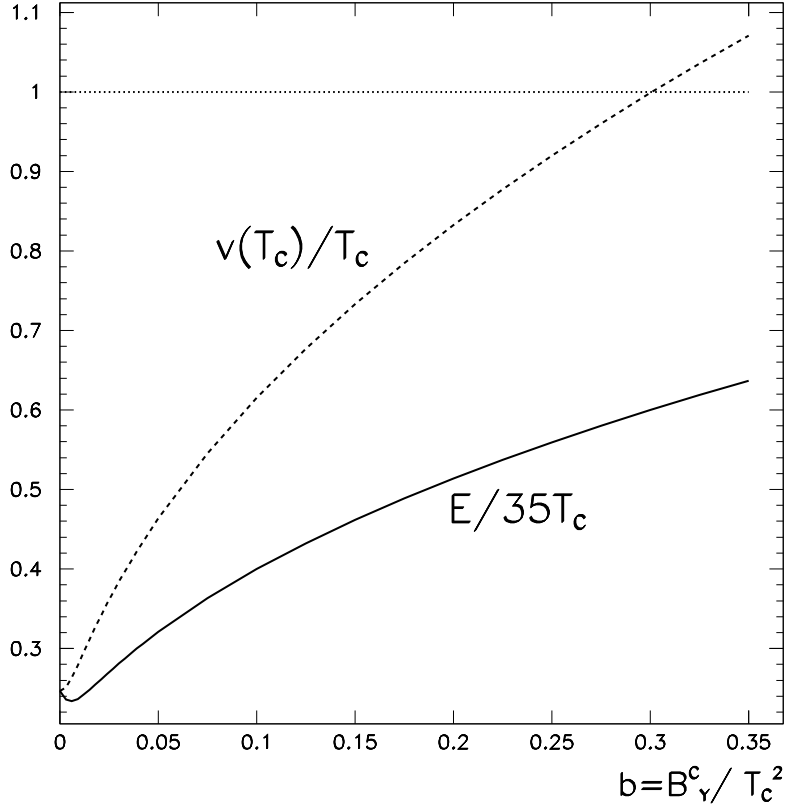,
width=12.cm,angle=0}}}
\vskip -0.5cm
\caption{The VEV at the critical temperature, $v(T_c)$, and the
sphaleron energy {\em vs.} the external magnetic field for
$M_h=M_W$. We see that even if $v(T_c)/T_c \simgeq 1$ the washout condition
$E/T_c \simgeq 35$ is far from being fulfilled. From Ref.\cite{labanda}}
\label{fig:washout}
\end{figure}

Comelli et al. \cite{labanda} considered what happens to the
sphaleron when an external hypercharge magnetic field,
$B_Y^{ext}$, is turned on. They found that the energy functional
is modified as
\be
E = E_0 - E_{\mathrm dip}\,,
\label{energy}
\ee
with
\be
\displaystyle
  E_0= \int d^3x \left[ \frac{1}{4} F^a_{ij}F^a_{ij} +
\frac{1}{4} f_{ij}f_{ij} + (D_i {\bf \Phi})^\dagger (D_i {\bf \Phi}) +
V({\bf \Phi}) \right]
\ee
and
\be
\label{intenergy} E_{\mathrm dip} = \int d^3x J_i A^Y_i = \half
\int d^3x f_{ij}f^c_{ij} \ee with $f_{ij} \equiv \partial_i A^Y_j
-  \partial_j A^Y_i$. A constant external hypermagnetic field
$B^{ext}_Y$ directed along the $x_3$ axis was assumed. In the
$\theta_w\to 0$ limit the sphaleron has no hypercharge
contribution and then $E_{\mathrm dip}^{(0)} = 0 $. At
$O(\theta_w)$, using (\ref{cur_0}) and (\ref{mu1}) the authors of
Ref.\cite{labanda} got a simple
magnetic dipole interaction energy
\be
E_{\mathrm dip}^{(1)} = \mu^{(1)} B^{ext}_Y \; .
\label{lino}
\ee
In order to assess the range of validity of the approximation
(\ref{lino}) one needs to go beyond the leading order in
$\theta_w$ and look for a nonlinear $B^{ext}_Y$--dependence of $E$.
This requires to solve the full set of  equations of motion for
the gauge fields and the Higgs in the presence of the external
magnetic field. Fortunately, a uniform  $B^{ext}_Y$ does not spoil the
axial symmetry of the problem. Furthermore, the equation of motion
are left unchanged ($\partial_i f^{ext}_{ij} =0$) with respect to the
free field case. The only modification induced by $B^{ext}_Y$ resides
in the boundary conditions since -- as $\xi \to \infty$ -- we now
have
\be
 f(\xi)\,,h(\xi) \to 1\,,\;\;\;\;\;\;\;\; f_3(\xi)\,,f_0(\xi) \to
 1-B^{ext}_Y \sin 2\theta_w \frac{\xi^2}{8 g v^2}
\ee
whereas the boundary condition for $\xi \to 0$ are left unchanged.

The solution of the sphaleron equation of motions with the
boundary conditions in the above were determined numerically by
the authors of Ref.\cite{labanda}. They showed that in the considered
$B^{ext}_Y$--range the corrections to the linear approximation
\[
\Delta E \simeq \mu^{(1)} \cos \theta_w B^{ext}_Y
\]
are less than $5 \%$. For larger values of $B^{ext}_Y$ non-linear
effects increase sharply. However, as we discussed in the previous
sections, for such large magnetic fields the broken phase of the SM
is believed to become unstable to the formation either of
$W$-condensates \cite{AmbOle} or of a mixed phase \cite{KajLPRS}.
In such situations the sphaleron solution  does not exist any
more. Therefore, it is safe to limit the previous analysis to
values $B^{ext}_Y \leq 0.4~T^2$.

The reduction of sphaleron energy due to the interaction with the field
$B^c_Y$ has relevant consequences on the sphaleronic transition rate
which is increased with respect to free field case. As a consequence,
in an external magnetic field the relation between the Higgs VEV and the
sphaleron energy is altered and Eq.(\ref{first}) does not imply
(\ref{washout}) any more. We can understand it by considering the linear
approximation to $E$,
\be
E\simeq E(B^c_Y=0) - \mu^{(1)} B^{ext}_Y \cos \theta_w \equiv \frac{4 \pi v}{g}
\left(\varepsilon_0 - \frac{\sin 2\theta_w}{g} \frac{B^{ext}_Y}{v^2} m^{(1)}
\right)
\label{newwash}
\ee
where $m^{(1)}$ is the $O(\theta_W)$ dipole moment expressed in
units of $e/\alpha_W M_W(T)$.
From the Fig.\ref{fig:washout} we see that even if $v(T_c)/T_c \geq 1$ the
washout condition $E/T_c \geq 35$ is far from being fulfilled.

It follows form the previous considerations, that even if a strong
\mfs might increase the strength of the EWPT, such an effect would not
help baryogenesis.


\chapter*{Conclusions}

\addcontentsline{toc}{chapter}{Conclusions}

In this review we have analyzed a large variety of aspects of
magnetic fields in the early Universe. Our exposition followed an
inverse-chronological order. In the first part of
Chap.\ref{chap:chap1} we discussed what observations tell us about
recent time fields and their evolution in galaxies and clusters of
galaxies. As we have seen, a final answer about the origin of
these fields is not yet available. Several arguments, however,
suggest that galactic and cluster fields were preexisting, or at
least contemporary to, their hosts. The main reasons in favor of
this thesis are: the ubiquity of the fields and the uniformity of
their strength; the theoretical problems with the MHD
amplification mechanisms, especially to explain the origin of
cluster fields; the observation of $\mu$G \mfs in high-redshift
galaxies.
It is reassuring that new ideas continuously appear to determine
these fields at all times. For example, a very recent one by Loeb and Waxman
proposes to look for fluctuations in the radio background from
intergalactic synchrotron emission of relativistic electrons interacting
with CMBR \cite{EliLoeb}.

A consistent and economical mechanism which may naturally explain
the early origin of both galactic and cluster magnetic fields, is
the adiabatic compression of a primeval field with strength in the
range $B_0 \sim 10^{-9} - 10^{-10}$ Gauss ($B_0$ is the intensity
that the primordial field would have today under the assumption of
adiabatic decay of the field due to the Hubble expansion). If this
was the case, two other interesting effects may arise: \ {\it a)}
\mfs may have affected structure formation perhaps helping to
solve some of the problems of the CDM scenario; \ {\it b)} \mfs
can have produced observable imprints in the CMBR. Given the
current theoretical uncertainties about the MHD of galaxies and
clusters, and the preliminary status of N-body simulations in the
presence of magnetic fields, the most promising possibility to
test the primordial origin hypothesis of cosmic \mfs comes from
the forthcoming observations of the CMBR anisotropies.

In Chap.\ref{chap:cmb} we reviewed several possible effects of
\mfs on the CMBR.  In first place, we showed that \mfs may affect
the isotropy of the CMBR. A \mf which is homogeneous through the
entire Hubble volume, would spoil Universe isotropy giving rise to
a dipole anisotropy in the CMBR. On the basis of this argument it
was shown that COBE measurements provide an upper limit on the
present time equivalent strength of a homogeneous cosmic \mf which
is roughly $3 \times 10^{-9}$ G. More plausibly, \mfs are tangled
on scales much smaller than the Hubble radius. In this case the
effect on the Universe geometry is negligible and much more
interesting effects may be produced on small angular scales. Some
of these effects arise as a consequence of MHD modes appearing in
the magnetized photon-baryon plasma in place of the usual acoustic
modes. The amplitude and velocity of the MHD modes depend on the
\mf intensity and spatial direction. Some of these modes are quite
different from standard scalar and tensor modes which are usually
considered in the theoretical analysis of the CMBR distortions.
For example, Alfv\'en waves have the peculiar property of not
beeing depleted by the Universe expansion inspite of their vectorial
nature. These modes are well suited to probe perturbations as
those generated by cosmic defects and primordial phase
transitions. Tangled magnetic fields, whose production is
predicted by several models and, which are observed in the
intercluster medium, are also expected to produce Alfv\'en waves.
Another interesting aspect of this kind of isocurvature
perturbations, is that they are not affected by Silk damping. The
polarization power spectrum of CMBR can also be affected by
primordial magnetic fields. This a consequence of the Faraday
rotation produced by the field on the CMB photons on their way
through the last scattering surface. Magnetic fields with strength
$B_0 \simgeq 10^{-9}$ G may have produced a detectable level of
depolarization. Furthermore, it was shown that because of the
polarization dependence of the Compton scattering, the
depolarization can feed-back into a temperature anisotropy. It was
concluded that the best strategy to identify the imprint of
primordial \mfs on the CMBR is probably to look for their
signature in the temperature and polarization anisotropies
cross-correlation. This method may probably reach a
sensitivity of $\sim 10^{-10}$ G for the present time equivalent
\mf strength when the forthcoming balloon and
satellite missions data is analyzed.
Some results from the Boomerang and Maxima experiments \cite{boomerang}
are already out with surprising results.
It is too early to decide the reasons why, if experimentally confirmed,
the second peak is low. We have verified, together with Edsj\"o \cite{Edsjo},
that magnetic fields can only
decrease the ratio of amplitude of the second peak with respect to the
first. Therefore the effect cannot be explained in terms of primordial
magnetic fields. Interesting constraints on the strength of these fields
will be available only when the amplitude of several peaks and the
polarization of CMBR will be measured.

Another period of the Universe history when primordial \mfs may
have produced observable consequences is the big-bang
nucleosynthesis. This subject was treated in Chap.\ref{chap:bbn}.
Three main effects have been discussed: the effect of the \mf
energy density on the Universe expansion; the modification
produced by a strong \mf of the electron-positron gas
thermodynamics; the modification of the weak processes keeping
neutrons and protons in chemical equilibrium. All these effects
produces a variation in the final neutron-to-proton ratio, hence
in the relative abundances of light relic elements. The effect of
the field on the Universe expansion rate  was showed to be
globally dominant, though the others cannot be neglected.
Furthermore, the non-gravitational effects of the \mf can exceed
that on the expansion rate in delimited regions where the \mf
intensity may be larger than the Universe mean value. In this case
these effects could have produced fluctuations in baryon to photon
ratio and in the relic neutrino temperature. Apparently, the
BBN upper bound on primordial magnetic fields, which is $B_0
\simleq 7 \times 10^{-7}$ G, looks less stringent than other
limits which come from the Faraday rotation measurements (RMs) of
distant quasars, or from the global isotropy of the CMBR. However,
we showed in Sec.\ref{sec:bbnconst} that this conclusion is not
correct if \mfs are tangled. The reason is that BBN probes length
scales which are of the order of the Hubble horizon size at BBN
time (which today corresponds approximately to 100 pc) whereas
CMBR and the RMs probe much larger scales. Furthermore,
constraints derived from the analysis of effects taking place at
different times may not be directly comparable if the \mf
evolution is not adiabatic.

In Chap.\ref{chap:generation} we reviewed some of the models which
predict the generation of \mfs in the early Universe. We first
discussed those models which invoke a first order phase
transition. This kind of transitions naturally provide some
conditions, as charge separation, out-of-equilibrium condition,
and high level of turbulence, which are known to be important
ingredients of magnetogenesis. We discussed the cases of the QCD
and the electroweak phase transitions (EWPT). In the case of the
EWPT some extension of the particles physics standard model has to
be invoked for the transition to be first order. Magnetic fields
may be generated during the EWPT from the non-trivial dynamics of
the gauge fields produced by the equilibration of the electrically
charged components of the Higgs field. This effect resembles the
Kibble mechanism for the formation of topological defects. It is
interesting that such a mechanism gives rise to magnetic fields, though
only on very small scales, even if the phase transition is second
order. In general, since the production of \mfs during a phase
transition is a causal phenomenon, the coherence length scale of
these fields at the generation time cannot exceed the horizon
radius at that time. Typically, once this length is adiabatically
re-scaled to present time, one gets coherence cell sizes which are
much smaller than those observed today in galaxies and the
inter-cluster medium. This problem may be eased by
the effect of the magnetic helicity which is expected to be
produced during primordial phase transitions.Helicity may help the
formation of large magnetic structures starting from small ones
(inverse cascade). Indeed, this is a quite common phenomenon in
MHD. Some estimates of the quantitative relevance of this effect
have been given,for example, in Sec.\ref{sec:evolution}.
We have seen that the QCD phase transition might indeed
give rise to phenomenological interesting values of the present
time \mf strength and coherence size but only assuming quite
optimistic conditions. Magnetic fields produced at the EWPT might
have played a role in the generation of galactic magnetic fields
only if they were amplified by a galactic dynamo. The problem with
the small coherence scale of magnetic fields produced in the early
Universe may be circumvented if the production mechanism was
not-causal. This may be possible if \mfs were produced during
inflation by the superadiabatic amplification of preexisting
quantum fluctuations of the gauge fields. This phenomenon,
however, can take place only if the conformal invariance of the
electromagnetic field is broken. In Sec.\ref{sec:inflation} we
have discussed several interesting mechanisms which have been
proposed in the literature to avoid this obstacle. Unfortunately,
although some results are encouraging, at the present status of
art, none of these model seems to offer any firm prediction.
Further work on the subject is, therefore, necessary.

Even if magnetic fields, produced during the electroweak phase transitions
or before, are nor the progenitor of galactic magnetic fields, they
may still have had other interesting cosmological consequences.
Perhaps the most intriguing possibility is that magnetic fields
played a role in the generation of the baryon asymmetry of the
Universe (BAU). The magnetic fields may influence electroweak
baryogenesis at two levels. At a first level, magnetic fields can
play an indirect role on electroweak baryogenesis by modifying the
free energy difference between the symmetric and broken phases,
the Higgs effective potential, and the rate of sphaleron, baryon
number violating, transitions. In Chap.\ref{chap:stability} we
showed, however, at this level, no significative  modifications
to arise with respect to the standard scenario. Magnetic fields, or
better their hypermagnetic progenitors, may have played a much
more direct role in the generation of the BAU if they possessed a
net helicity. Indeed, it is well known from field theory that the
hypermagnetic helicity coincides with the Chern-Simon number which
can be converted into baryons and leptons numbers by the Abelian
anomaly. The origin of the primordial magnetic helicity is still
matter of speculation. Among other possibilities which we have
review in Chap.\ref{chap:generation}, one of the more discussed in
the literature is that a net hypermagnetic helicity of the
Universe arise by an anomalous coupling of the gauge fields to an
oscillating pseudoscalar field. The existence of pseudoscalar
fields of this kind is required by several extensions of the
particle physics standard model.
However, it must be admitted that the mechanisms for generation
of fields that are large and extended at the same time are far
from being fully understood.
 \vskip .1cm
Large magnetic fields would also have a
profound effect on chirality  but it is also quite
far from observability. As a rule of the thumb, particle physics
effects will appear at the earliest at B =$~m_\pi^2$, the $\pi$
being the lightest hadron. This makes these effects difficult to
test. Large fields are required. The only hope is the existence of
superconductive strings.
\vskip .1cm
Are the observed fields, so widespread, of early origin
or some seeds fields were rapidly enhanced by a dynamo mechanism?
This question remains unanswered but high precision CMBR acoustic
peak measurements may very well provide a breakthrough.

\chapter*{Acknowledgements}

\addcontentsline{toc}{chapter}{Acknowledgements}

The authors thank J.~Adams, M.~Bander, P.~Coles, D.~Comelli, U.~Danielsson,
J.~Edsj\"o, F.~De Felice, A.~Dolgov, D.~Harari, A.~Loeb, M.~Pietroni,
G.~Raffelt, A.~Riotto, M.~Shaposhnikov, A.~Schwimmer, O.~T\"ornkvist,
T.~Vachaspati and E.~Waxman for several useful discussions.

This project was suggested to us by David Schramm. His untimely death
left us without his advice.


\end{document}